# Thioethyl-porphyrazine/Nanocarbon Hybrids for Photoinduced Electron Transfer

Sandra Belviso,*,a,b Andrea Capasso,c Ernesto Santoro,a Leyla Najafi,c,d Francesco Lelj,a,b Stefano Superchi,a Daniele Casarini,a Claudio Villani,e Davide Spirito,c,f Sebastiano Bellani,c Antonio Esau Del Rio Castillo,c and Francesco Bonaccorso*,c

aUniversità degli Studi della Basilicata, Dipartimento di Scienze, via dell'Ateneo Lucano, 10; 85100 Potenza, Italy

bLASCAMM, CR-INSTM Unità della Basilicata

cIstituto Italiano di Tecnologia, Graphene Labs, Via Morego 30; 16163 Genova, Italy

dUniversità degli Studi di Genova, Dipartimento di Chimica e Chimica Industriale, Via Dodecaneso 3; 16163 Genova, Italy

eUniversità degli Studi di Roma "La Sapienza", Dipartimento di Chimica e Tecnologie del Farmaco, Piazzale Aldo Moro, 5; 00185 Roma, Italy

fIstituto Italiano di Tecnologia, Nanochemistry Department, Via Morego 30; 16163 Genova, Italy

Corresponding authors
sandra.belviso@unibas.it; francesco.bonaccorso@iit.it

**ABSTRACT**

We designed a novel pyrene-substituted thioethyl-porphyrazine (**PzPy**) and the formation of supramolecular assembly with nanocarbons demonstrating photoinduced electron transfer ability. As revealed by spectroscopic and electrochemical studies, **PzPy** displays wide spectral absorption in the visible range, charge separation upon photoexcitation, as well as bandgap and highest occupied/lowest unoccupied molecular orbital (HOMO/LUMO) energy values, matching the key requirements of organic optoelectronic. Moreover, the presence of a pyrene moiety promotes attractive interactions with π-conjugated systems. In particular, theoretical calculations show that in the **PzPy** the HOMO and LUMO are localized on different positions of the molecule, *i.e.*, the HOMO on the pyrene moiety and the LUMO on the macrocycle. Therefore, HOMO-LUMO excitation gives rise to a charge separation, preventing excitons recombination. Two kinds of non-covalent hybrid composites are prepared by mixing the **PzPy** with single wall carbon nanotubes (SWNTs) and graphene nanoflakes (GNFs), respectively, and used for photocurrent generation through charge transfer processes between **PzPy** and nanocarbons. Photoconduction experiments show photocurrent generation upon visible light irradiation of both **PzPy**/SWNT and **PzPy**/GNF composites (0.78 and 0.71 mA/W at 500 nm, respectively), demonstrating their suitability for optoelectronic applications and light harvesting systems.

## 1. Introduction

The growing interest towards the development of new materials and technologies for optoelectronic application has fueled the research in organic materials, due to their low-cost high-throughput film manufacturing by solution-processed techniques (spin casting, inkjet printing, roll-to-roll and large-area deposition processes).[1-3] Moreover, organic films often features low-weight and high-flexibility, which are of interest in view of both scaling up prospective and device integrations, respectively.[1-5] Typically, photoactive organic films are based on electron donor-acceptor interactions, where photoexcitation of the donor promotes a charge separation accompanied by an electron transfer to the acceptor species.[1,5-7] The donors are usually semiconducting poly-aryl[6,7] and –heteroaryl[8] polymers or small molecules,[9,10] displaying high molar extinction coefficients -ε- (>$10^4$ M$^{−1}$cm$^{−1}$).[1-3] The acceptor systems are, instead, usually based on carbon nanomaterials.[11] In particular, fullerene derivatives such as [6,6]-phenyl-C$_{61}$-butyric acid methyl ester (PCBM) have been widely used for this purpose.[12] More recently, carbon nanotubes (CNTs) have also been reported as acceptors,[13-15] and the use of graphene is rapidly raising too.[16-21] In particular, single wall carbon nanotubes (SWNTs) have been proposed as acceptor material since they possess nanoscale diameter and large aspect ratio in combination with favorable electrical transport properties.[14,19] Furthermore, SWNTs show unique optoelectronic properties, electrochemical stability, and energy levels that can be suitably coupled to those of conjugated polymers with long-range charge transport properties to make efficient composites for optoelectronic devices.[19] Similarly, graphene has high electron and hole mobility at room temperature, with reported values exceeding 15000 $cm^2$ $V^{-1}$ $s^{-1}$.[20-22] Graphene is also electrochemically stable, with a very high specific surface area ($\sim$ 2630 $m^2$ $g^{-1}$), surpassing that of both SWNTs and graphite.[23,24] Moreover, graphene has an optical transmittance over the visible range of 97.7%,[25] and its work function ($\sim$ 4.5 eV)[26] can be matched to the energy levels of organic donor materials.[22] Moving from these motivations, both SWNTs and functionalized graphene have been used as electron acceptors in combination with conjugated polymer donors, such as poly(3-octylthiophene) (P3OT), or poly(p-phenylenevinylenes), and poly(3-hexylthiophene) (P3HT),[14,19,27,28] especially in organic bulk heterojunction (BHJ) solar cell devices.[19]

Amongst the non-polymeric donors a prominent role is occupied by tetrapyrrole macrocycles such as porphyrins[29] and phthalocyanines,[29,30] which, since the first organic solar cell reported by Tang in 1986 based on Cu-phthalocyanine,[31] have been widely used as p-type photoelectron donor dyes transferring charges to an n-type acceptor systems.[32-35] Recently, photoactive hybrid materials constituted by a tetrapyrrolic macrocycle donor linked to a nanocarbon acceptor (*e.g.*, CNTs) have been reported,[30,36,37] with the linkage occurring by either covalent bonding[38-43] or supramolecular interactions with the CNT surface.[44-62] The latter is the most promising approch, avoiding chemical modification of the CNT surface, which determines a modification of the CNT $sp^2$ network and, consequently, of its electronic structure.[63] Non covalent interactions between the tetrapyrrole dye and the CNT are usually mediated by pyrene units, which interact with the aromatic surface of the nanocarbon through π-π stacking bonding.[51,52,54-56] Notably, in the existing literature the pyrene is not directly linked to the macrocycle, but it is, instead, linked through an alkyl chain spacer[52,55] or benzo-fused to the macrocycle itself.[62] Recently, photoactive hybrid materials composed by porphyrins and phthalocyanines with graphene,[64-70] and its derivatives (*e.g.*, graphene oxide,[71] and reduced graphene oxide[72,73]) have also been described. In some of these examples,[64-66] the supramolecular interactions between the macrocycle and graphene[74] are mediated by the presence of a pyrene unit linked to the macrocycle through an alkyl chain spacer. These results highlight the fact that nanoassemblies of tetrapyrroles with either CNTs or graphene represent promising materials for the development of novel optoelectronic devices.

Seeking for new types of tetrapyrrole donors, there is a huge potential in porphyrazines,[75-76] which can be considered as structural hybrids of porphyrins and phthalocyanines, but have never been investigated for photocurrent generation so far. In particular, thioalkyl porphyrazines, having thioethereal moieties at the β-position of the macrocycle, are rather interesting. In fact, the presence of sulfur atoms imparts outstanding optical and electrochemical properties[77-80,81-83] to these compounds, promoting coordination sites out and above the molecular plane, favoring the presence of columnar discotic liquid-crystalline mesophases.[77-80] This is a relevant feature because highly ordered materials such as liquid crystals assist both exciton and charge carrier mobilities.[1,84-86] One of the key research effort in optoelectronic devices is devoted to increase the spectral response of the photoactive layer, obtaining panchromatic light absorbers, which can be ideal candidates for both photodetecting and energy harvesting applications. In the UV-visible light range porphyrins and phthalocyanines display narrow and strong ($\varepsilon \sim 3 \div 6 \times 10^5$ $M^{-1}$ $cm^{-1}$) absorption bands at about 400 nm and 680 nm, respectively, but only 10% of the overall absorption lays in the 450-600 nm range, where the solar emission reaches its maximum.[29,87] On the contrary, the absorption spectrum of thioalkyl porphyrazines displays bands having lower ε (~$1 \div 5 \times 10^4$ $M^{-1}$ $cm^{-1}$), but being much wider in the 300-800 nm range.[78] A higher total absorbance then results in the wavelength range corresponding to the maximum of the solar emission spectrum.[81,88,89]

Recently, we prepared several aryl and arylethynyl-substituted derivatives, in which both electron-donating and electron-withdrawing groups strongly perturbed the molecular charge distribution.[83] Computational Density Functional Theory (DFT) and Time Dependent Density Functional Theory (TDDFT) studies indicated that while in symmetrical porphyrazines both the highest occupied molecular orbital (HOMO) and the lowest unoccupied molecular orbital (LUMO) are localized on the macrocycle, in compounds with electron donating groups the HOMO is localized on the aryl substituent and LUMO on the macrocycle, while the opposite holds in the presence of electron withdrawing groups.[83] In the aryl-substituted systems, the macrocycle then displays an ambivalent character, behaving both as electron-acceptor and electron-donor depending on the substituents, giving rise to an unconventional "push-pull" system suitable for non-linear optics.[90] These features can have a relevant role also for photoactive materials because the presence of charge transfer HOMO-LUMO transitions and the location of these orbitals on different moieties of the molecules can, in principle, hinder exciton recombination enhancing its lifetime, then making these compounds promising electron donors. Moreover, all these aryl and arylethynyl-substituted porphyrazines show HOMO and HOMO-LUMO bandgap energies compatible with nanocarbons as acceptors.[91]

Taking into account the charge transfer properties of these compounds and the possibility to functionalize the porphyrazine macrocycles, we designed and prepared a novel pyrene-substituted thioalkyl porphyrazine **PzPy** (Scheme 1) to be exploited for photocurrent generation. In this compound, a macrocycle, which is a strong light absorber, and a pyrene unit, which is able to interact with the nanocarbons, are present simultaneously. Moreover, the direct bond between the macrocycle and pyrene moieties can, in principle, allow an electron exchange between them and, hopefully, promote a charge separation upon photoexcitation likewise the above mentioned aryl-substituted porphyrazines.[83] For these reasons, this compound is particularly suited for the formation of nanohybrids with CNTs and graphene. The preparation of **PzPy**/nanocarbon hybrids, the study of the supramolecular interactions within the composites, and the results of photoconduction experiments are reported herein. We prepared and fully characterized two different non-covalent hybrid composites by mixing the **PzPy** with either SWNTs or graphene nanoflakes (GNFs). We have then fabricated photodetectors, which show photocurrent generation upon visible light irradiation reaching 0.78 and 0.71 mA/W at 500 nm, for **PzPy**/SWNT and **PzPy**/GNF composites, respectively. To the best of our knowledge, these systems are the first example of hybrid composed by nanocarbons and alkyl porphyrazines and the first case in which graphene and pyrene-substituted tetrapyrroles nanohybrids are used to assemble photoactive devices. In fact, the only two examples reported in the literature so far concern the use of graphene and phthalocyanine poly(p-phenylene vinylene (PPV) oligomers,[69,70] contrary to our case in which single molecules of tetrapyrrole are used. Moreover, the use of graphene as electron acceptor in tetrapyrrole-based optoelectronics represents a breakthrough for the development of novel organic optoelectronic devices.

## 2. Results and Discussion

**2.1 Synthesis and structural characterization.** The pyrene-substituted thioethyl porphyrazine **PzPy** is prepared in a three-step procedure starting from the parent symmetrically substituted thioethyl porphyrazine **1** (Scheme 1).[92] The first step is an hydrogen replacement of one alkylsulfanyl tail leading to the non-symmetrical β-H-substituted porphyrazine **2**.[81] Such approach affords a very clean reaction, easy purification procedures, yielding non-symmetrically substituted product. The subsequent bromination of **2** leads to the bromo derivative **3**,[82] which, through Pd-catalysed Suzuki[93,94] coupling reaction,[79,83] provides the pyrene-substituted thioethyl porphyrazine **PzPy** in 45% yield.

Supramolecular interactions between aromatic systems and nanocarbons are usually determined by noncovalent π-π interactions[68,95] which, being highly directional,[96] are strictly dependent on shape and mutual orientation of the interacting molecules. Therefore, **PzPy** is fully characterized for its structural, conformational, and stereochemical features. The **PzPy** stereochemical aspects are particularly interesting, because in this molecule hindered rotation around the pyrene-macrocycle bond is expected to induce a twist of the pyrene moiety from the porphyrazine plane and then the presence of atropisomerism. This structural feature is relevant for the possible interactions with SWNTs. In fact, the presence of an axially chiral structure in **PzPy** could favour the interaction with semiconducting chiral CNTs, which are known to be optically resolved by chiral biaryls and bis-porphyrins.[97-102] Besides, the **PzPy** twisted arrangement should not impair the interaction with the 2D graphene structure, given that actractive arene-arene interactions are often edge-to-face and offset stacked, rather than face-to-face stacked.[103,104]

Density Functional Theory computations at M06/6-311G(d,p) level of theory, taking into account the solvent environment by IEFPCM solvation model,[105,106] clearly revealed the twist of the pyrene ring in the two most stable conformers, having a dihedral angle of 57° and 65° between the pyrene and macrocycle, respectively (see Figure 1 and Table S1 in Supplementary Information, S.I.). An arene-arene rotational barrier of 23.44 kcal mol$^{-1}$ was also calculated. Experimental measurements by variable temperature nuclear magnetic resonance (NMR) and variable temperature high performance liquid chromatography (HPLC) analyses[107-109] provided a rotational barrier of 22.0 ± 0.1 kcal mol$^{-1}$,[110] *i.e.*, lower than that obtained by calculations, a value which does not allow isolation of the single enantiomers at room temperature.[111,112] However, even if **PzPy** cannot be isolated in its enantiomeric forms at room temperature, its molecular axial chirality is nevertheless important because it can favour the interaction also with semiconducting racemic chiral nanotubes.

**2.2 Spectroscopic and electrochemical study.** The electronic properties of **PzPy** are investigated by experimental and theoretical analysis of its UV-vis spectrum. In Figure 2 are reported the experimental UV-vis spectrum of **PzPy** in $CH_2Cl_2$ together with those of the parent porphyrazine **1** and of 1-pyrene boronic acid, taken as a model of unconjugated pyrene chromophore. The UV-vis spectrum of **PzPy** shows the typical features of non-aggregated 'free-base' thioalkyl-porphyrazines, displaying a $Q_x(0, 0)$ band at 709 nm (14104 $cm^{-1}$)[89] and a broad $Q_y(0, 0)$ band at 640 nm (15625 $cm^{-1}$),[89] a Soret band near 350 nm (28571 $cm^{-1}$)[89] and in between a band at 513 nm (19493 $cm^{-1}$) associated to $n_{sulfur} \rightarrow \pi^*$ transitions.[83,88,89] Moreover, an absorption band peaked at 277 nm (44052 $cm^{-1}$), corresponding to one of the typical bands of the pyrene chromophore,[113,114] is clearly visible. Such band is allied to a long-axis polarized pyrene transition and the fact that it appears less resolved than in unconjugated pyrene indicates an electronic connection between the two aromatic moieties.[113] The spectral comparison with the parent unsubstituted porphyrazine **1** (Figure 2) evidences a marked increase of the Soret band, which cannot be simply ascribed to a contribution of the underneath characteristic pyrene features located at 330 and 345 nm (30303 and 28985 $cm^{-1}$),[113] which are far less intense, but can be associated to intramolecular delocalization.[113]

To get further insight into the electronic structure of **PzPy** a TDDFT computational investigation is carried out to provide an interpretation of the UV-vis spectral features. The UV-vis computed spectrum at the M06/6-311G(d,p)/$CH_2Cl_2$ level of theory in the 23500÷12500 $cm^{-1}$ (435-800 nm) range is characterized by twelve transitions $S_0 \rightarrow S_n$ (n=1,12) in case of conformers 1 and 2 (Figure 1). Conformer 2 showing transitions hypsochromically shifted compared to conformer 1 (see Figure 3). Figure 4 reports the main molecular orbital (MO) levels composition and the corresponding orbital energies for conformer 1, while a full interpretation of the MO's electronic transitions allied to the main spectral features is reported in S.I.. As inferred from Table S3 in S.I. almost all the transitions have multi-determinantal character with at least two configurations contributing to their composition. As a general trend, all the transitions in the aforementioned range are generated by excitations between occupied orbitals in the range HOMO ÷ HOMO-5 and empty ones LUMO and LUMO+1, these two being localized on the porphyrazine core. In particular, $S_1$ and $S_2$ involve excitations from HOMO to LUMO and LUMO+1. The composition of these three MO levels gives to the transitions a well defined charge transfer (pyrene → porphyrazine) character. This proces, which determines upon excitation the separation of electrons and holes on different parts of the molecule, can be relevant for both the exciton formation and lifetime. The presence of charge transfer HOMO-LUMO transitions and the location of these orbitals on different moieties of the molecules, reducing the kinetics of the radiative relaxation processes, prevents exciton recombination, thus making this compound promising dye as photoactive material for optoelectronics.

The UV-vis spectrum of **PzPy** is also recorded in DMF (Figure S5 in S.I.), the solvent chosen for the preparation of nanocomposites with SWNTs and GNFs. The spectrum recorded in anhydrous DMF is markedly different from the one in $CH_2Cl_2$, rather closely resembling that of a thioalkyl porphyrazine metal complex[83] with a single Q band located at 665 nm (15037 cm$^{-1}$) (Figure S6 in S.I.). This effect, also observed in substituted phthalocyanines,[55,115,116] can be ascribed to a deprotonation of the macrocyle inner cavity which produces more symmetric anionic structures having $D_{4h}$ symmetry like the metal complexes.[117] This is also confirmed by basic titration of phthalocyanines, which shows a pH dependent transition from the typical 'free-base' UV-vis spectrum to that of mono- and dianionic species, both identical to those of the corresponding metal complexes.[118,119] Contrary to what is reported in literature,[55,117] our understanding is that the macrocycle deprotonation cannot be ascribed to DMF, which is not basic enough ($pK_b$ 15.2)[120] to remove the inner pyrrolic protons ($pK_a$ 12.5),[118,119] but to dimethylamine traces ($pK_b$ 3.27)[121] usually present as impurities in commercial DMF.[122] Experimental and computational studies reported in S.I. further confirmed such hypothesis. Notably, a typical 'free-base' UV-vis spectrum, similar to that obtained in $CH_2Cl_2$, is instead observed in non anhydrous DMF (Figure S5 in S.I.), where the water present in the solvent competes as acid with the porphyrazine in the acid-base equilibrium with the amine. In summary, the shape of the **PzPy** UV-vis spectrum in DMF depends on the water content of the solvent: a complex-like spectrum appears in anhydrous DMF, while a spectrum due to superimposition of anionic and 'free-base' species appears in a wet solvent, as demostated by the presence of three isosbestic points at 446 nm, 620 nm, and 690 nm (Figure S5 in S.I.).[55]

The cyclic voltammetry (CV) and differential pulse voltammetry (DPV) measurements of **PzPy** in $CH_2Cl_2$ are reported in Figure 5 and the electrochemical data for both solvents are collected in Table 1. The redox behavior of **PzPy** appears as the typical one of thioalkyl-porphyrazines. In fact, two sequential one-electron reduction processes are present in the cathodic region, which can be assigned to the formation of a porphyrazine π-anion radical and a porphyrazine dianion, respectively.[83,123,124] The redox processes in $CH_2Cl_2$ are characterized by a cathodic peak and its anodic counterpart displaying half-wave potentials $E_{1/2}(\Delta E_p)$ = −0.900 V (0.096) e $E_{1/2}(\Delta E_p)$ = − 1.232 (0.095) (*vs* Fc/Fc$^+$, used as internal standard), respectively. Both processes can be considered as *quasi*-reversible because the conditions for the reversibility are not rigorously fulfilled (as inferred from $\Delta E_p = |E_a - E_c|$ values). Furthermore, CV of compound **PzPy** has shown a *quasi*-reversible oxidation wave with $E_{1/2}(\Delta E_p)$ = +0.657 V (0.108) (*vs* Fc/Fc$^+$). HOMO and LUMO energies obtained from electrochemical data are reported in Table 1, by assuming the energy level of ferrocene/ferrocenium at − 4.8 eV.[125-127] The HOMO energies are also estimated according to a recent procedure, taking into account solvation and electrode surface effects.[128] As inferred from the electrochemical data in DMF (Table 1), strong anodic shifts for all the potential values with respect to $CH_2Cl_2$ are observed. In this case, it must be recalled that the potential values can be affected not only by the solvent but also by the presence of anionic species together with the 'free base' porphyrazine (*vide supra*). However, these factors are expected to have a minor influence, because only small potential changes are observed retaining electrochemical data in both anhydrous and not anhydrous DMF. Notably, the HOMO and LUMO energy values make **PzPy** suitable for fabrication of BHJ cells with nanocarbons as acceptors. In fact, with acceptors such as PCBM, SWNT, and graphene, having work functions [26,129-131] in the -4.3÷-4.9 eV range, donors with HOMO energy values lower than − 5.20 eV and a bandgap energy range of 1.30 ÷ 1.80 eV are required.[132]

**Table 1.** Summary of the peak potentials of **PzPy** $E_{1/2}$ ($\Delta E_p = |E_a - E_c|$) (Volts *vs* Fc/Fc$^+$).[a]

| Solvent | Technique | Oxidation | Reduction | | $E_{HOMO}$ | | | $E_{LUMO}$ | |
|---|---|---|---|---|---|---|---|---|---|
| | | | I | II | Exptl.[b] | Exptl.[c] | Comp.[d] | Exptl.[b] | Comp.[d] |
| CH$_2$Cl$_2$ | (CV) | 0.657 (0.108) | -0.900 (0.096) | -1.232 (0.095) | -5.46 | -5.52±0.18 | | -3.90 | |
| | (DPV)$_{ox}$ | 0.648 | -0.902 | -1.240 | -5.45 | -5.51±0.18 | -5.85 | -3.90 | -3.48 |
| | (DPV)$_{red}$ | 0.650 | -0.896 | -1.234 | -5.45 | -5.51±0.18 | | -3.90 | |
| DMF[e] | (CV) | n.d. | -0.738 (0.064) | -1.130 (0.060) | n.d. | n.d. | -5.87 | -4.06 | -3.50 |
| | (DPV)$_{ox}$ | 0.716 | -0.752 | -1.152 | -5.52 | -5.60±0.18 | | -4.05 | |

[a] Measured 10$^{-3}$ M solution at a glassy carbon working electrode. [b] Values (eV) referred to first oxidation and first reduction, and calculated assuming the energy level for the Ferrocene at − 4.8 eV [see ref. 126]. [c] Values (eV) obtained by the equation $E_{HOMO} = -(1.4 \pm 0.1) \times q \times V_{CV} - (4.6 \pm 0.08)$ [see ref. 128]. [d] Computations in CH$_2$Cl$_2$ and DMF by PCM solvation model as described in experimental section and at the M06/6-311G(d,p) level of theory. [e] Anhydrous DMF solvent.

The DFT computed values for the HOMO and LUMO energies at the M06/6-311G(d,p) level of theory in CH$_2$Cl$_2$ are -5.85 and -3.48 eV, respectively, and -5.87 and -3.50 eV in DMF. These data are in good agreement with values calculated from experimental ones. Notably, the HOMO-LUMO transition retains also in DMF a charge transfer character (see Figure S9 in S.I.). The long-range corrected xc-functional CAM-B3LYP gives values of -6.62 and -2.77 eV in CH$_2$Cl$_2$, suggesting that the hypsochromic shift in the low energy part of the absorption spectrum is due to a too large HOMO-LUMO energy gap engendered by this xc-functional.

**2.3 Hybrid PzPy/nanocarbons.** **PzPy** is used to prepare non covalent bonded nanohybrid composites with two classes of nanocarbons, SWNTs and graphene. As described in the Methods, commercially available SWNTs (diameter range of 0.7-1.1 nm) are characterized and then dispersed in DMF, and subsequently mixed with **PzPy** for the composites realization. DMF is preferred with respect to *N*-methyl-2-pyrrolidone, one of the best solvent for the dispersion of nanocarbons,[133,134] for its lower boiling point (152°C *vs* 202°C), which helps its removal by evaporation during the device preparation.[135] Raman spectroscopy is used for the nanotubes structural characterization and to ascertain the batch composition (see discussion in S.I.). The Raman analysis performed with an excitation wavelength at 532 nm confirms that the SWNTs batch is composed, at least, by a mixture of nanotubes, with chiralities (10,3), (9,2), (7,6), and (7,5), which are resonant to 532 nm excitation wavelength.[136]

Electron microscopy is used to study the interaction between SWNTs and **PzPy**. Figure 6a reports a scanning electron microscope (SEM) micrograph taken on pristine SWNTs deposited onto a Si/SiO$_2$ substrate. The SWNTs appear strongly entangled and forming bundles. By comparison, the **PzPy**/SWNT nanohybrid (Figure 6c) evidences an intermixed structure between the composite materials. The transmission electron microscopy (TEM) micrograph of SWNTs in Figure 6b shows the presence of large aggregates of nanotubes with length in the order of hundreds of nm (see supporting information for statistical analysis). The comparison with the TEM image of the nanohybrid composite (Figure 6d) reveals the interaction between **PzPy** and SWNTs, since the nanotubes appear here debundled for the effect of π-π interactions between the **PzPy** and the SWNT sidewalls, which overcome nanotube-nanotube van der Waals attraction.[137] In Figure 6d the presence of a molasses-type film of **PzPy** covering the nanotubes is also clearly visible, further confirming the expected strong supramolecular interactions between the two materials.

To investigate more in details the **PzPy**/SWNT nanohybrid, UV-vis-NIR absorption and fluorescence spectroscopies are performed. The spectrum of the pristine SWNTs (Figure 7a) confirms the presence of different SWNTs chiralities.[100] The steady-state absorption spectroscopic measurements of **PzPy**/SWNT nanocomposites show some diagnostic changes in the macrocycle absorption features upon nanotubes enrichment (Figure 7b). The SWNTs dispersion contributes to the increase in optical density, especially in the blue region of the spectra, giving rise to an increase of the 350 nm absorption, related to the porphyrazine Soret band, with respect to the Q bands. Moreover, a splitting and broadening of the Soret band is observed, revealing interaction between the macrocycle and the added SWNTs.[63,98] Upon excitation, compound **PzPy** displays emission spectra significantly more intense than **1**, lacking of the pyrene moiety. Analysis of emission spectra at two different excitation wavelenghths clearly shows that addition of SWNTs to the porphyrazine dispersion gives rise to a strong quenching of the macrocycle fluorescence (Figure 7c and 7d). Exciting at 350 nm (Figure 7d), the emission bands allied to the pyrene chromophore proportionally decrease upon SWNTs addition. On the contrary, at 610 nm excitation wavelength (Figure 7c), the fluorescence ascribed to the macrocycle immediately drops, completely quenching at a 1:1

**PzPy**/SWNT weight ratio. These effects suggest the occurrence of excited-state events such as electron transfer and energy transfer,[45,54] which highlight the presence of a porphyrazine-SWNT interaction. The preparation of nanocomposite of **PzPy** with GNFs (**PzPy**/GNF) is also carried out. Graphene nanoflakes are obtained by ultrasonication of pristine graphite[133,138] and centrifugation in DMF (see Method for details). The GNFs sample is characterized by Raman spectroscopy, a fast and non-destructive technique used to identify number of layers, defects, doping, disorder and possible chemical modifications.[139,140] The Raman spectrum of the obtained GNFs (Figure 8) shows the typical D, D',G, and 2D bands of graphene. See S.I. for the Physical description of these Raman modes. The 2D peak position is at ~2696 $cm^{-1}$ and its full width at half maximum is ~60 $cm^{-1}$. The I(2D)/I(G) intensity ratio ranges from 0.6 to 0.7. This is consistent with the samples being a combination of single-layer graphene (SLG) and few-layer graphene (FLG) flakes.[141] The Raman spectra show significant D (~1345 $cm^{-1}$) and D' (~1615 $cm^{-1}$) peaks intensity, with I(D)/I(G) ranging from 0.9 to 1.3. This is attributed to the edges of sub-micrometre flakes[142] rather than to the presence of a large amount of structural defects. Indeed, if a large amount of defects were present in the basal plane of graphene the G (1582 $cm^{-1}$) and D' peak should merge into a broader band,[143] which is not the case for the spectra of our dispersions.[141]

The **PzPy**/GNF composites are prepared by mixing and stirring the porphyrazine and the GNFs dispersions in DMF at different weight ratios. The **PzPy**/GNF composite is studied by electron microscopy. Figure 9a shows a SEM micrograph of GNFs deposited onto a Si/SiO$_2$ substrate: the GNFs cluster in stacks forming a rather uniform film. When isolated, as in the TEM micrograph in Figure 9b, it is possible to evaluate their size distribution in lateral size peaking at 150 nm see Figure S11 for the statistical analysis. The SEM micrograph in Figure 9c and TEM micrograph in Figure 9d provide an insight into the morphology of the **PzPy**/GNF nanohybrid, where the GNF appears as a **PzPy** scaffold uniformly distributed.

Comparison of the Raman spectra of the porphyrazine and the graphene composite evidences a difference in the signal in the 900-1100 $cm^{-1}$ range allied to the porphyrazine macrocycle (Figure S10 in S.I.). The comparison with the pristine graphene shows that in the composite the I(D)/I(G) and I(D')/I(G) ratios increases, indicating an increment of $sp^3$ hybridized carbon atoms, which may be related with the presence of the alkyl chains of the **PzPy** molecules.

The absorption spectra of the composites do not show any appreciable changes with respect to the spectra of the individual components (Figure 10b). In fact, maxima of the UV-vis region for the **PzPy**/GNF DMF dispersions are identical with the one of **PzPy** in DMF.

On the contrary, fluorescence experiments for **PzPy**/GNF hybrid exciting at two different wavelenghts show that, increasing the graphene/porphyrazine ratio, the fluorescence of the macrocycle is proportionally reduced. It reveals that also in this case an energy transfer process between the two species occurs, confirming the hybrid formation (Figure 10c and d). These quenching experiments show that in the hybrid with nanotubes a more efficient energy transfer process occurs with respect to the one with graphene. In fact, by comparison of Figure 7c and Figure 10c it can be noticed that, while a complete emmision quenching is observed in a 1:1 **PzPy**/SWNT hybrid, a higher percentage of GNF with respect to the **PzPy** is necessary to obtain the same effect. This suggests a stronger interaction between the **PzPy** and SWNT with respect to that between **PzPy** and GNF. In fact, the computed molecular conformations of the porphyrazine present a dihedral angle between the two aryl moieties, which makes the compound suitable to wrap the SWNT, thus allowing π−π interactions with both the aromatic units (Figure 11). A similar effect has recently been observed in molecular dinamics simulation of interactions between SWNTs and chiral binaphthyl dendrimers.[144] On the contrary, the planar structure of graphene prevents the simultaneous interaction of both the macrocycle and the pyrene moieties.

**2.4 Photoconduction experiments on PzPy nanohybrids.** To test the optoelectronics properties of the as-synthesized **PzPy** nanohybrids and measure their photoconductivity under visible light illumination, we fabricated photodetectors by depositing films of the pristine **PzPy**, the composite **PzPy**/GNF, and **PzPy**/SWNT on interdigitated gold electrodes through drop casting method (see Experimental section, *Device fabrication and measurement*). Pristine GNF and SWNT films were also tested, however the resulting devices were electrically shortened (resistance between 100-1000 Ω), thus yielding high dark current densities without any photoresponse, therefore their data are not reported in the following discussion.

The device architecture is depicted in Figure 12a. Under monochromatic illumination at a wavelength of 500 nm (with intensity of 0.39 mW cm$^{-2}$), the **PzPy** film shows a linear photocurrent in the range 0-1 V, reaching 20 nA at 1 V bias voltage. The **PzPy**/GNF device presents a higher photocurrent, with respect to the **PzPy** one, throughout the whole range (Figure 12c): The I/V curve is linear between 0 and ~0.4 V, where a steep increase occurs, bringing the current up to 92 nA at 1 V. In the case of the **PzPy**/SWNT device, the overall I/V curve shape is analogous but with a smaller current (13.5 nA vs 18.0 nA) at a bias voltage of 0.2 V. The two curves cross at 36.5 nA for 0.54 V. However, beyond 0.55 V, the current increase up to 112 nA at 1.0 V. Concerning the responsivity (Figure 12d), the photodetector with **PzPy**/GNF as active material has a 2.6, 3.5 and 4.8 fold increase at 0.2, 0.6 and 1.0 V, respectively, compared to pristine **PzPy**. The photodetector with **PzPy**/SWNT instead has a 1.9, 3.5 and 5.7 fold increase at 0.2, 0.6 and 1.0 V, respectively, with respect to pristine **PzPy**. Under chopped 1 SUN illumination the **PzPy**/GNF device shows a photocurrent around 2.5 μA at 1 V of applied DC bias.

The photodetectors show a flat spectral responsivity over all the 400-750 nm visible range, with a 4 to 5 fold increase for the **PzPy**/GNF with respect to **PzPy**. The maximum response is observed towards the lowest wavelengths, reaching 0.23 and 0.80 mA W$^{-1}$ for **PzPy** and **PzPy**/GNF-based devices, respectively. By comparison with the absorbance measurement in Figure 2, we note that the photoresponse spectrum of the **PzPy** composites has an analogous trend to the absorption one. The photocurrent is thus more intense at 400-450 nm (near to the Soret band), at 470-500 nm (in correspondence to the n$_{sulphur}$→π$^*$ transitions), at 630-650 nm (broad Q$_y$(0, 0) band absorption), and beyond 700 nm (Q$_x$(0, 0) band). These observations, together with the quenching of photoluminescence (Figures 7 and 10), are a signature of photoexcited charge tranfer from the **PzPy** molecule to the nanocarbons, as depicted in Figure 12b. The transferred charges are then extracted thanks to the high conductivity of the nanocarbon materials, yielding the enhanced photoresponse with respect to the bare **PzPy**.[145,146]

Transient currents under chopped 1 SUN illumination of **PzPy**/GNF are also acquired with sampling period of 0.4 s, and light switching period of 5 s (Figure 13). After light illumination, the photocurrent is about 55% higher than before the illumination and a reversible rise/decay of the photocurrent in response to several on/off illumination cycles is observed, indicating the photocurrent stability of the nanocomposite.

From a comparison between the two types of nanocarbons used, it appears that the **PzPy**/GNF device has in general a slightly lower photoresponse than the **PzPy**/SWNT one (0.71 vs 0.78 mA W$^{-1}$ at 500 nm). This can be explained considering the favored π−π interactions occurring between the SWNTs and the aromatic units of the **PzPy**, as previously discussed and modeled in Figure 11. However, in our view, such a small difference in performance is strongly offset by a much greater flexibility in production, and processing that GNFs offer in comparison to the SWNTs. The GNFs are easier and more economic when produced by LPE even in large volumes,[138,147] whereas SWNTs are costly (1000$ per gram)[148] and the production is cumbersome, requiring further steps of purification and sorting[136,149,150] to isolate the batch of nanotubes with the desired properties. Therefore, a proof of principle of the capability of graphene to act as electron acceptor and charge transport component in **PzPy**-based photoactive nanohybrids is significant. Although these nanohybrids do not match the photocurrent values obtained by devices based on high-quality CVD graphene,[151,152] or fully optimized organic detectors,[153] the results are very promising if compared with the ones obtained by other solution-processed photoconductors attaining photocurrents in the order of $10^{-4}$ mA W$^{-1}$ and $10^{-2}$ mA W$^{-1}$.[154,155]

## 3. Conclusions

Non covalent bonded nanohybrids composed by the novel pyrene-substituted thioethyl-porphyrazine **PzPy** and SWNTs or GNFs have been synthesized and used for photocurrent generation. The **PzPy** structure has been specifically designed to finely tune the structural and electronic properties in order to: i) promote the non covalent interactions with π-conjugated nanocarbons, ii) enhance the spectral absorption in the visible range, iii) provide charge separation upon excitation, iv) obtain HOMO/LUMO energy values and bandgap suitable to fit the requirements of donor materials. In particular, the thioalkyl-porphyrazine core guarantees an intense absorption in the 300-800 nm range and the presence of a pyrene moiety promotes π-π attractive interactions with the nanocarbons. In **PzPy**, the hindered rotation between the arene-arene bond gives rise to a twist of the pyrene moiety from the porphyrazine plane and, eventually, to chirality due to atropisomerism. This is an important structural feature of **PzPy** given that molecular chirality, which is a key factor in supramolecular interactions[156] and organic optoelectronics,[157] is now emerging as a relevant issue also for the development of photoactive materials.[158-160] The presence of an axially chiral structure in **PzPy** could in fact favour the interaction with semiconducting chiral nanotubes. Spectroscopic and electrochemical measurements together with TDDFT calculations provide an HOMO-LUMO energy bandgap of 1.55 eV, which makes **PzPy** suitable as absorber donor coupled with nanocarbons acceptors. Calculations show that HOMO and LUMO are localized on the pyrene and the macrocycle moiety, respectively. This gives rise to a charge separation upon HOMO-LUMO excitation, eventually preventing charge recombination. SEM and TEM analysis elucidated that in both hybrids the nanocarbons surface appeared covered by a **PzPy** film and SWNTs debundling is observed in the **PzPy**/SWNT composite. This is a clear signature that π-π interactions between **PzPy** and the SWNT sidewalls overcome nanotube-nanotube van der Waals attraction. Emission spectroscopy experiments have also shown a quench of **PzPy** fluorescence upon addition of both SWNTs and GNFs, demonstrating the occurrence of energy transfer processes. Measurements of the **PzPy**/SWNT and **PzPy**/GNF nanocomposites photoconductivity under visible light illumination have shown a higher photocurrent generation for both **PzPy**/SWNT and **PzPy**/GNF-based devices with respect to the one of pristine **PzPy**. Moreover, the devices based on the hybrids **PzPy**/nanocarbons have shown a spectral photoresponse 4 to 5 fold higher than **PzPy** and peaked in agreement to the **PzPy** absorption features. This confirms the presence of photoexcited charge tranfer from the **PzPy** molecule to the nanocarbons. Although the **PzPy**/GNF device has a slightly lower photoresponse than the **PzPy**/SWNT one, graphene presents several advantages over SWNTs, in term of cost, preparation, processability and environmental compatibility. The presented results opens the way to the use of such hybrid materials in devices such as photodiodes and solar cells, even on flexible substrates.

## 4. Experimental Section

*General:* Chemicals and solvents are of reagent grade (Aldrich). Solvents are dried and distilled before use according to standard procedures. Solvents used in physical measurements are of spectroscopic or HPLC grade. Silica gel used for chromatography is Merck Kieselgel 60 (70-230 mesh). Commercially available SWNTs (carbon ~90%, >77% as SWNT, 0.7<d<1.1 nm) (Aldrich, code n. 704121) are used for the nanohydrids preparation after structural characterization as reported in S.I. Not-anhydrous spectroscopic grade DMF is used as solvent for the preparation of the composites of **PzPy** with SWNTs and GNFs. $^1$H NMR spectra are recorded on a 400, 500 or 600 MHz Varian spectrometer with SiMe$_4$ as internal standard. IR spectra are recorded using KBr disks on a JASCO J460 FT-IR spectrometer. Mass spectra are acquired in positive reflectron mode at 20 kV using Ettan MALDI-TOF Pro mass spectrometer (Amersham Biosciences) equipped with an UV nitrogen laser (337 nm) with delayed extraction mode and low mass rejection. For each spectrum 256 single shots are accumulated. Spectra are externally calibrated using two standard peptides (ile- Ang III, M+H 897.531 and hACTH 18-39, M+H 2465.199, monoisotopics). The matrix is prepared by mixing 1% (w/v) alpha-cyano-4-hydroxy-cinnamic acid solution, 50% acetonitrile (v/v), and 0.5% (v/v) trifluoroacetic acid. Samples for mass spectrometric analysis are prepared by dissolving 4 μL of porphyrazine solution (<10$^{-4}$ M in CH$_2$Cl$_2$) directly in 4 μL of the matrix. 0.4 μL of this mixture is deposited on the probe tip and allowed to evaporate. The 'free-base' thioethyl porphyrazine **1**[92] its non-symmetrically H-substituted counterpart **2** [81] and the Br-substituted porphyrazines **3**[82,83] are prepared according to previously described procedures.[77,78,161]

**2-(1-Pyrene)-3, 7, 8, 12, 13, 17, 18-heptakis(ethylthio)-5, 10, 15, 20- 21H, 23H-porphyrazine (PzPy).** Brominated porphyrazine **3** (100 mg, 0.123 mmol) is dissolved in a dry DMF/toluene (2:3, v/v) mixture (20 mL) heating at 100°C under N$_2$. Then, potassium carbonate (136 mg, 0.984 mmol), tetrakis(triphenylphosphine) palladium(0) (10% mol, 14.0 mg, 0.0123 mmol) and 1-pyrene boronic acid (121 mg, 0.492 mmol) are added in sequence and the mixture heated at reflux. The reaction is monitored by thin-layer chromatography (TLC) and stopped after 8 h. After cooling to room temperature water (30 mL) is added and the solution is extracted with CHCl$_3$ (35 mL). The organic fractions are collected, dried over anhydrous Na$_2$SO$_4$ and filtered. After removal of the solvent under reduced pressure, the crude product is purified by column chromatography on silica gel (hexane:CH$_2$Cl$_2$ 6:4 v/v), recovering **PzPy** as a dark blue solid (R$_f$ = 0.19, 45% yield). $^1$H-NMR (600 MHz, CDCl$_3$) δ/ppm: -1.10 (s, 2H, N$_p$-*H*); 0.51 (t, *J* = 7.4Hz, 3H); 1.25 (t, *J* = 7.4Hz, 3H); 1.44 (t, *J* = 7.4Hz, 3H); 1.54 (t, *J* = 7.4Hz, 3H); 1.58 (t, *J* = 7.4Hz, 6H); 1.66 (t, *J* = 7.4Hz, 3H); 3.06 (dq, *J* = 12.5, 7.4Hz, 1H); 3.20 (dq, *J* = 12.5, 7.4Hz, 1H); 3.29 (dq, *J* = 12.5, 7.4Hz, 1H); 3.33 (dq, *J* = 12.5, 7.4Hz, 1H); 3.91(dq, *J* = 7.4, 1.7Hz, 2H); 4.13 (m, 6H); 4.29 (dq, *J* = 7.4, 2.6Hz, 2H); 7.97 (d, *J* = 9.0Hz, 1H); 8.07 (t, *J* = 7.5Hz, 1H); 8.09 (d, *J* = 9.0Hz, 1H); 8.19 (d, *J* = 7.5Hz, 1H); 8.27 (d, *J* = 9.0Hz, 1H); 8.31 (d, *J* = 9.0Hz, 1H); 8.32 (d, *J* = 7.5Hz, 1H); 8.51 (d, *J* = 7.5Hz, 1H); 8.61 (d, *J* = 7.5Hz, 1H). $^{13}$C-NMR (125 MHz, CDCl$_3$) δ/ppm: 14.13, 14.30, 15.10, 15.64, 15.70, 15.72, 15.79, 27.84, 27.88, 29.51, 29.53, 29.61, 29.67, 29.71, 124.35, 124.76, 124.82, 125.46, 125.60, 126.13, 126.26, 127.59, 127.92, 128.16, 128.97, 130.10, 131.00, 131.01, 131.46, 131.69, 135.43, 138.97, 140.42, 140.66, 140.86, 140.90, 142.64, 143.29, 146.09, 146.40, 147.85, 149.56, 149.66, 150.43, 151.39, 152.75. MALDI-MS: *m/z* 935.31 [M+H]$^+$ (calcd. for C$_{46}$H$_{47}$N$_8$S$_7$, 935.20). UV-vis, CH$_2$Cl$_2$, λ$_{max}$/nm (log ε): 277 (4.33), 350 (4.63) Soret, 513 (4.11), 640 (4.29) and 709 (4.42) Q bands. FTIR (KBr $\tilde{v}_{max}$/cm$^{-1}$): 3290 (w, N$_p$H), 3040 (w), 2959 (m), 2923 (m), 2856 (m), 1641 (br, m), 1482 (m), 1437 (m), 1375 (w), 1066 (br, s), 799 (m), 744 (m), 700 (w), 691 (w).

*PzPy characterization:* The ¹H NMR spectra at variable temperatures are recorded with a spectrometer operating at 600 MHz. UV-vis spectra are recorded in the 250-800 nm range by UV-vis-NIR 05E Cary spectrophotometer in 1 cm path length quartz cells (the concentration of the dispersions is *ca.* $10^{-6}$M in compound). The cyclic voltammetry (CV) and differential pulse voltammetry (DPV) experiments are performed with an EG & G Princeton Applied Research Model 263A Potentiostat/Galvanostat. Data are collected and analyzed by the Model 270 electrochemical analysis system software on a PC computer.[162] A standard three-electrode arrangement is used. The working electrode is a glassy carbon button ($\varnothing$ = 3 mm). A platinum wire served as counter electrode and a home-made AgCl/Ag electrode containing saturated KCl is used as the reference electrode. All the oxidation and reduction potentials are reported relative to the ferrocene/ferrocenium (Fc/Fc$^+$) potential scale, using the voltammetric oxidation of Fc as an internal reference. The reproducibility of individual potential values is within ± 5 mV. All the electrochemical measurements are carried out using Schlenk techniques (N$_2$) at room temperature. The concentration of the supporting electrolyte [N(C$_4$H$_9$-*n*)$_4$BF$_4$] is typically of 0.15 M. Cyclic voltammograms are recorded by scanning the potential at 200 mVs$^{-1}$. DPV measurements are performed at 5 mVs$^{-1}$ with a pulse height of 50 mV and 50 ms pulse width.

*Computational Method:* The Gaussian 09, Revision D.01 software [163] is used for all the computations. The DFT using the M06 meta-hybrid xc functional [164] with the 6-311G(d,p) basis set as implemented in the program[165,166] has been used for geometry optimization and computation of the Kohn-Sham orbital energies simulating the solvation effects by SCRF (Self Consistent Reaction Field) IEFPCM (Integral Equation Formalism Polarizable Continuum Model) as implemented in the program.[105,106] Default gradient and displacement thresholds are used for the geometry optimization convergence criteria. In calculations of **PzPy** it is assumed that the thioethyl side chains conformation are similar to that already found in similar heptakis substituted porphyrazines.[83] All the reported geometries are relative minima of the molecular potential energy surface (electronic energy in the Born-Oppenheimer approximation), as confirmed by the analytical computation of the Hessian matrix at the same level of approximation. The TDDFT[167,168] is applied for computing the excitation wavelengths, oscillator strengths and associated excited state composition in terms of monoelectronic excitations between occupied and unoccupied Kohn-Sham orbitals at the level of theory. The Gaussian 09 default approach is used for these computations.[169] The Kohn-Sham orbitals are drawn using the program MolDen 4.9.[170] Molecular orbital isosurfaces refer to wavefunction computed at 0.02 (e/a.u.$^3$)$^{1/2}$.

*Nanohybrid preparation:* **(1) PzPy/SWNT.** SWNTs (Sigma SWNT: carbon ~90%, >77% as SWCNT, 0.7<d<1.1 nm) are dispersed in DMF to make three dispersions with different initial concentration (0.1, 0.2, 0.4 mg/mL, respectively). The dispersions are ultrasonicated for 1h in ice water and then centrifuged at 1500g for 1h. 1 mL of the supernatant of each dispersion is taken (named $C_1$, $C_2$, $C_3$, respectively). A **PzPy** solution in DMF with concentration of 0.03 mg/mL (c = 3.21x10$^{-5}$ M) is prepared (named P1). Three **PzPy**/SWNT nanocomposites at different wt ratios are prepared by mixing 3 mL of **PzPy** solution with 1 mL of $C_1$, $C_2$ and $C_3$, respectively. The composites (named $PC_1$, $PC_2$, $PC_3$) are stirred for 10 hours and then sonicated 1 minute before device fabrication. The three solutions contained the same **PzPy** concentration of 0.0225 mg/mL (c = 2.41x10$^{-5}$ M).

**(2) PzPy/GNF.** Graphene nanoflakes are produced by liquid phase exfoliation of graphite.[133] Graphite flakes (1 g) (Sigma Aldrich) is dispersed in 100 mL of DMF and ultrasonicated (Branson® 5800) for 6 hours. The obtained dispersion is ultracentrifuged at 12300g (in a Beckman Coulter Optima™ XE-90 with a SW41Ti rotor) for 30 mins at 15 °C, exploiting sedimentation-based separation (SBS) to remove thick flakes and un-exfoliated graphite.[171,172] After the ultracentrifugation process, we collected the supernatant by pipetting. Composites are prepared by mixing graphene flakes with **PzPy** in DMF at different wt ratios and stirring for 1 day. Three composites with **PzPy**/GNF wt ratios of 1:1, 3:4 and 3:7, respectively, are produced. The three solutions contained the same **PzPy** concentration.

*Nanohybrid characterization:* The SWNT and GNF dispersions as well as the nanohybrids are characterized by optical absorption spectroscopy in the range 300–800 nm with a Cary Varian 6000i UV–vis–NIR spectrometer. The absorption spectra are acquired using a 1 mL quartz glass cuvette. The solvent baseline is subtracted to the recorded spectrum. Photoluminescence emission spectra of the nanohybrids are recorded with a Horiba FluoroMax4 fluorometer with excitation at 350 and 610 nm. Raman measurements are collected with a Renishaw inVia confocal Raman microscope with excitation line of 532 nm (2.33 eV) with a 50× objective lens, and an incident power of ~1 mW on the samples. The GNF and SWNT dispersions are drop cast onto Si/SiO$_2$ wafer (300 nm thermally grown SiO$_2$ - LDB Technologies Ltd.) and dried under vacuum. 20 spectra are collected for each sample; Lorentzian functions are used to fit the peaks. GNF and SWNT are characterized morphologically by transmission electron microscopy (TEM - JOEL JEM 1011 - operated at 100kV). The materials are dropped with a pipette on holey carbon 200 mesh grids and dried under vacuum overnight. Field-emission scanning electron microscopy (FE-SEM) is also performed on samples drop cast onto Si wafer (JoelJSM-7500FA).

*Device fabrication and measurement:* Interdigitated gold electrodes are fabricated on Si/SiO$_2$ substrates by UV lithography with a SUSS MicroTec MJB4 mask aligner. The lithography and development of the positive photoresist are followed by thermal evaporation of chromium/gold (thickness 9 nm/40 nm) and lift-off. 20 μL **PzPy** or **PzPy**/GNF (weigth ratio of 1:1.3), 0.5 mg/mL chlorobenzene dispersions, or **PzPy**/SWNT (weigth:ratio of 1:1.3), 0.5 mg/mL DMF dispersion, are drop cast on gold electrodes. Finally, the devices are annealed at 70 °C for 30 min under inert atmosphere. Photodetection experiments are performed in ambient conditions. Monochromatic light is obtained from a xenon lamp coupled to a monochromator (Spectral Products CM110), and brought to the sample with a system of lenses and mirrors in order to illuminate the device area. In this condition, the spectral responsivity, is simply calculated as the ratio between the measured photocurrent density and the optical input power density for each light wavelength. Light power is measured using a calibrated Si photodiode (Thorlabs S120VC). DC bias and current measure is provided by a Keithley 2612 sourcemeter. Light is modulated with a mechanical chopper, and AC photocurrent is acquired using a current preamplifier (DL 1211) and lock-in amplifier (Signal Recovery 7265). Spectral responsivity and IV curves of the photodetectors are acquired at 857 Hz. The detection limit of the measurement set-up is equal to 1 nA.

Transient currents under chopped 1 SUN illumination (xenon light source equipped with AM 1.5 G filters) are acquired with sampling period of 0.4 s, and light switching period of 5s.

**Acknowledgements**
This project has received funding from the European Union's Horizon 2020 research and innovation program under grant agreement No. 696656—GrapheneCore1, and by Regione Basilicata – 'Basilicata Innovazione' grant (Project J41H12000070001).

[1]  S. –S. Sun, N. S. Sariciftci, Eds. *Organic Photovoltaics: Mechanisms, Materials, and Devices* Taylor & Francis: Boca Raton, FL, **2005**.

[2]  C. Brabec, U. Scherf, V. Dyakonov, Eds. *Organic Photovoltaics: Materials, Device Physics, and Manufacturing Technologies 2nd Ed.* Wiley-VCH: Weinheim, Germany, **2014**.

[3]  B. P. Rand, H. Richter, Eds. *Organic Solar Cells: Fundamentals, Devices, and Upscaling* Pan Stanford Publishing, CRC Press- Taylor & Francis: Boca Raton, FL, **2014**.

[4]  B. Kippelen, J. -L. Brédas, *Energy Environ. Sci.* **2009**, *2*, 251.

[5]  K. A. Mazzio, C. K. Luscombe, *Chem. Soc. Rev.* **2015**, *44*, 78.

[6]  G. Yu, J. Gao, J. C. Hummelen, F. Wudl, A. J. Heeger, *Science* **1995**, *270*, 1789.

[7]  J. J. M. Halls, C. A. Walsh, N. C. Greenham, E. A. Marseglia, R. H. Friend, S. C. Moratti, A. B. Holmes, *Nature* **1995**, *376*, 498.

[8]  P. Schilinsky, C. Waldauf, C. J. Brabec, *Appl. Phys. Lett.*, **2002**, *81*, 3885.

[9]  Y. Sun, G. C. Welch, W. L. Leong, C. J. Takacs, G. C. Bazan, A. J. Heeger, *Nat. Mater.* **2012**, *11*, 44.

[10] A. Mishra, P. Baüerle, *Angew. Chem., Int. Ed.* **2012**, *51*, 2020.

[11] D. Jariwala, V. K. Sangwan, L. J. Lauhon, T. J. Marks, M. C. Hersam, *Chem. Soc. Rev.* **2013**, *42*, 2824.

[12] Y. He, Y. Li, *Phys. Chem. Chem. Phys.* **2011**, *13*, 1970.

[13] S. Cataldo, P. Salice, E. Menna, B. Pignataro, *Energy Environ. Sci.* **2012**, *5*, 5919.

[14] L. Wang, H. Liu, R. M. Konik, J. A. Misewich, S. S. Wong, *Chem. Soc. Rev.* **2013**, *42*, 8134.

[15] V. Sgobba, D. M. Guldi, *Chem. Soc. Rev.* **2009**, *38*, 165.

[16] A. C. Ferrari, F. Bonaccorso, V. Fal'ko, K. S. Novoselov, S. Roche, P. Bøggild, S. Borini, F. H. L. Koppens, V. Palermo, N. Pugno, J. A. Garrido, R. Sordan, A. Bianco, L. Ballerini, M. Prato et.al., *Nanoscale* **2015**, *7*, 4598.

[17] X. Wan, G. Long, L. Huang, Y. Chen, *Adv. Mater.* **2011**, *23*, 5342.

[18] I. V. Lightcap, P. V. Kamat, *Acc. Chem. Res.* **2013**, *46*, 2235.

[19] D. M. Guldi, V. Sgobba, *Chem. Commun.* **2011**, *47*, 606.

[20] P. V. Kamat, *J. Phys. Chem. Lett*. **2011**, *2*, 242.

[21] F. Bonaccorso, N. Balis, M. M. Stylianakis, M. Savarese, C. Adamo, M. Gemmi, V. Pellegrini, E. Stratakis, E. Kymakis, *Adv. Funct. Mater*. **2015**, *25*, 3870.

[22] F. Bonaccorso, L. Colombo, G. Yu, M. Stoller, V. Tozzini, A. C. Ferrari, R. S. Ruoff, V. Pellegrini, *Science* **2015**, *347*, 1246501.

[23] H.-J. Choi, S.-M. Junga, J.-M. Seo, D. W. Chang, L. Dai, J.-B. Baek, *Nano Energy* **2012**, *1*, 534.

[24] M. Pumera, *Chem. Rec*. **2009**, *9*, 211.

[25] R. R. Nair, P. Blake, A. N. Grigorenko, K. S. Novoselov, T. J. Booth, T. Stauber, N. M. R. Peres, A. K. Geim, *Science* **2008**, *320*, 1308.

[26] J.-H. Kim, J. H. Hwang, J. Suh, S. Tongay, S. Kwon, C. C. Hwang, J. Wu, J. Y. Park, *Appl. Phys. Lett*. **2013**, *103*, 171604.

[27] Q. Liu, Z. Liu, X. Zhang, L. Yang, N. Zhang, G. Pan, G. S. Yin, Y. Chen, J. Wei, *Adv. Funct. Mater.* **2009**, *19*, 894.

[28] Z. Liu, Q. Liu, Y. Huang, Y. Ma, S. Yin, X. Zhang, W. Sun, Y. Chen, *Adv. Mater*. **2008**, *20*, 3924.

[29] M. G. Walter, A. B. Rudine, C. C. Wamser, *J. Porphyrins Phthalocyanines* **2010**, *14*, 759.

[30] G. Bottari, G. de la Torre, D. M. Guldi, T. Torres, *Chem. Rev.* **2010**, *110,* 6768.

[31] C. W. Tang, *Appl. Phys. Lett*. **1986**, *48*, 183.

[32] P. Peumans, S. R. Forrest, *Appl. Phys. Lett.* **2001**, *79*, 126.

[33] R. F. Bailey-Salzman, B. P. Rand, S. R. Forrest, *Appl. Phys. Lett.* **2007**, *91*, 013508.

[34] I. Kim, H. M. Haverinen, Z. Wang, S. Madakuni, Y. Kim, J. Li, G. E. Jabbour, *Chem. Mater.* **2009**, *21*, 4256.

[35] K. Takahashi, T. Goda, T. Yamaguchi, T. Komura, K. Murata, *J. Phys. Chem. B* **1999**, *103,* 4868.

[36] M. R. Axet, O. Dechy-Cabaret, J. Durand, M. Gouygou, P. Serp, *Coord. Chem. Rev*. **2016**, *308*, 236.

[37] G. Bottari, G. de la Torre, T. Torres, *Acc. Chem. Res*. **2015**, *48*, 900.

[38] D. Baskaran, J. W. Mays, X. P. Zhang, M. S. Bratcher, *J. Am. Chem. Soc.* **2005**, *127*, 6916.

[39] D. He, Y. Peng, H. Yang, D. Ma, Y. Wang, K. Chen, P. Chen, J. Shi, *Dyes and Pigments* **2013**, *99*, 395.

[40] M. E. Lipińska, S. L. H. Rebelo, M. F. R. Pereira, J. L. Figueiredo, C. Freire, *Mat. Chem. Phys.* **2013**, *143*, 296.

[41] B. Ballesteros, S. Campidelli, G. de la Torre, C. Ehli, D. M. Guldi, M. Prato, T. Torres, *Chem. Commun.* **2007**, *28*, 2950.

[42] B. Ballesteros, G. de la Torre, C. Ehli, G. M. A. Rahman, F. Agulló-Rueda, D. M. Guldi, T. Torres, *J. Am. Chem. Soc.* **2007**, *129*, 5061.

[43] S. Campidelli, B. Ballesteros, A. Filoramo, D. Diaz Díaz, G. de la Torre, T. Torres, G. M. A. Rahman, C. Ehli, D. Kiessling, F. Werner, V. Sgobba, D.M. Guldi, C. Cioffi, M. Prato, J. P. Bourgoin, *J. Am. Chem. Soc.* **2008**, *130*, 11503.

[44] K. Qu, H. Xu, C. Zhao, J. Ren, X. Qu, *RSC Adv.* **2011**, *1*, 632.

[45] D. M. Guldi, G. M. A. Rahman, S. Quin, M. Tchoul, W. T. Ford, M. Marcaccio, D. Paolucci, F. Paolucci, S. Campidelli, M. Prato, *Chem. Eur. J.* **2006**, *12*, 2152.

[46] S. K. Das, N.. K. Subbaiyan, F. D'Souza, A. S. D. Sandanayaka, T. Hasobe, O. Ito, *Energy Environ. Sci.* **2011**, *4*, 707.

[47] R. Chitta, A. S. D. Sandanayaka, A. L. Schumacher, L. D'Souza, Y. Araki, O. Ito, F. D'Souza, *J. Phys. Chem. C* **2007**, *111*, 6947.

[48] D. M. Guldi, G. M. A. Rahman, M. Prato, N. Jux, S. Qin, W. Ford, *Angew. Chem. Int. Ed.* **2005**, *44*, 2015.

[49] V. Sgobba, G. M. A. Rahman, D. M. Guldi, N. Jux, S. Campidelli, M. Prato, *Adv. Mater.* **2006**, *18*, 2264.

[50] T. Hasobe, S. Fukuzumi, P. V. Kamat, *J. Phys. Chem. B* **2006**, *110*, 25477.

[51] J. Bartelmess, A. R. M. Soares, M. V. Martinez-Diaz, M. G. P. M. S. Neves, A. C. Tomé, J. A. S. Cavaleiro, T. Torres, D. M. Guldi, *Chem. Comm.* **2011**, *47*, 3490.

[52] E. Maligaspe, A. S. D. Sandanayaka, T. Hasobe, O. Ito, F. D'Souza, *J. Am. Chem. Soc.* **2010**, *132*, 8158.

[53] H. Murakami, T. Nomura, N. Nakashima, *Chem. Phys. Lett.* **2003**, *378*, 481.

[54] S. Kyatskaya, J. R. G. Mascaros, L. Bogani, F. Hennrich, M. Kappes, W. Wernsdorfer, M. Ruben, *J. Am. Chem. Soc.* **2009**, *131*, 15143.

[55] J. Bartelmess, B. Ballesteros, G. de la Torre, D. Kiessling, S. Campidelli, M. Prato, T. Torres, D. M. Guldi, *J. Am. Chem. Soc.* **2010**, *132*, 16202.

[56] M. Ince, J. Bartelmess, D. Kiessling, K. Dirian, M. V. Martínez-Díaz, T. Torres, D. M. Guldi, *Chem. Sci.* **2012**, *3*, 1472.

[57] R. A. Hatton, N. P. Blanchard, V. Stolojan, A. J. Miller, S. R. P. Silva, *Langmuir* **2007**, *23*, 6424.

[58] S. K. Das, N. K.; Subbaiyan, F. D'Souza, A. S. D. Sandanayaka, T. Wakahara, O. Ito, *J. Porphyrins Phthalocyanines* **2011**, *15*, 1033.

[59] A. S. D. Sandanayaka, N. K. Subbaiyan, S. K. Das, R. Chitta, E. Maligaspe, T. Hasobe, O. Ito, F. D'Souza, *ChemPhysChem* **2011**, *12*, 2266.

[60] J. Bartelmess, C. Ehli, J.-J. Cid, M. Garcia-Iglesias, P. Vazquez, T. Torres, D. M. Guldi, *Chem. Sci*. **2011**, *2*, 652.

[61] J. Bartelmess, C. Ehli, J.-J. Cid, M. Garcia-Iglesias, P. Vazquez, T. Torres, D. M. Guldi, *J. Mater. Chem*. **2011**, *21*, 8014.

[62] Q. Zhong, V. V. Diev, S. T. Roberts, P.D. Antunez, R. L. Brutchey, S. E. Bradforth, M.E. Thompson, *ACS Nano* **2013**, *7*, 3466.

[63] E. Bekyarova, S. Sarkar, F. Wang, M. E. Itkis, I. Kalinina, X. Tian, R. C. Haddon, *Acc. Chem. Res*. **2013**, *46*, 65.

[64] B. Wang, V. Engelhardt, A. Roth, R. Faust, D. M. Guldi, *Nanoscale* **2017**, *9*, 11632.

[65] A. Roth, M.-E. Ragoussi, L. Wibmer, G. Katsukis, G. de la Torre, T. Torres, D. M. Guldi, *Chem. Sci.* **2014**, *5*, 3432.

[66] C. K. C. Bikram, S. K. Das, K. Ohkubo, S. Fukuzumi, F. D'Souza, *Chem. Comm*. **2012**, *48*, 11859.

[67] M.-E. Ragoussi, G. Katsukis, A. Roth, J. Malig, G. de la Torre, D. M. Guldi, T. Torres, *J. Am. Chem. Soc.* **2014**, *136*, 4593.

[68] M.-E. Ragoussi, J. Malig, G. Katsukis, B. Butz, E. Spiecker, G. de la Torre, T. Torres, D. M. Guldi, *Angew. Chem., Int. Ed*. **2012**, *51*, 6421.

[69] J. Malig, N. Jux, D. Kiessling, J.-J. Cid, P. Vazquez, T. Torres, D. M. Guldi, *Angew. Chem., Int. Ed*. **2011**, *50*, 3561.

[70] L. Brinkhaus, G. Katsukis, J. Malig, R. D. Costa, M. Garcia- Iglesias, P. Vazquez, T. Torres, D. M. Guldi, *Small* **2013**, *9*, 2348.

[71] X. Zhang, Y. Feng, S. Tang, W. Feng, *Carbon* **2010**, *48*, 211.

[72] B. Mondal, R. Bera, S. K. Nayak, A. Patra, *J. Mater. Chem. C,* **2016**, *4*, 6027.

[73] Y. Yang, R. Suna, M. Tanga, S. Rena, *Physica E* **2017**, *86*, 76.

[74] V. Georgakilas, J. N. Tiwari, K. C. Kemp, J. A. Perman, A. B. Bourlinos, K. S. Kim, R. Zboril, *Chem. Rev*. **2016**, *116*, 5464.

[75] C. C. Leznoff, A. B. P. Lever, Eds. *Phthalocyanines, Properties and Applications,* Vols. 1-4, VCH: New York, **1989-1996**.

[76] P. A. Stuzhin, C. Ercolani, Porphyrazines with Annulated Heterocycles. In *The Porphyrin Handbook*, *Vol 15. Phthalocyanines: Synthesis.* Kadish, K. M.: Smith, K. M.; Guilard, R. Eds.; Academic Press: New York, **2003**; pp. 263-365.

[77] F. Lelj, G. Morelli, G. Ricciardi, A. Roviello, A. Sirigu, *Liq. Cryst*. **1992**, *12***,** 941.

[78] S. Belviso, G. Ricciardi, F. Lelj, *J. Mater. Chem.* **2000**, *10,* 297.

[79] S. Belviso, M. Amati, M. De Bonis, F. Lelj, *Mol. Cryst. Liq. Cryst*. **2008**, *481*, 56.

[80] S. Belviso, F. Cammarota, R. Rossano, F. Lelj, *J. Porphyrins Phthalocyanines* **2016**, *20*, 223.

[81] S. Belviso, G. Ricciardi, F. Lelj, L. Monsù Scolaro, A. Bencini, C. Carbonera, *J. Chem. Soc., Dalton Trans.* **2001**, 1143.

[82] S. Belviso, A. Giugliano, M. Amati, G. Ricciardi, F. Lelj, L. Monsù Scolaro, *Dalton Trans.* **2004**, 305.

[83] S. Belviso, M. Amati, R. Rossano, A. Crispini, F. Lelj, *Dalton Trans*. **2015**, *44*, 2191.

[84] J. Nelson, *Science* **2001**, *293*, 1059.

[85] L. Schmidt-Mende, A. Fechtenkotter, K. Mullen, E. Moons, R. H. Friend, J. D. MacKenzie, S*cience* **2001**, *293*, 1119.

[86] Q. Sun, L. Dai, X. Zhou, L. Li, Q. Li, *Appl. Phys. Lett.* **2007**, *91,* 253505.

[87] J. Mack, M. J. Stillman, Electronic Structure of Metal Phthalocyanine and Porphyrin Complexes from Analysis of the UV-Visible Absorption and Magnetic Circular Dichroism Spectra and Molecular Orbital Calculations. In *The Porphyrin Handbook*, *Vol 16. Phthalocyanines: Spectroscopic and Electrochemical Characterization*, Kadish, K. M.; Smith, K. M.; Guilard, R., Eds.; Academic Press: New York, **2003**; pp. 43-116.

[88] M. Gouterman, *J. Mol. Spectrosc.* **1961**, *6*, 138.

[89] M. Gouterman, Optical Spectra and Electronic Structure of Porphyrins and Related Rings. In *The Porphyrins*; Dolphin, D., Ed.; Academic Press: New York, **1978**; Vol. 3, pp. 1-165.

[90] G. de la Torre, P. Vázquez, F. Agulló-López, T. Torres, *Chem. Rev.* **2004**, *104,* 3723.

[91] H. Xu, R. Chen, Q. Sun, W. Lai, Q. Su, W. Huang, X. Liu, *Chem. Soc. Rev*. **2014**, *43*, 3207.

[92] G. Ricciardi, S. Belviso, F. Lelj, S. Ristori, *J. Porphyrins Phthalocyanines*, **1998**, *2*, 177.

[93] N. Miyaura, A. Suzuki, *Chem. Rev.* **1995**, *95*, 2457.

[94] C. M. Muzzi, C. J. Medforth, L. Voss, M. Cancilla, C. Lebrilla, J.-G. Ma, J. A. Shelnutt, K. M. Smith, *Tetrahedron Lett*. **1999**, *40*, 6159.

[95] S. M. Kozlov, F. Vines, A. Gorling, *Carbon* **2012**, *50*, 2482 .

[96] C. A. Hunter, K. R. Lawson, C. Perkins, C. J. Urch, *J. Chem. Soc. Perk. T.* **2001**, *2*, 651.

[97] X.-Q. Chen, X.-Y. Liao, J.-G. Yu, F.-P. Jiao, X.-Y. Jiang, *Nano* **2013**, *8*, 1330002.

[98] G. Liu, F. Wang, S. Chaunchaiyakul, Y. Saito, A. Bauri, T. Kimura, Y. Kuwahara, N. Komatsu, *J. Am. Chem. Soc*. **2013**, *135*, 4805.

[99] G. Liu, A. F. M. M. Rahman, S. Chaunchaiyakul, T. Kimura, Y. Kuwahara, N. Komatsu, *Chem. Eur. J*. **2013**, *19*, 16221.

[100] G. Liu, T. Yasumitsu, L. Zhao, X. Peng, F. Wang, A. K. Bauri, S. Aonuma, T. Kimura, N. Komatsu, *Org. Biomol. Chem.* **2012**, *10*, 5830.

[101] F. Wang, K. Matsuda, A. F. M. M. Rahman, T. Kimura, N. Komatsu, *Nanoscale* **2011**, *3*, 4117.

[102] X. Peng, N. Komatsu, S. Bhattacharya, T. Shimawaki, S. Aonuma, T. Kimura, A. Osuka, *Nat. Nanotech.* **2007**, *2*, 361.

[103] C. A. Hunter, *Chem. Soc. Rev.* **1994**, 101.

[104] S. Superchi, M. I. Donnoli, G. Proni, G. P. Spada, C. Rosini, *J. Org. Chem.* **1999**, *64*, 4762.

[105] J. Tomasi, B. Mennucci, R. Cammi, *Chem. Rev*. **2005**, *105*, 2999.

[106] G. Scalmani, M. J. Frisch, *J. Chem. Phys.* **2010**, *132*, 114110.

[107] F. Ceccacci, G. Mancini, P. Mencarelli, C. Villani, *Tetrahedron:Asymmetry* **2003**, *14*, 3117.

[108] S. Levi Mortera, R. Sabia, M. Pierini, F. Gasparrini, C. Villani, *Chem. Commun*. **2012**, *48*, 3167.

[109] R. Sabia, A. Ciogli, M. Pierini, F. Gasparrini, C. Villani, *J. Chromatogr. A* **2014**, *1362*, 144.

[110] S. Belviso, E. Santoro, F. Lelj, D. Casarini, C. Villani, R. Franzini, S. Superchi, unpublished results.

[111] E. L. Eliel, S. H. Wilen, L. N. Mander, *Stereochemistry of Organic Compounds*. Wiley: New York, NY, **1994**. pp. 1142-1148.

[112] M. Oki, Recent Advances in Atropisomerism in *Topics in Stereochemisty*, Eliel, E. L., Allinger, N. L., Wilen, S. H. Eds. John Wiley & Sons: New York, **1983**; Vol. 14, pp. 3-4.

[113] S. A. Asher, *Anal. Chem*. **1984**, *56*, 720.

[114] H. H. Jaffé, M. Orchin, *Theory and Applications of Ultraviolet Spectroscopy*; John Wiley & Sons: New York, **1962**; pp. 334-335.

[115] C. Nitschke, S. M. O'Flaherty, M. Kroll, W. J. Blau, *J. Phys. Chem. B* **2004**, *108*, 1287.

[116] A. A. Esenpinar, M. Bulut, *Dyes Pigments* **2008**, *76*, 249.

[117] M. J. Stillman, T. Nyokong, In *Phthalocyanines: Properties and Applications*; Leznoff, C. C.; Lever, A. B. P., Eds.; VCH: Weinheim, **1989**; p 139.

[118] H. Isago, H. Fujita, *J. Porphyrins Phthalocyanines* **2013**, *17*, 447.

[119] H. Isago, *Optical Spectra of Phthalocyanines and Related Compounds*; Springer: Japan, **2015**; pp. 79-80.

[120] H. M. Grant, P. Mctigue, D. G. Ward, *Australian J. Chem.* **1983**, *36*, 2211.

[121] *CRC Handbook of Chemistry and Physics, 91st Edition*; Haynes, W. H. Ed.; CRC Press, Taylor & Francis: Boca Raton, FL, **2010**. p. 8-42.

[122] W. L. F. Armarego, C. L. L. Chai, *Purification of Laboratory Chemicals 5$^{th}$ Edition;* Butterworth-Heinemann: Burlington, MA, **2003**. pp. 215-216.

[123] L. A. Bottomley, W. H. Chiou, *J. Electroanal. Chem.* **1986**, *198*, 331.

[124] E. A. Ough, K. A. M. Creber, M. Stillman, *Inorg. Chim. Acta* **1996**, *246*, 361.

[125] The number of exchanged electrons in each of these redox processes is also confirmed by analyzing the width of the peaks at half height in the corresponding differential small pulse amplitude voltammograms. Considering that in our pulse voltammograms obtained using a pulse amplitude of 50 mV, the measured values of the peak half widths are greater than 90.4 mV, a one–electron reduction processes has to be associated with all the studied compounds. See: E. P. Perry, R. A. Osteryoung, *Anal. Chem.* **1965**, *37*,1634.

[126] Y. Liu, M. S. Liu, X-C. Li, A. K-Y. Jen, *Chem. Mater.* **1998**, *10*, 3301.

[127] J. Pommerehne, H. Vestweber, W. Guss, R. F. Mahrt, H. Bassler, M. Porch, J. Daub, *Adv. Mater*. **1995**, *7*, 551.

[128] B. W. D. Andrade, S. Datta, S. R. Forrest, P. Djurovich, E. Polikarpov, M. E. Thompson, *Org. Electron.* **2005**, *6*, 11.

[129] R. J. Davis, M. T. Lloyd, S. R. Ferreira, M. J. Bruzek, S. E. Watkins, L. Lindell, P. Sehati, M. Fahlman, J. E. Anthony, J. W. P. Hsu, *J. Mater. Chem.* **2011**, *21*, 1721.

[130] S. Suzuki, C. Bower, Y. Watanabe, O. Zhou, *Appl. Phys. Lett*. **2000**, *76*, 4007.

[131] J. P. Sun, Z. X. Zhang, S. M. Hou, G. M. Zhang, Z. N. Gu, X. Y. Zhao, W. M. Liu, Z. Q. Xue, *Appl. Phys. A* **2002**, *75*, 479.

[132] H. Xu, R. Chen, Q. Sun, W. Lai, Q. Su, W. Huang, X. Liu, *Chem. Soc. Rev.* **2014**, *43*, 3259.

[133] Y. Hernandez, V. Nicolosi, M. Lotya, F. M. Blighe, Z. Sun, S. De, I. T. McGovern, B. Holland, M. Byrne, Y. K. Gun'Ko, J. J. Boland, P. Niraj, G. Duesberg, S. Krishnamurthy, R. Goodhue, J. Hutchison, V. Scardaci, A. C. Ferrari, J. N. Coleman, *Nat. Nanotech*. **2008**, *3*, 563.

[134] T. Hasan, P. H. Tan, F. Bonaccorso, A. G. Rozhin, V. Scardaci, W. I. Milne, A. C. Ferrari, *J. Phys. Chem. C* **2008**, *112*, 20227.

[135] S. Casaluci, M. Gemmi, V. Pellegrini, A. Di Carlo, F. Bonaccorso, *Nanoscale*, **2016**,*8*, 5368.

[136] F. Bonaccorso, T. Hasan, P. H. Tan, C. Sciascia, G. Privitera, G. Di Marco, P. G. Gucciardi, A. C. Ferrari, *J. Phys. Chem. C* **2010**, *114*, 17267.

[137] C. Ehli, G. M. A. Rahman, N. Jux, D. Balbinot, D. M. Guldi, F. Paolucci, M. Marcaccio, D. Paolucci, M. Melle-Franco, F. Zerbetto, S. Campidelli, M. Prato, *J. Am. Chem. Soc.* **2006**, *128*, 11222.

[138] F. Bonaccorso, A. Bartolotta, J. N. Coleman, C. Backes, *Adv. Mater.* **2016**, *28*, 6136.

[139] A. C. Ferrari, J. C. Meyer, V. Scardaci, C. Casiraghi, M. Lazzeri, F. Mauri, S. Piscanec, D. Jiang, K. S. Novoselov, S. Roth, A. K. Geim, *Physical Review Letters* **2006**, *97*, 187401.

[140] A. C. Ferrari, J. Robertson, *Physical Review B* **2001**, *64*, 075414.

[141] J. Hassoun, F. Bonaccorso, M. Agostini, M. Angelucci, M. G. Betti, R. Cingolani, M. Gemmi, C. Mariani, S. Panero, V. Pellegrini, B. Scrosati, *Nano Lett.* **2014**, *14*, 4901.

[142] C. Casiraghi, A. Hartschuh, H. Qian, S. Piscanec, C. Georgi, A. Fasoli, K. S. Novoselov, D. M. Basko, A. C. Ferrari, *Nano Lett.* **2009**, *9*, 1433.

[143] A. C. Ferrari, D. M. Basko, *Nat. Nanotech.* **2013**, *8*, 235.

[144] Z. Mo, X. Zhu, G. Wang, W. Han, R. Guo, *J. Mater. Res.* **2014**, *29*, 2156.

[145] G. Konstantatos, M. Badioli, L. Gaudreau, J. Osmond, M. Bernechea, F. P. G. de Arquer, F. Gatti, F. H. L. Koppens, *Nat. Nanotech.* **2012**, *7*, 363.

[146] Z. Sun, Z. Liu, J. Li, G. Tai, S.-P Lau, F. Yan, *Adv. Mater.* **2012**, 24, 5878.

[147] F. Bonaccorso, A. Lombardo, T. Hasan, Z. Sun, L. Colombo, A. C. Ferrari, *Mater. Today*, **2012**, *15*, 564.

[148] http://www.sigmaaldrich.com/catalog/product/aldrich/775533?lang=it®ion=IT.

[149] M. S. Arnold, A. A. Green, J. F. Hulvat, S. I. Stupp, M. C. Hersam, *Nature Nanotech*. **2006**, *1*, 60.

[150] M. S. Arnold, S. I. Stupp, M. C. Hersam, *Nano Lett.*, **2005**, *5*, 713.

[151] S.-Y. Chen, Y.-Y. Lu, F.-Y. Shih, P-H Ho, Y.-F. Chen, C.-W. Chen, Y.-T Chen, W.-H. Wang, *Carbon* **2013**, *63*, 23.

[152] For a review see: F. H. L. Koppens, T. Mueller, P. Avouris, A. C. Ferrari, M. S. Vitiello, M. Polini, *Nat. Nanotech.* **2014**, *9*, 780.

[153] K.-J. Baeg, M. Binda, D. Natali, M. Caironi, Y.-Y. Noh, *Adv. Mater.* **2013**, *25*, 4267.

[154] D. J. Finn, M. Lotya, G. Cunningham, R. J. Smith, D. McCloskey, J. F. Donegan, J. N. Coleman, *J. Mater. Chem.* C **2014**, *2*, 925.

[155] F. Withers, H. Yang, L. Britnell, A. P. Rooney, E. Lewis, A. Felten, C. R. Woods, V. Sanchez Romaguera, T. Georgiou, A. Eckmann et al., *Nano Lett.* **2014**, *14*, 3987.

[156] F. R. Keene, Ed. *Chirality in Supramolecular Assemblies: Causes and Consequences*, Wiley: Chichester, UK, **2016**.

[157] V. Tasco, M. Esposito, F. Todisco, A. Benedetti, M. Cuscunà, D. Sanvitto, A. Passaseo, *Appl. Phys. A.*, **2016**, *122*, 1.

[158] K. Sugawara, N. Nakamura, Y. Yamane, S. Hayase, T. Nokami, T. Itoh, *Green Energy & Environment.* **2016**, *1*, 149.

[159] C. Liu, G. Yang, Y. Si, Y. Liu, X. Pan, *J. Mater. Chem. C*, **2017**, *5*, 3495.

[160] P. Josse, L. Favereau, C. Shen, S. Dabos-Seignon, P. Blanchard, C. Cabanetos, J. Crassous, *Chem. Eur. J.* **2017**, *23*, 6277.

[161] G. Ricciardi, S. Belviso, G. Giancane, R. Tafuro, T. Wagner, L. Valli, *J. Phys. Chem. B,* **2004**, *108*, 7854.

[162] A. J. Bard, L. R. Faulkner, *Electrochemical Methods, Fundamentals and Appplications*; John Wiley & Sons: New York, **1980**.

[163] M. J. Frisch, G. W. Trucks, H. B. Schlegel, G. E. Scuseria, M. A. Robb, J. R. Cheeseman, G. Scalmani, V. Barone, B. Mennucci, G. A. Petersson, H. Nakatsuji, M. Caricato, X. Li, H. P. Hratchian, A. F. Izmaylov, J. Bloino, G. Zheng, J. L. Sonnenberg, M. Hada, M. Ehara et al., Gaussian 09, Revision D.01. Gaussian, Inc., Wallingford CT, **2013**.

[164] Y. Zhao, D. G. Truhlar, *Theor. Chem. Acc.* **2008**, *120*, 215.

[165] J. S. Binkley, J. A. Pople, W. J. Hehre, *J. Am. Chem. Soc*. **1980**, *102*, 939.

[166] M. S. Gordon, J. S. Binkley, J. A. Pople, W. J. Pietro, W. J. Hehre, *J. Am. Chem. Soc.* **1982**, *104,* 2797.

[167] M. Casida, Time dependent density functional response theory for molecules. In *Recent advances in Density Functional Methods*; Chong, D. P., Eds.; World Scientific: Singapore, **1995**; Vol. 1, p 155.

[168] M. E. Casida, C. Jamorski, K. C.; Casida, D. R. Salahub, *J. Chem. Phys.* **1998**, *108*, 4439.

[169] R. E. Stratmann, G. E.; Scuseria, M. J. Frisch, *J. Chem. Phys.* **1998**, *109*, 8218.

[170] G. Schaftenaar, J. H. Noordik, *J. Comput.-Aided Mol. Design* **2000**, *14*, 123.

[171] A. Capasso, A. E. Del Rio Castillo, H. Sun, A. Ansaldo, V. Pellegrini, F. Bonaccorso, *Solid State Comm.* **2016**, *224*, 53.

[172] A. C. Ferrari, J. Robertson, *Phys. Rev. B* **2000**, *61*, 14095.

# Figures and Schemes

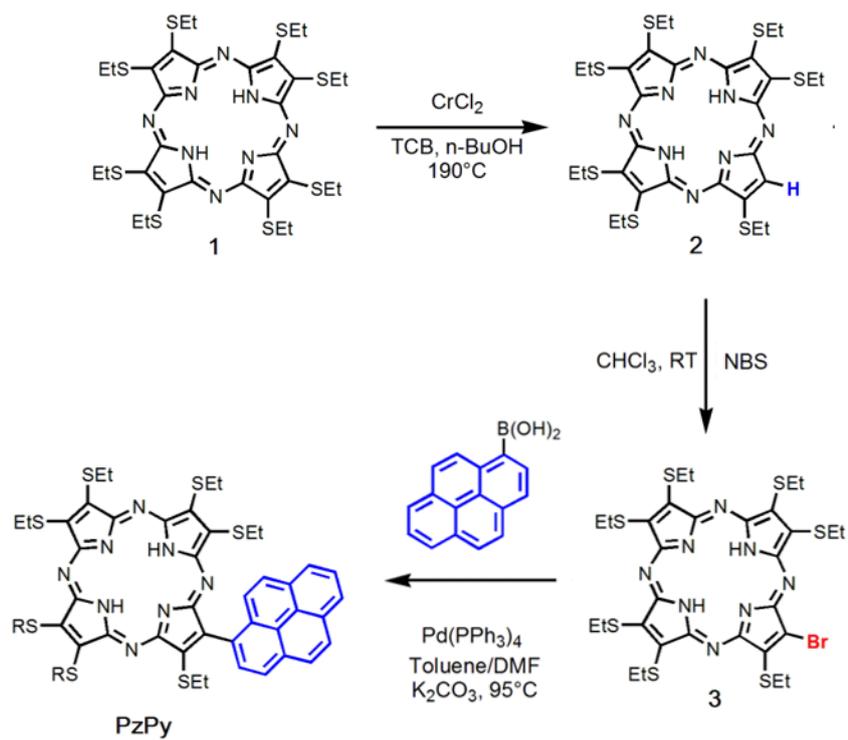

**Scheme 1**

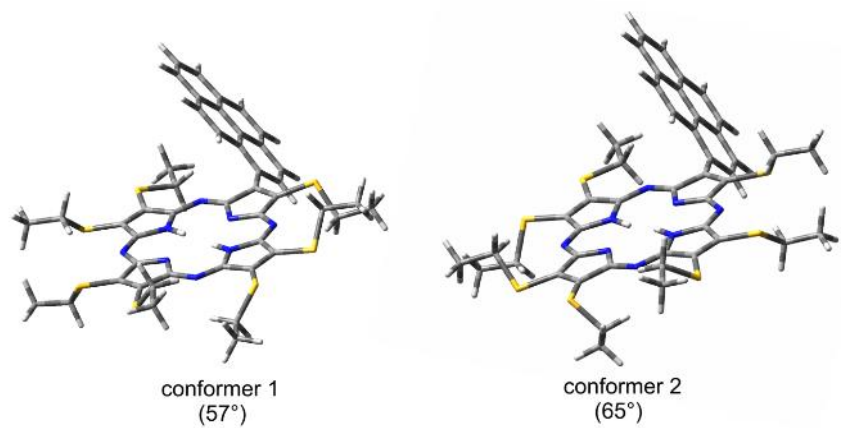

**Figure 1.** Computed (DFT/M06/6-311G(d,p)/CH$_2$Cl$_2$) most stable conformers of **PzPy**.

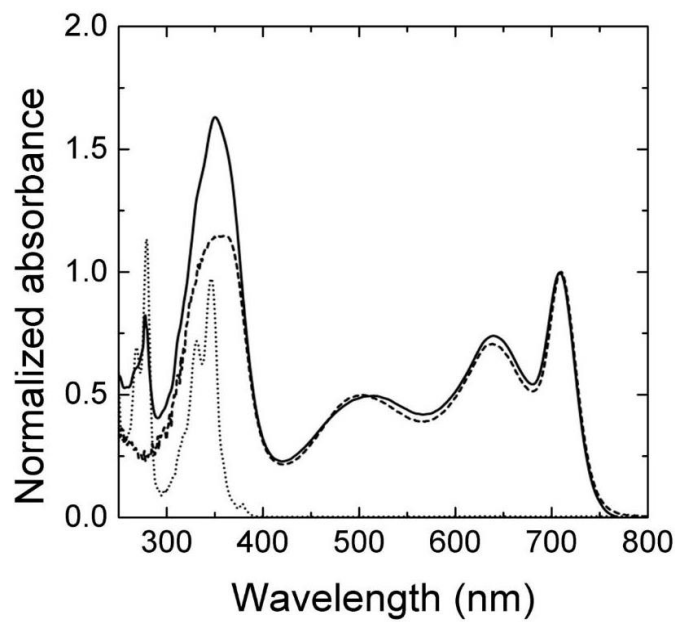

**Figure 2.** Experimental (CH$_2$Cl$_2$) UV-vis spectra for **PzPy** (solid line), porphyrazine **1** (dashed line), and 1-pyrene boronic acid (dotted line).

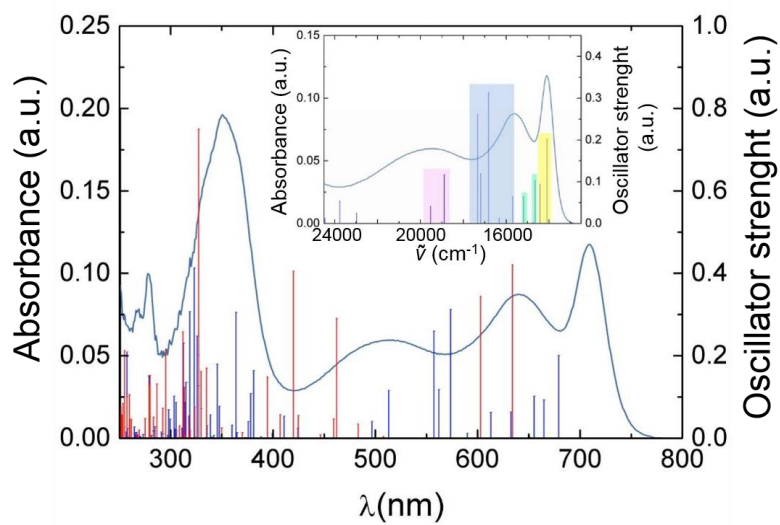

**Figure 3.** TD-DFT Kohn-Sham first 75 singlet states computed at the M06/6-311G(d,p)/CH$_2$Cl$_2$ level of theory for conformers 1 and 2 (blue and red sticks respectively); continuous line: UV-vis **PzPy** spectrum in CH$_2$Cl$_2$. *Inset:* singlet states in the 23500÷12500 cm$^{-1}$ (425-800 nm) range for conformer 1. Energies of the transitions in the inset have been bathochromically shifted by 0.077 eV (621 cm$^{-1}$). Shaded areas have the same coloring as shaded areas on the occupied MO levels involved in the transitions shown in Figure 4. All reported excitations encompass LUMO and LUMO+1 ("pale red" shaded on the same Figure 4) and their composition is detailed in Table S3 in SI.

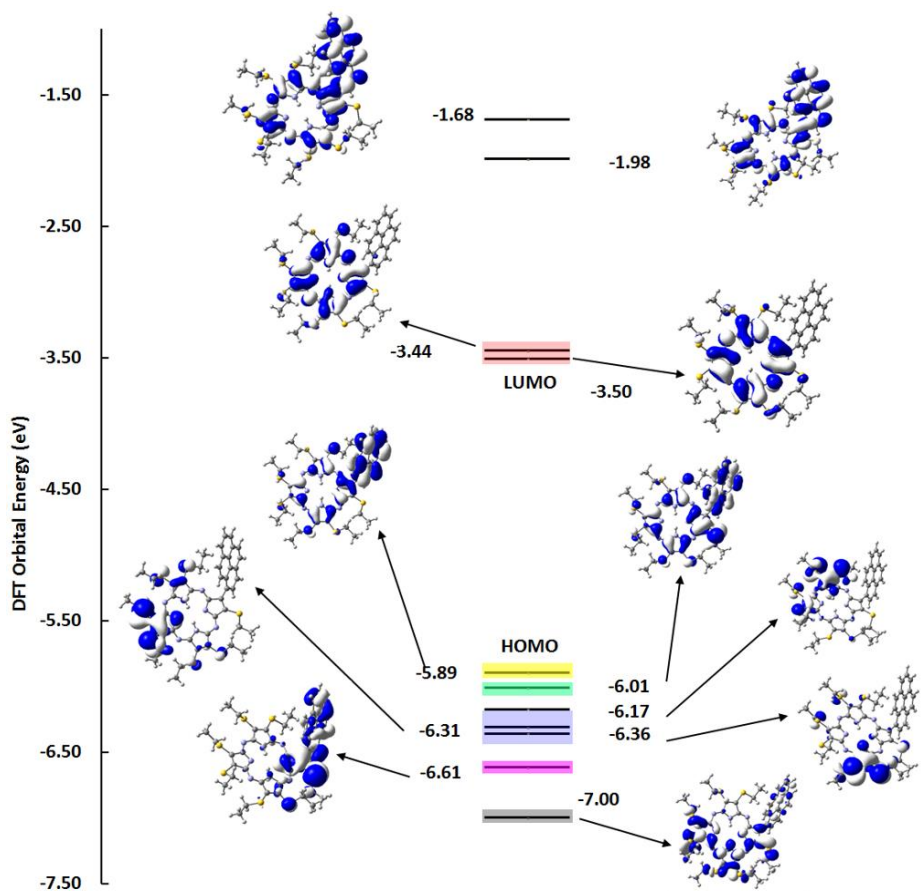

**Figure 4.** DFT Kohn-Sham orbital energies at the M06/6-311G(d,p)/CH$_2$Cl$_2$ level of theory involving the excitations contributing to the states in the 23000÷12500 cm$^{-1}$ (435-800 nm) range computed for conformer 1 of **PzPy**. Shaded areas have the same coloring as the transitions in the spectrum of Figure 3 inset indicating the MO's from where transitions start. Surface contours are drawn at 0.02(e/au$^3$)$^{1/2}$.

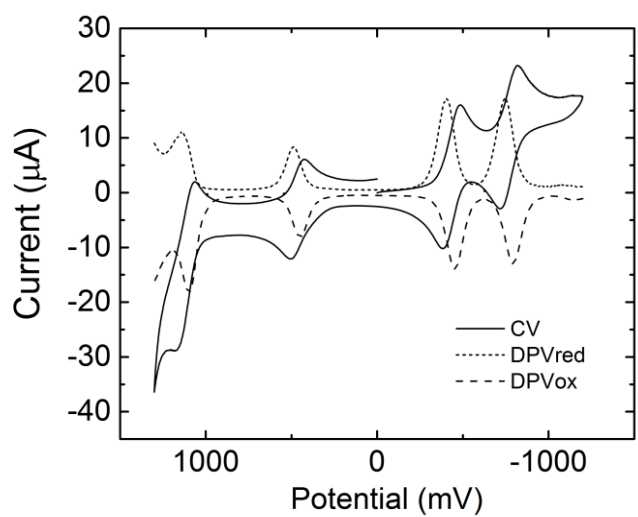

**Figure 5.** Cyclic voltammogramm and differential pulse voltammogramms of **PzPy** in CH$_2$Cl$_2$.

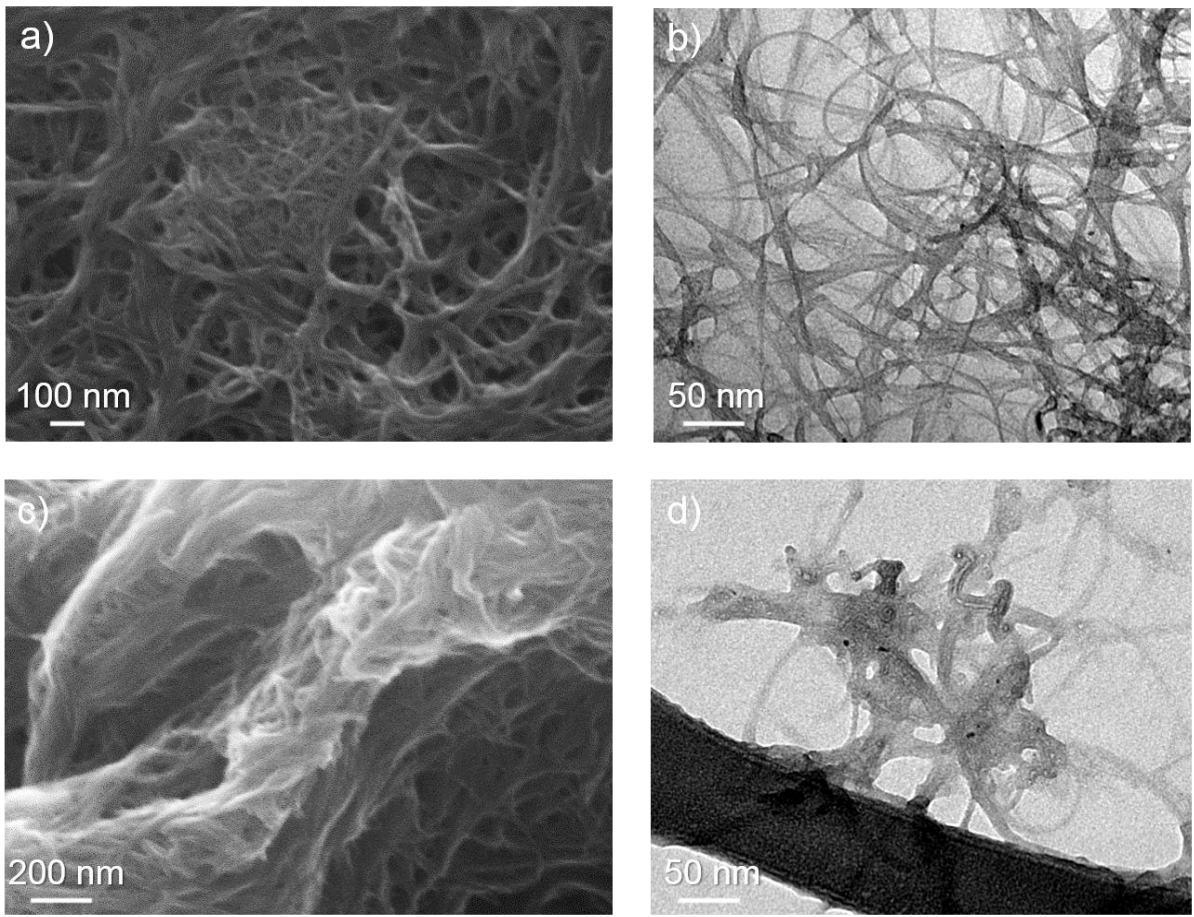

**Figure 6.** Electron microscopy of SWNTs and **PzPy**/SWNT nanohybrid. SEM images of a) SWNTs and c) **PzPy**/SWNT nanohybrid deposited on a Si/SiO$_2$ substrate. TEM images of b) SWNTs and d) **PzPy**/SWNT nanohybrid drop cast from a DMF solution on a lacey carbon TEM grid.

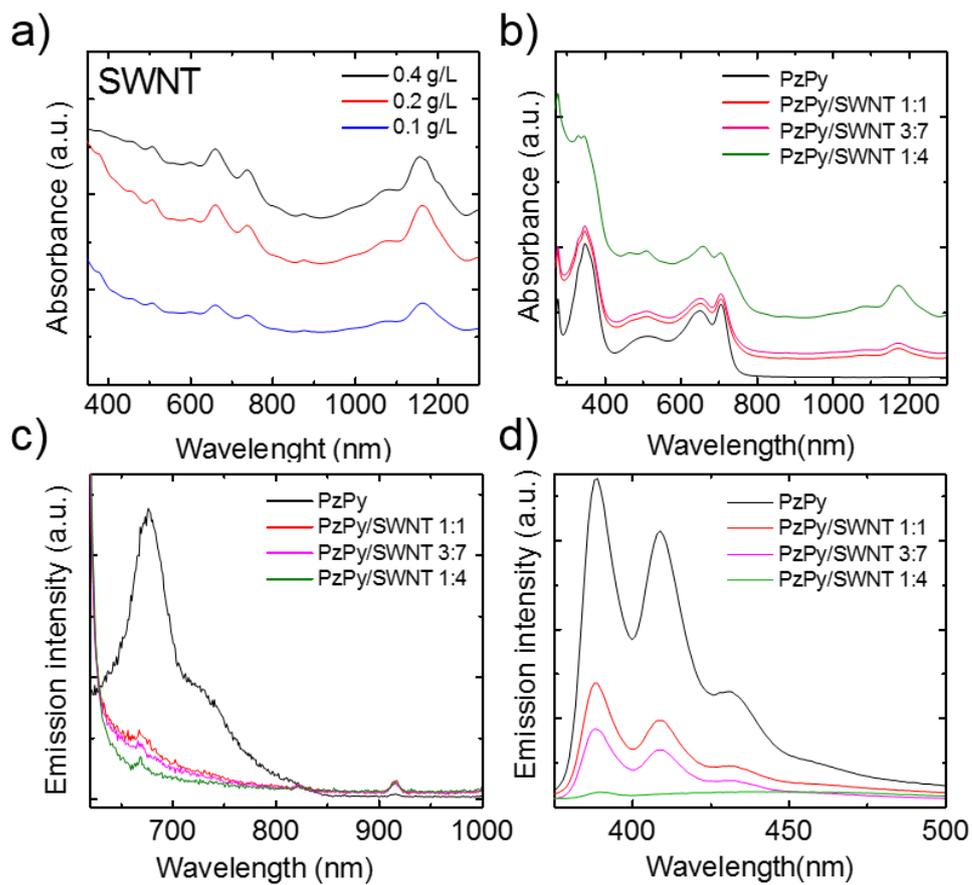

**Figure 7.** UV-vis absorption spectra of DMF dispersions of a) SWNTs and b) **PzPy**/SWNT nanohybrids in DMF at various weight ratios (see Experimental). Emission spectra of **PzPy**/SWNT nanohybrids with excitation at c) 610 nm and d) 350 nm.

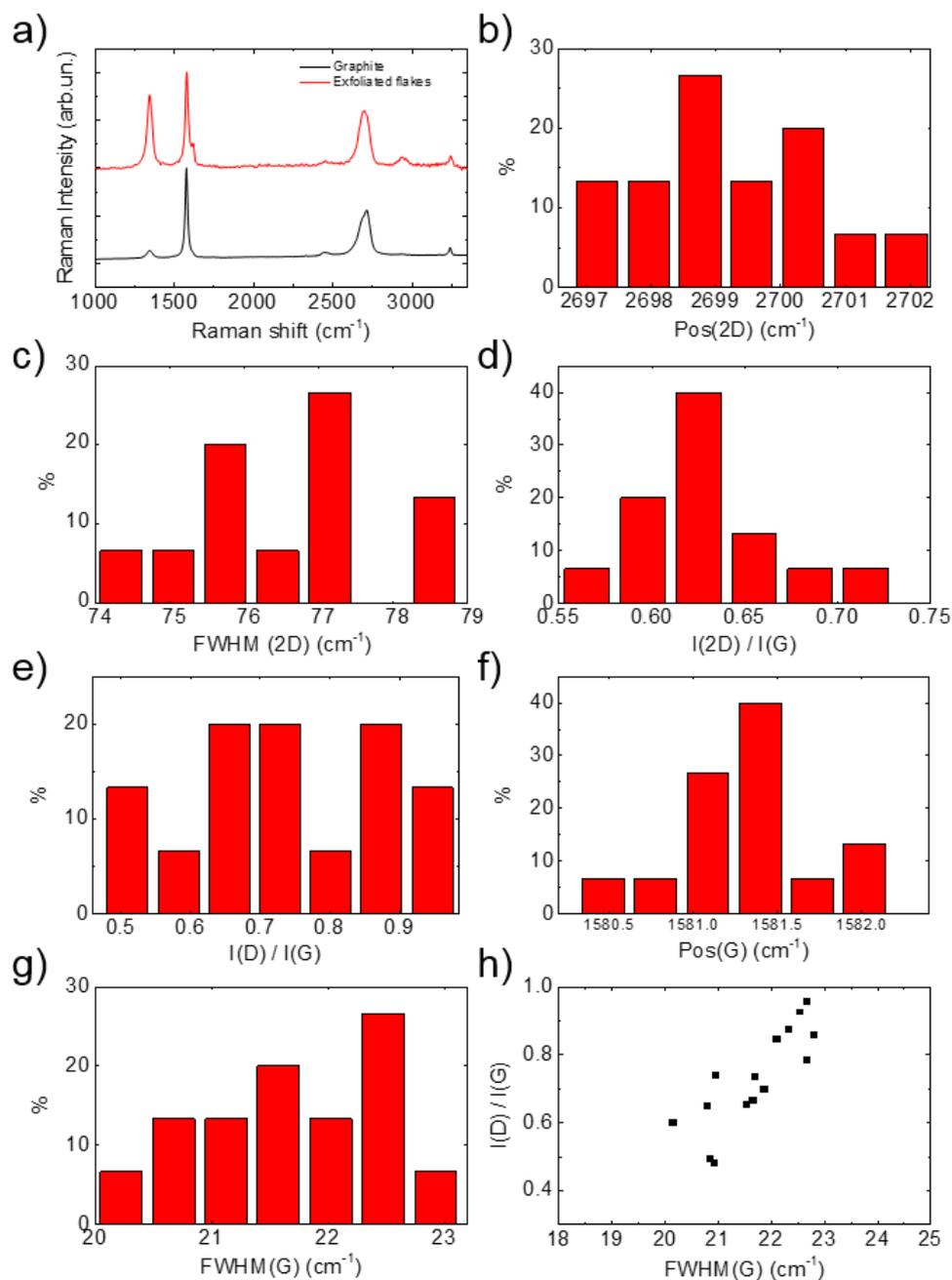

**Figure 8.** Raman spectra and statistics of the GNF drop cast from a DMF dispersion onto a Si/SiO$_2$ substrate (excitation at λ = 532 nm).

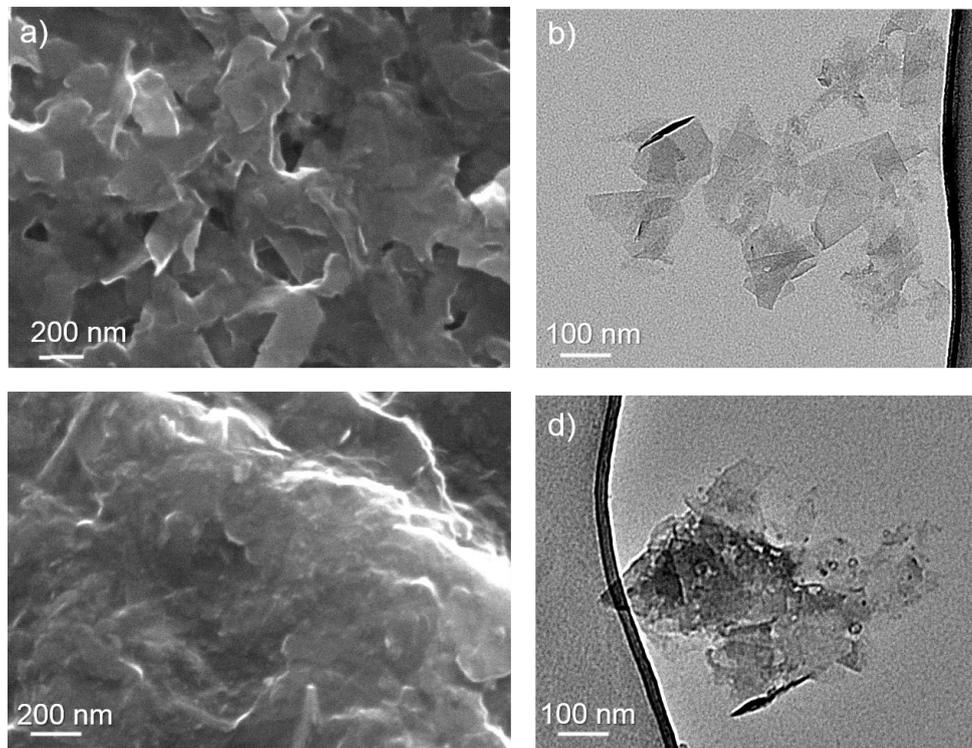

**Figure 9.** Electron microscopy of GNF and **PzPy**/GNF. SEM images of a) GNF and c) **PzPy**/GNF nanohybrid deposited on a Si/SiO$_2$ substrate. TEM images of b) GNF and d) **PzPy**/GNF nanohybrid drop cast from a DMF dispersion on a lacey carbon TEM grid.

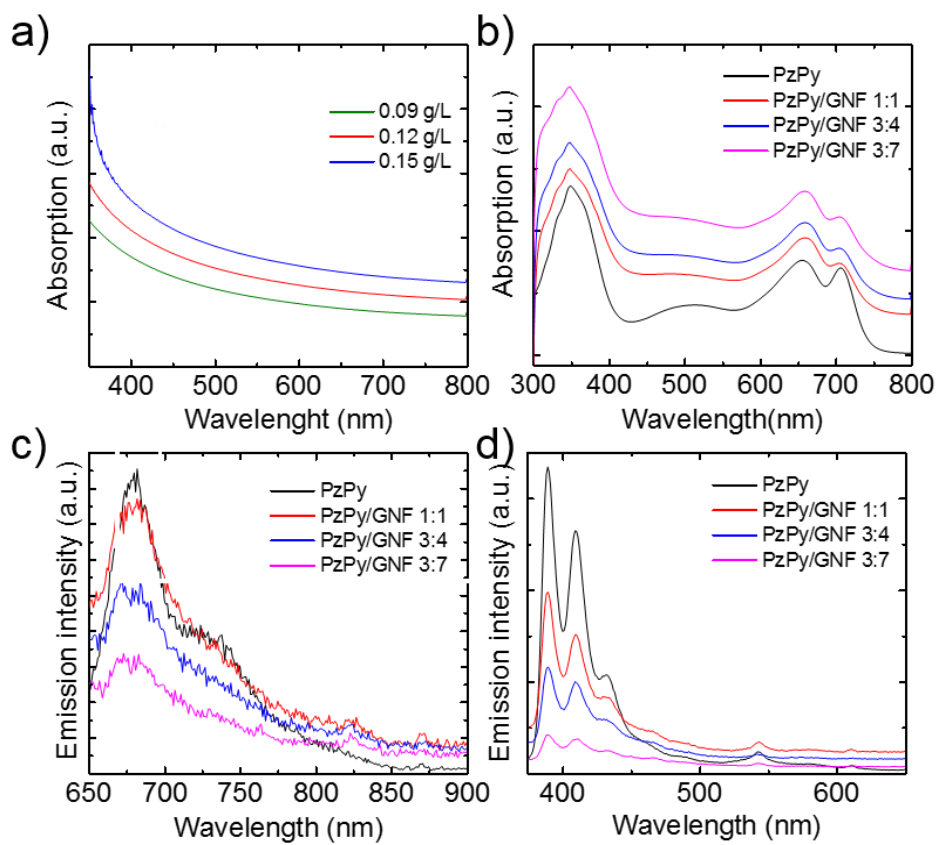

**Figure 10.** UV-vis absorption spectra of DMF dispersions of a) GNF and b) **PzPy**/GNF nanohybrid at various weight ratios. Emission spectra of DMF dispersions of **PzPy**/GNF nanohybrid at various weight ratios with excitation at c) 610 nm and d) 350 nm.

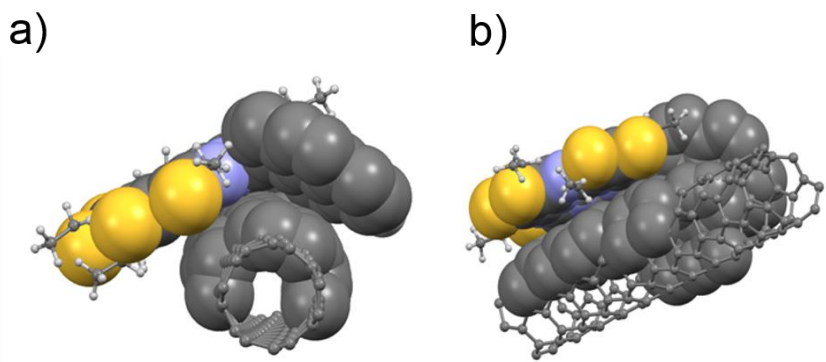

**Figure 11.** Molecular models of the **PzPy**/SWNT nanohybrid.

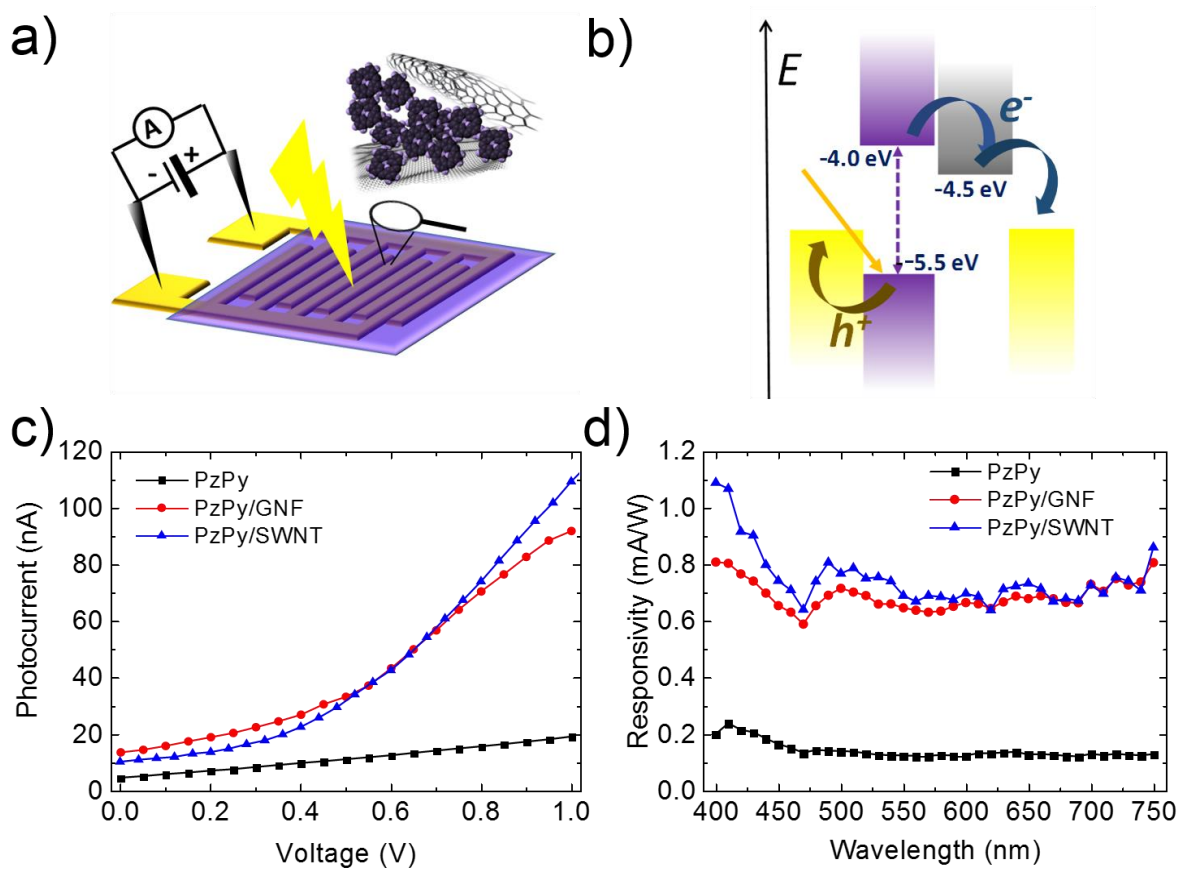

**Figure 12.** (a) Sketch of devices architecture and film composition. Interdigitated area: 0.5 × 0.5 cm². Material mass loading of the active film: 10 μg cm$^{-2}$. (b) Energy level diagram of the materials and photodetector working principle, showing exitation by light in **PzPy** followed by electron transfer to the nanocarbon acceptor and hole transfer to the metal electrode. (c) Photocurrent in **PzPy** and **PzPy** nanohybrids under 500 nm light. (d) Spectral responsivity in **PzPy** and **PzPy/** nanohybrids measured at 1 V.

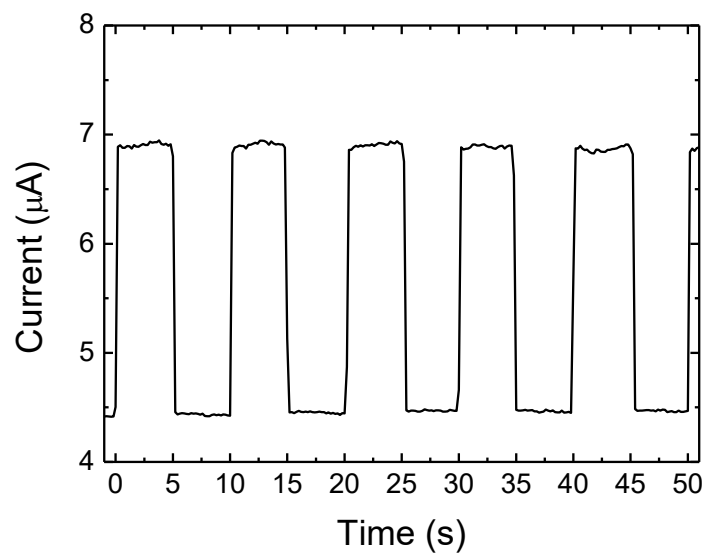

**Figure 13. PzPy/**GNF current response (at 1 V bias) to on/off cycles under 1 SUN illumination.

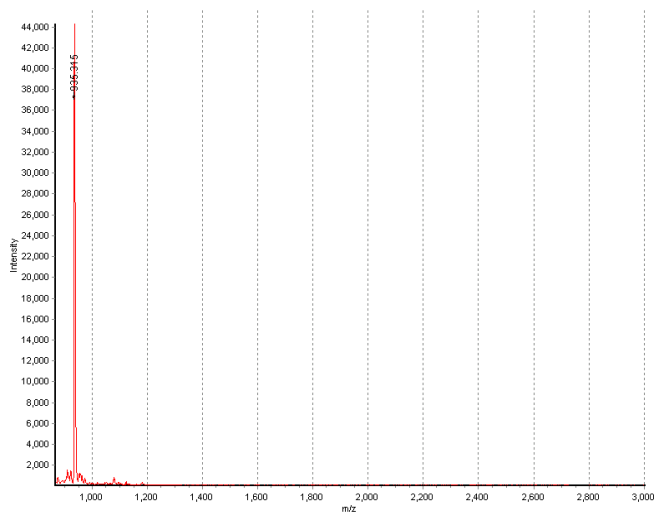
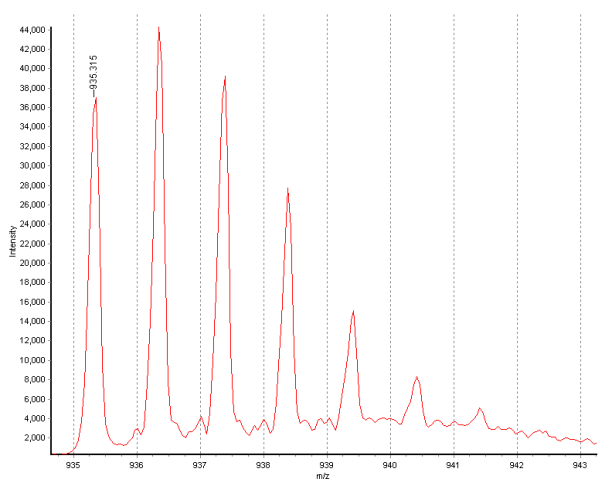

**Figure S1.** MS-MALDI spectrum of compound **PzPy**.

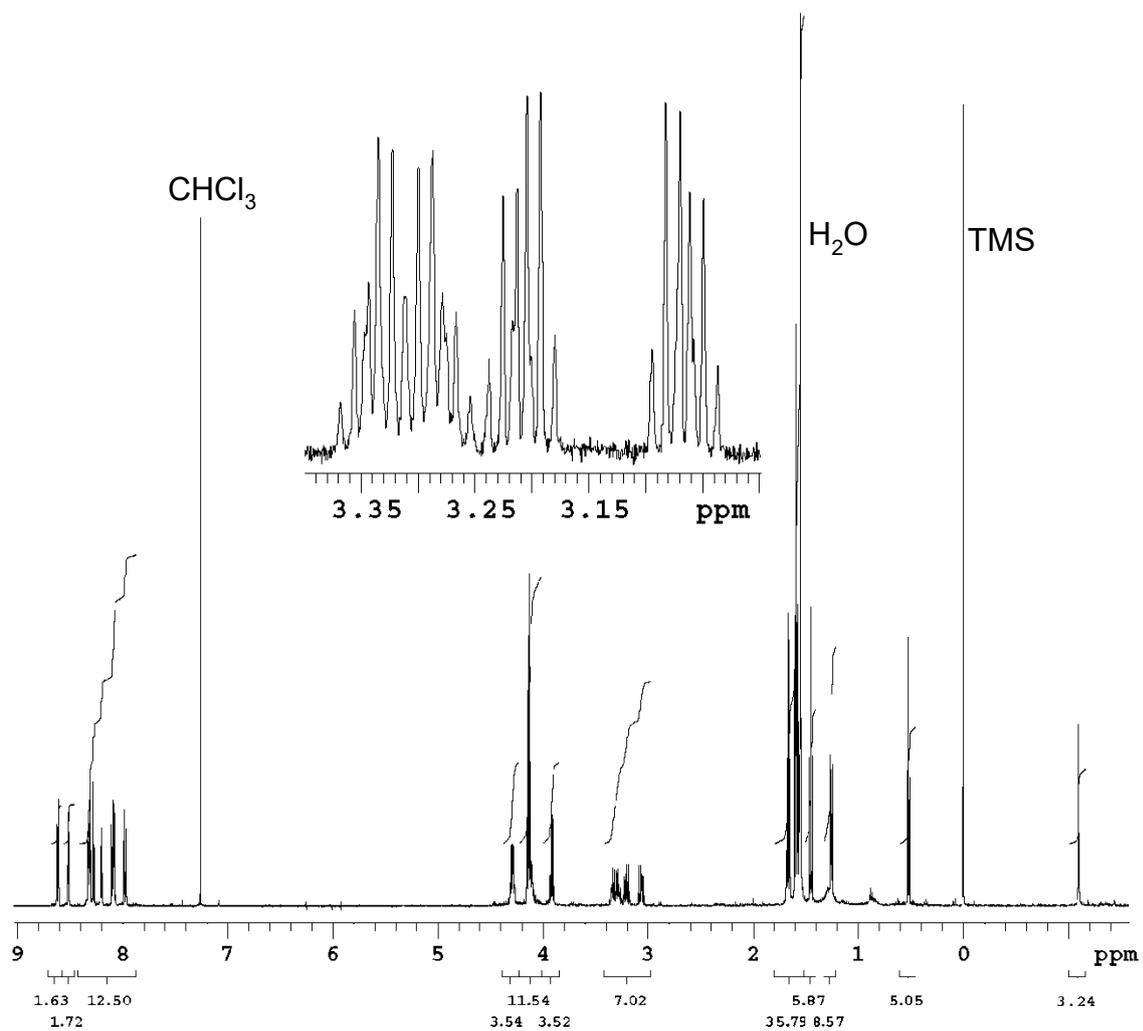

**Figure S2.** $^1$H NMR (600 MHz) spectrum of **PzPy** in CDCl$_3$.

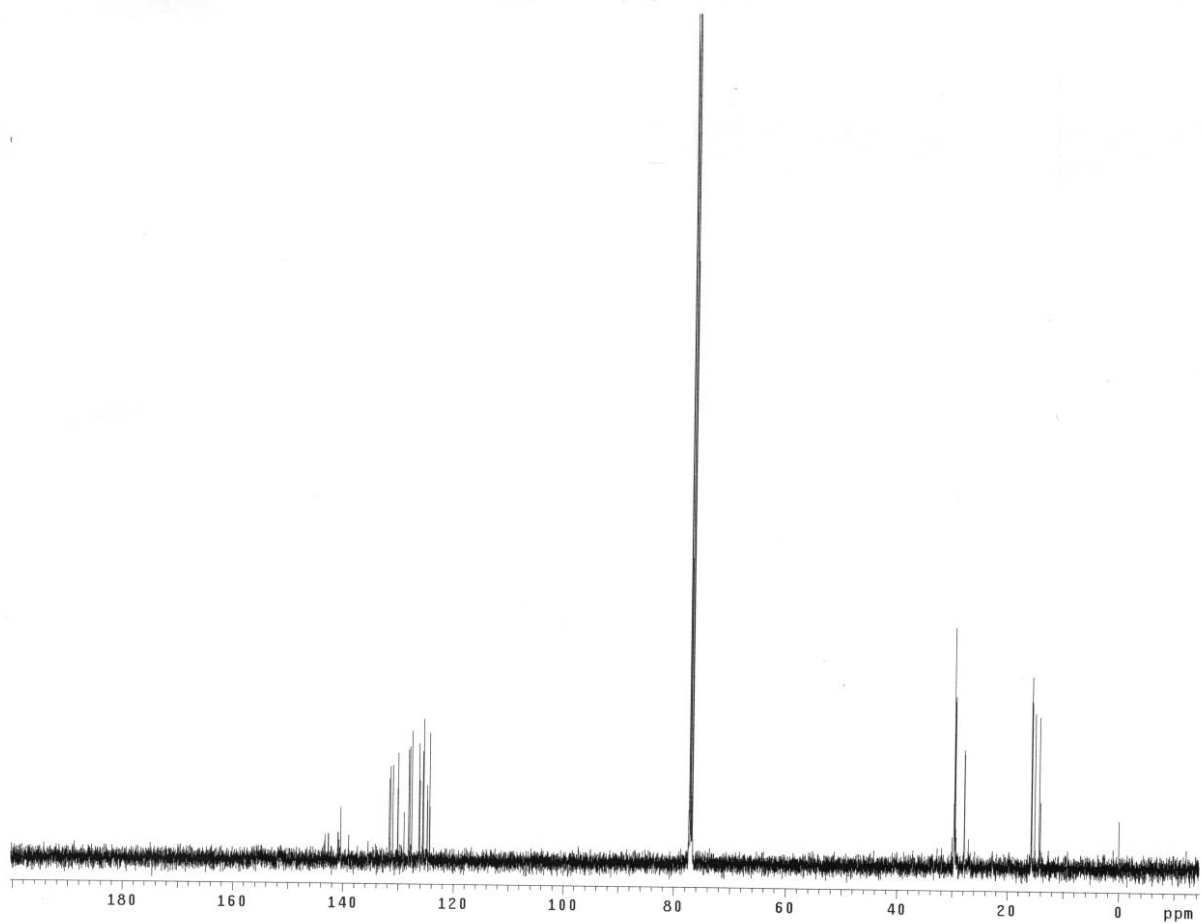

**Figure S3.** $^{13}$C NMR (150 MHz) spectrum of **PzPy** in CDCl$_3$.

**Table S1.** Computed conformers Boltzmann distribution of **PzPy**.

|  | M06/6-311G(d,p) (CH$_2$Cl$_2$) | | M06/6-311G(d,p) (n-hexane) | | M06/6-311G(d,p) (DMF) | |
| --- | --- | --- | --- | --- | --- | --- |
| Conformer | ΔG (Kcal/mol) | Pop (%) | ΔG (Kcal/mol) | Pop (%) | ΔG (Kcal/mol) | Pop (%) |
| **1** | 0.00 | 64.7 | 0.00 | 69.8 | 0.00 | 91.4 |
| **2** | 0.36 | 35.3 | 0.50 | 30.2 | 1.40 | 8.6 |

**Computational study on the PzPy molecular orbitals (MO) composition, energies, and allied electronic transitions.**

The UV-vis computed spectrum at the M06/6-311G(d,p)/CH$_2$Cl$_2$ level of theory in the 23500÷12500 cm$^{-1}$ (435-800 nm) range is characterized by twelve transitions S$_0$ → S$_n$ (n=1,12) in case of conformer 1 and 2. Conformer 2 showing transitions hypsochromically shifted compared to conformer 1 (see Figure 3). The band at ≈14104 cm$^{-1}$ (709 nm) appears to be originated by the S$_1$ with contributions from S$_2$ and S$_3$. Band at ≈15625 cm$^{-1}$ (640 nm) has contribution from $S_5$ and $S_6$ and a contribution from $S_1$ of conformer 2 and, toward the blue, from $S_8$-$S_{10}$ of conformer 1. The wide band centered at ≈19493 cm$^{-1}$ (513 nm) is contributed by the $S_{11}$-$S_{13}$ of conformer 1 and $S_8$-$S_{12}$ of conformer 2. Figure 4 reports the main molecular orbital (MO) levels composition and the corresponding orbital energies for conformer 1. As reported in Table S3 in S.I. almost all the transitions have multi-determinantal character with at least two configurations contributing to their composition. As a general trend, all the transitions in the aforementioned range are engendered by excitations involving occupied orbitals in the range HOMO ÷ HOMO-5 and the empty ones LUMO and LUMO+1, which are both localized on the porphyrazine core. HOMO and HOMO-1 are characterized by the combination of porphyrazine and pyrene HOMO levels; the three MO levels, (HOMO-2) ÷ (HOMO-4), have larger contribution from each of the pyrrole π orbitals and the lone pairs (n$_\pi$) of sulfur atoms linked to the corresponding C$_\beta$ atoms. HOMO-5 is characterized by the combination of the same pyrrole unit orbitals but now including large contribution from the pyrene HOMO. This behavior is unlike the one found in case of previously studied aryl derivatives,[83] where the reciprocal perturbation is almost negligible because the large dihedral angle between the planes of the substituted hepta(thioethyl)-porphyrazine and the aryl substituent. Hence, the aryl orbital contribution is almost absent. In the present case, although the dihedral angle is even larger because of the bulkiness of the pyrene, there is an "accidental" almost degeneracy (*i.e.*, -5.87 and -5.97 eV at the M06/6-311G(d,p)/CH$_2$Cl$_2$ level of theory for the H-substituted hepta(thioethyl)porphyrazine **2** (Scheme 1) and pyrene, respectively) between the two HOMO levels (Figure S4 in S.I.) that enhances the mixing. However, the small overlap between the two orbitals makes the reciprocal perturbation small. Analogous behavior is found in the case of the LUMO levels of the two moieties (-1.75 and -1.74 eV for the macrocycle **2** and pyrene, respectively) that delocalize the LUMO+2 and LUMO+3 on both (see Figure 4 and Figure S4 in S.I.). $S_1$ and $S_2$ involve excitations from HOMO to LUMO and LUMO+1. The composition of these three MO levels gives to the transitions a well defined charge transfer (pyrene → porphyrazine) character. This, leading upon excitation to separation of electrons and holes on different parts of the molecule, can be relevant for both the exciton formation and lifetime. Besides, $S_3$ has a less marked multi-determinantal character being mainly (81%) described by (HOMO-1)→LUMO excitation. $S_4$ has negligible intensity whilst $S_5$ is described by a (HOMO-1) to (LUMO+1) excitation with two minor components (HOMO-3) to LUMO (11%) and (HOMO-2) to (LUMO+1) (12%). The $S_6$-$S_{10}$ states have multi-determinantal character involving excitations from MO levels in the range (HOMO-2) ÷ (HOMO-4). The band centered at ≈19493 cm$^{-1}$ (513 nm) gets further contributions from $S_{11}$-$S_{12}$ and its higher energy side from $S_{13}$-$S_{14}$. The former state excitations start mainly from HOMO-5 whilst the latter from HOMO-6 and HOMO-7, all showing definite contributions from the pyrene moiety atomic orbitals.

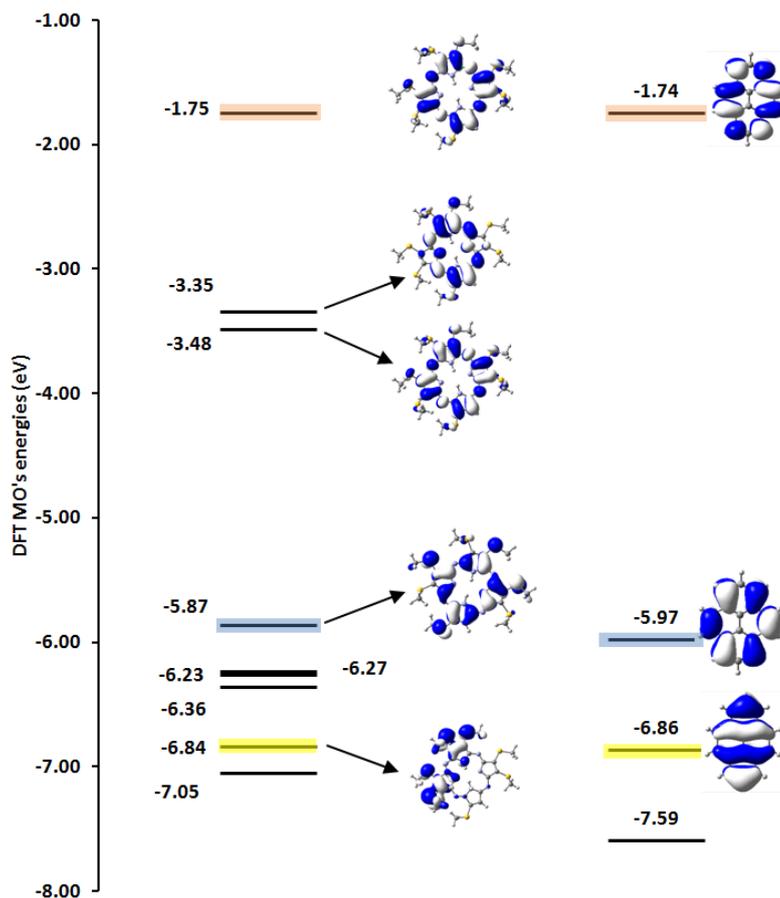

**Figure S4.** Kohn-Sham DFT energies of the MO's computed at the M06/6-311G(d,p) level of theory for the hepta(thioethyl)-porphyrazine **2** and pyrene. Shaded areas "blue" and "red" refers to almost degenerate energy levels of the two fragments significantly mixing in the composite molecule **PzPy**. "Yellow" shaded levels are almost degenerate as well, but the two orbitals do not mix because the lack of any relevant contribution to the $\pi_\perp$ orbital of the porphyrazine $C_\beta$ and the smaller contribution of pyrene C1 where the two fragments are joined. Surface contours are drawn at $0.02(e/au^3)^{1/2}$

## Study of diethylamine effect on PzPy spectra.

The role of diethylamine is confirmed by adding an excess of this base to a **PzPy** solution in not anhydrous DMF. A spectrum identical to that in anhydrous DMF, with disappearance of the lower-energy Q band at 709 nm (Figure S7 in S.I.) was then observed. Computations confirmed that DMF itself is not responsible for the pyrroles deprotonation. In fact, modelling of the interaction of one DMF molecule with **PzPy** at the DFT/M06/6-31G(d)/DMF level shows that the DMF molecule does not give H bonds with the inner pyrrolic hydrogens (Figure S8 in S.I.) and the corresponding TDDFT computed spectrum in DMF is close to the experimental one in non anhydrous solvent.

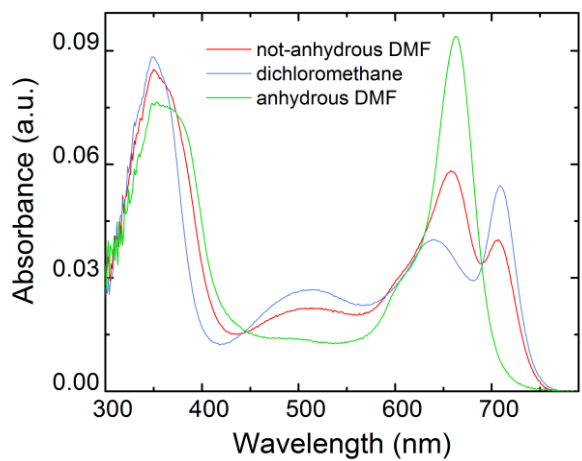

**Figure S5.** UV-vis spectra of compound **PzPy** in in the 300-800 nm range in different solvents.

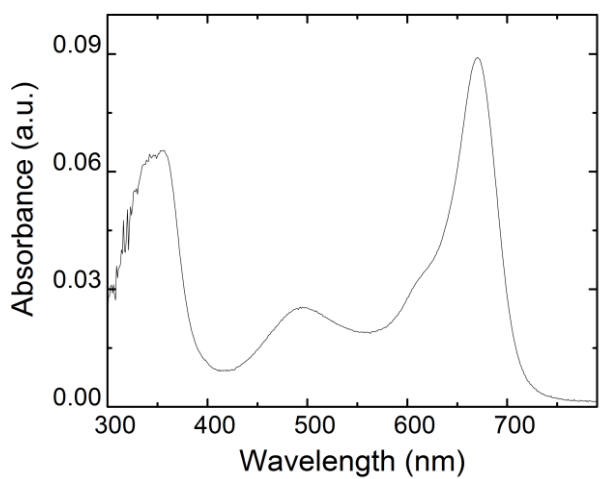

**Figure S6.** Typical UV-vis spectrum of thioalkyl porphyrazine metal complex: case of octakis-(thioethyl) porphyrazinate Copper (II) complex in $CH_2Cl_2$.

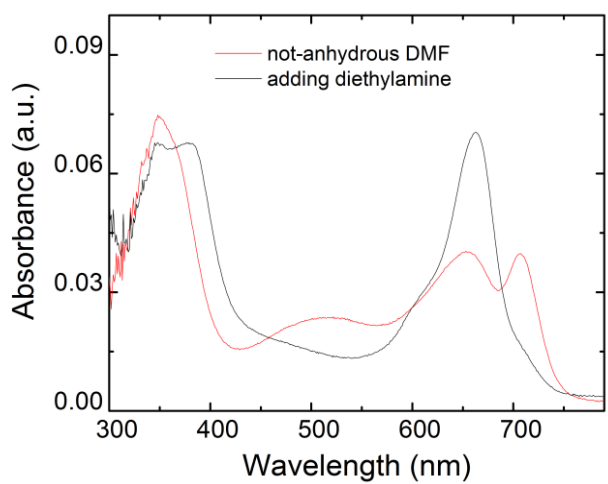

**Figure S7.** Change of the UV-vis spectrum of compound **PzPy** in not-anhydrous DMF upon addition of excess of diethylamine.

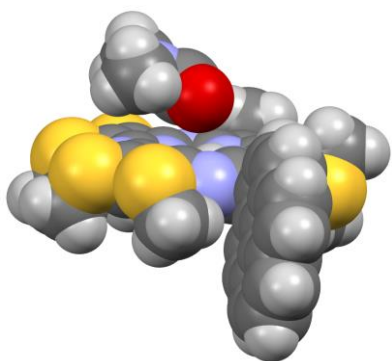

**Figure S8.** Optimized structure of the interaction of one DMF molecule with the S-Me substituted **PzPy** model at the DFT/M06/6-31G(d)/DMF level of theory. No specific interaction exists between the DMF molecule and the "inner" **PzPy** hydrogen atoms.

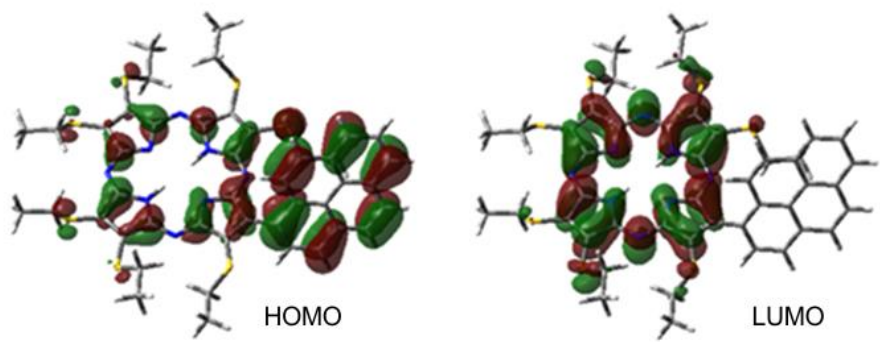

**Figure S9.** Frontier MO calculated at DFT/M06/6-311G(d,p)/IEFPCM(DMF) level of theory for major conformer of **PzPy**. Surface contour are drawn at 0.02 $(e/au^3)^{1/2}$.

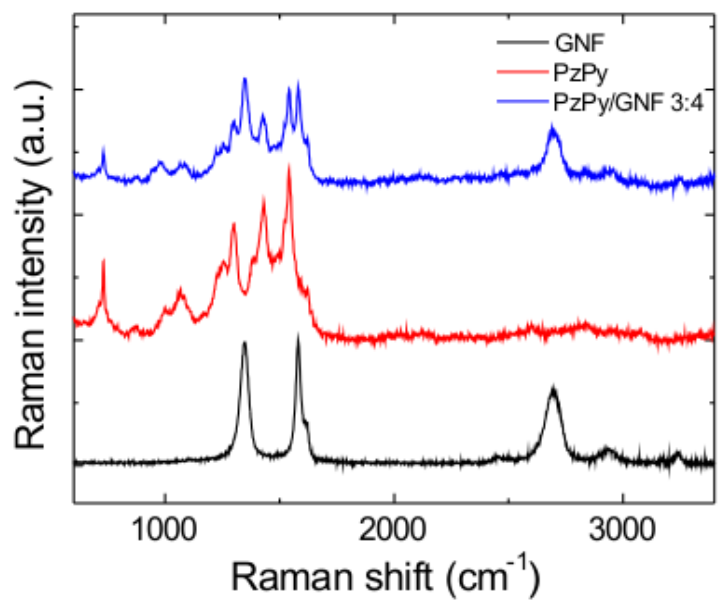

**Figure S10.** Raman spectra of **PzPy**, GNFs and **PzPy**/GNF 3:4 nanohybrid drop casted from a DMF dispersions onto silicon oxide wafer and excited at λ = 532 nm.

## Physical description of GNFs Raman modes

The G peak corresponds to the $E_{2g}$ phonon at the Brillouin zone centre.[1] The D peak is due to the breathing modes of $sp^2$ rings and requires a defect for its activation by double resonance.[2,3,4] The 2D peak is the second order of the D peak.[6] This is a single peak in monolayer graphene, whereas it splits in multi-layer graphene, reflecting the evolution of the band structure. The 2D peak is always seen, even when no D peak is present, since no defects are required for the activation of two phonons with the same momentum, one backscattered from the other. Double resonance can also happen as intra-valley process, *i.e.* connecting two points belonging to the same cone around K or K'.[2] This process gives rise to the D' peak. The 2D' is the second order of the D'.

## Structural characterization of SWNTs/Graphene.

Commercially available single walled carbon nanotubes (SWNT) (Aldrich, code n. 704121) used in nanohybrid preparation were structurally characterized by Raman spectroscopy and Transmission electron microscopy as reported below.

**(specifications: carbon ~90%, >77% as SWCNT, 0.7<d<1.1 nm)**

The length analysis over 200 SWNTs shows a distribution centered around 250 nm, Figure S11a. The statistical distribution of graphene flakes was analyzed measuring the longest side of the flakes on TEM. The Figure S11b shows the statistical distribution.

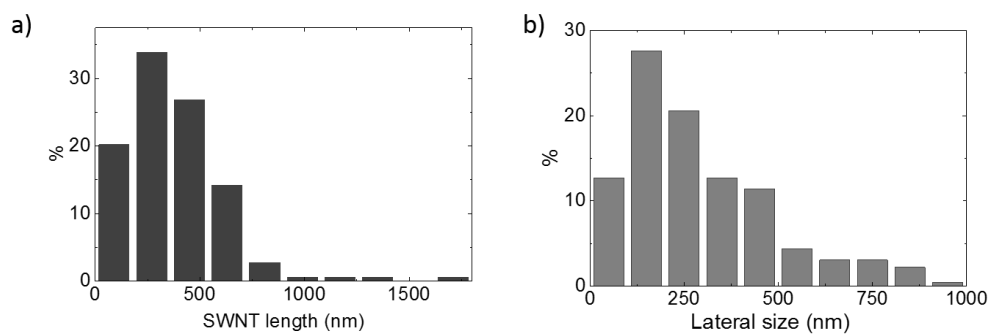

**Figure S11.** (a) Nanotubes length and (b) graphene lateral size distribution.

The analysis performed using 532 nm and 633 nm show that the sample is enriched with nanotubes with diameters in the range from 0.78 to 1.1 nm. This estimation is made in accordance with the Kataura plot[5] and experimental measurements.[6,7,8]. By measuring the Raman shift of the radial breathing modes ($\omega_{RBM}$) and using the equation below, one can estimate the diameter ($d_t$) and hence the chirality of the SWNTs in the batch:[9]

$$\omega_{RBM} = \left(\frac{227}{d_t}\right)\sqrt{1 + (0.082 \pm 0.009)d_t^2}$$

| 532 nm | | |
|---|---|---|
| $\omega_{RBM}$ (cm$^{-1}$) | Chiral vector | Diameter (nm) |
| 275.6 | (9,3) | 0.8 |
| 275.6 | (9,2) | 0.8 |
| 633 nm | | |
| $\omega_{RBM}$ (cm$^{-1}$) | Chiral vector | Diameter (nm) |
| 220 | (12,3) | 1.1 |
| 258 | (10,3) | 0.9 |
| 264 | (7,6) | 0.9 |
| 265 | (7,5) | 0.85 |
| 285.5 | (8,3) | 0.78 |

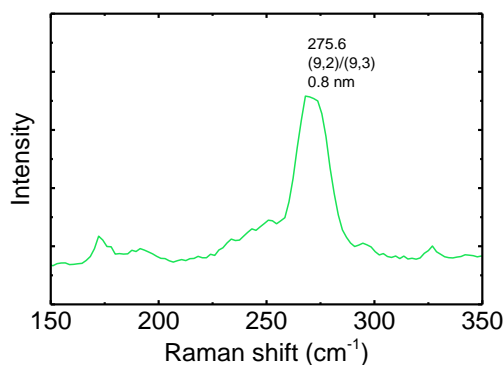
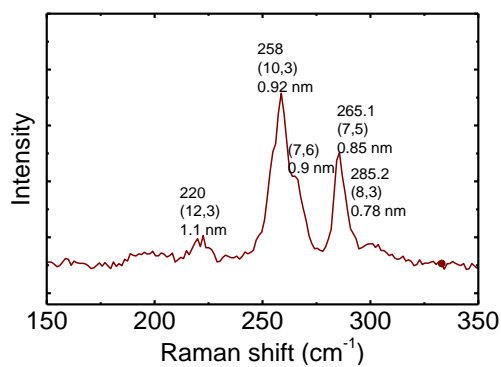

**Table S2.**
(a) Cartesian coordinates of the TS at the DFT/M06/6-311G(d,p)/CH$_2$Cl$_2$ level of theory for **PzPy**.

[173]H  -2.0950363839   0.8225336021  -0.3405657762

[174]N  -2.3992765168   1.7848458435  -0.2494154987

[175]C  -1.5635541668   2.8621597652  -0.18365157

[176]C  -3.7053732023   2.1771684734  -0.1896181426

[177]C  -2.4007769191   4.0441317532  -0.1144193589

[178]C  -3.7114498415   3.6218115926  -0.0952646876

[179]N  -0.2569249944   2.8600992496  -0.1389781531

[180]N  -4.7688531764   1.4125812424  -0.1994312196

[181]C   0.4867466307   1.7601646129  -0.1544916419

[182]C  -4.717122174    0.0864619826  -0.1972327804

[183]C   1.9523506565   1.8421170906  -0.1597206311

[184]C  -5.9394587601  -0.7247381471  -0.1449665569

[185]C   2.4164534805   0.5636590107  -0.4349173358

[186]C  -5.5175106966  -2.0188863334  -0.0695461863

[187]C   1.1631147155  -0.258640106   -0.3026825665

[188]C  -4.0514923239  -1.9580015182  -0.0773466007

[189]N  -3.6091980875  -0.6853838553  -0.1896037885

[190]N   0.0511506147   0.4918607915  -0.2025360536

[191]N   1.1850994407  -1.57655177    -0.153549607

[192]N  -3.3139385264  -3.055473466    0.0296150049

[193]C  -2.0040038967  -3.0459861967   0.0602179534

[194]C   0.1285551552  -2.3339217857  -0.0037591609

[195]C  -1.1535993999  -4.2141893252   0.1442253609

[196]C   0.1525484768  -3.7779193834   0.1152990375

[197]N  -1.1822814441  -1.9568496714   0.0118635811

[198]H  -1.5041842412  -0.9966161482   0.0024997612

[199]S  -1.8344535975   5.6831952264   0.1216626808

[200]S  -5.1373308335   4.6389750795  -0.1298231857

[201]S  -7.58717941    -0.1554926525   0.0114720593

[202]S   2.7073982866   3.3994526841   0.1864364218

[203]S  -6.5153898398  -3.4308822736   0.2087202815

[204]S   1.5809194169  -4.787625379    0.052704764

[205]S  -1.6831720238  -5.8816480175   0.0932234519

[206]C  -6.212975655   3.8829539407   1.1626796747

[207]C  -0.5047784837  5.843920108   -1.1448609824

[208]H  -6.2004080533  2.8012691785   1.0189814709

[209]H  -5.7841538051  4.1237185864   2.1389255527

[210]H   0.1228744947  4.9536108518  -1.0855826099

[211]H  -0.98348973    5.8897568409  -2.1265743093

[212]C   1.7949289234  3.9840816829   1.6784363003

[213]H   0.833985337   4.4007719515   1.3713634363

[214]H   1.6151374442  3.1129171252   2.3143625079

[215]C   2.6807832742 -3.9744631184   1.2889136466

[216]C  -3.014139408  -5.9234137536   1.3671815349

[217]H   2.8680036528 -2.9520183332   0.9559613916

[218]H   2.1444979597 -3.9530246457   2.2414401516

[219]H  -3.7452200373 -5.1494374829   1.1259655382

[220]H  -2.5564324006 -5.6939110286   2.3330830031

[221]C  -5.8654799268 -4.6338252306  -1.0257575441

[222]C  -7.7044008516  1.1336654919  -1.2982625045

[223]H  -6.0799316133 -4.2389199371  -2.022444988

[224]H  -4.7835495835 -4.7039649331  -0.8952467073

[225]H  -6.9884061119  1.9255609847  -1.0702647952

[226]H  -7.42583878    0.6688590848  -2.2480363515

[227]C   3.7185364269 -0.0712973358  -0.7131242252

[228]C   5.0188410871  0.4987781304  -0.5421533147

[229]C   6.1511697751   -0.3622575915   -0.4143245163

[230]C   6.0341700669   -1.7561924941   -0.6449309172

[231]C   4.8138497106   -2.2286922111   -1.125773796

[232]C   3.7086153139   -1.4109167665   -1.1596543241

[233]C   5.3012860627    1.8963566089   -0.5953069639

[234]C   7.4307831068    0.1682015093   -0.0837650984

[235]H   4.7287686815   -3.2617808767   -1.4546087526

[236]H   2.7852168469   -1.8432362219   -1.5090539106

[237]C   6.526115689     2.4103384457   -0.3280021757

[238]H   4.5288723314    2.5638749395   -0.9384982614

[239]H   6.6957490843    3.4813049529   -0.4046616512

[240]C   7.6234472353    1.5684142243    0.0127718372

[241]C   7.1603561517   -2.6093828343   -0.4453349364

[242]C   8.353669206    -2.1077729169   -0.0489197953

[243]H   7.0324101998   -3.6751562287   -0.6174419398

[244]H   9.2055071242   -2.762277938     0.1157138681

[245]C   8.5340279773   -0.7011655308    0.1314540805

[246]C   9.7705739234   -0.1593988829    0.4911211178

[247]C   9.940011817     1.2122299351    0.6070161204

[248]H  10.6053197079   -0.8319841392    0.6698246804

[249]H  10.9099567662    1.6149479426    0.8810036556

[250]C   8.8818880229    2.0705893934    0.3562518665

[251]H   9.0196394548    3.1468719607    0.4172366742

[252]C  -9.1274488836   1.645248209   -1.320169888

[253]H  -9.410495132   2.0727031565   -0.3519278222

[254]H  -9.2312734263   2.4309707531   -2.0742301454

[255]H  -9.8400565643   0.8503199928   -1.5605041976

[256]C  -7.6067474839   4.4476410745   1.0055095004

[257]H  -8.2672725733   4.0291110729   1.7707139359

[258]H  -7.6205845296   5.5371198278   1.1115375466

[259]H  -8.0250545345   4.197282196   0.0244723748

[260]C  2.6655901101   5.0104124126   2.3687643987

[261]H  2.1417656621   5.4163759796   3.2392097647

[262]H  3.6097343677   4.5757474691   2.7096592515

[263]H  2.8984893927   5.8489455267   1.7037222581

[264]C  0.2836031162   7.0984042999   -0.8440867049

[265]H  1.0586876508   7.2416405922   -1.6023032113

[266]H  -0.3511482848   7.9904005033   -0.8431294124

[267]H  0.7753364471   7.0305226095   0.1322085406

[268]C  -6.5477177602   -5.9624638961   -0.7890244318

[269]H  -7.634435948   -5.8855203564   -0.895363176

[270]H  -6.191954586   -6.7017504307   -1.5126955383

[271]H  -6.3315274427   -6.343742101   0.2150933531

[272]C  -3.6281523226   -7.3051094551   1.3490156349

[273]H  -2.8891448031   -8.0814246284   1.570611188

[274]H  -4.4200135191   -7.3717893771   2.1008196783

[275]H  -4.0725259537  -7.5274990869   0.3726594026

[276]C   3.9598479279  -4.7748595428   1.3827638451

[277]H   4.4743881657  -4.8169687825   0.4164658554

[278]H   4.64068123    -4.3035567818   2.0979109375

[279]H   3.775461324   -5.8009329558   1.7157316065

[280]

(b) Cartesian coordinates of conformer 1 of **PzPy** at the DFT/M06/6-311G(d,p)/CH$_2$Cl$_2$ level of theory.

| | | | |
|---|---:|---:|---:|
| H |  1.93004700 |  1.15918600 | -0.61531300 |
| N |  2.21000300 |  2.13186700 | -0.55839800 |
| C |  1.34537300 |  3.18838700 | -0.59135200 |
| C |  3.49743800 |  2.55950800 | -0.43497000 |
| C |  2.14240400 |  4.39070400 | -0.48131100 |
| C |  3.46481700 |  4.00604400 | -0.40957000 |
| N |  0.04238700 |  3.14234500 | -0.71428800 |
| N |  4.57272300 |  1.81808700 | -0.31022000 |
| C | -0.64517400 |  2.01136400 | -0.81072800 |
| C |  4.55139300 |  0.49283500 | -0.29425300 |
| C | -2.11146600 |  2.01470300 | -0.90612900 |
| C |  5.79017600 | -0.28228200 | -0.13618100 |
| C | -2.48806400 |  0.70650400 | -0.91908900 |
| C |  5.41850900 | -1.59010400 | -0.22491300 |
| C | -1.23688000 | -0.05426100 | -0.86827900 |
| C |  3.95670000 | -1.57505500 | -0.39005400 |
| N |  3.47560700 | -0.31335400 | -0.43428700 |
| N | -0.15517600 |  0.75159600 | -0.80711400 |
| N | -1.25610700 | -1.38457500 | -0.91800400 |
| N |  3.26508100 | -2.70453400 | -0.44938800 |
| C |  1.95934800 | -2.75074500 | -0.57376400 |
| C | -0.18826000 | -2.12958500 | -0.82824100 |
| C |  1.16955700 | -3.95343700 | -0.64886400 |
| C | -0.14924600 | -3.58327500 | -0.82241100 |
| N |  1.09986100 | -1.69812900 | -0.67453100 |
| H |  1.37220800 | -0.72255800 | -0.63960500 |
| S |  1.54554000 |  6.03217000 | -0.59977200 |
| S |  4.83452300 |  5.05973300 | -0.12996200 |
| S |  7.42024000 |  0.35314600 | -0.06796700 |
| S | -3.16478300 |  3.40912000 | -0.76843800 |
| S |  6.42923700 | -2.99306700 |  0.04712300 |
| S | -1.46931200 | -4.72018300 | -0.87481000 |
| S |  1.79575700 | -5.59041600 | -0.68601700 |
| C |  6.10887100 |  4.40278500 | -1.28585900 |
| C |  0.10992700 |  6.03575600 |  0.55540800 |
| H |  6.26062300 |  3.34395500 | -1.06727600 |
| H |  5.72147300 |  4.50498500 | -2.30314600 |
| H | -0.56989000 | 5.23275500 | 0.26381900 |
| H | 0.49080400 | 5.83419800 | 1.56026000 |
| C | -2.48717200 | 4.55111900 | -2.04376200 |
| H | -1.41673300 | 4.66790100 | -1.85991700 |
| H | -2.63209200 | 4.08562400 | -3.02308500 |
| C | -2.88855400 | -3.81118000 | -1.60570200 |
| C | 2.90272500 | -5.62616000 | 0.78573200 |
| H | -3.11731000 | -2.95315000 | -0.97711700 |
| H | -2.58947600 | -3.44651900 | -2.59274500 |
| H | 3.64641600 | -4.83409300 | 0.67120000 |
| H | 2.29254900 | -5.41759400 | 1.66865700 |
| C | 5.94610300 | -4.13224000 | -1.31628100 |
| C | 7.31130100 | 1.63362200 | 1.25135700 |
| H | 4.85767100 | -4.21471500 | -1.31834300 |
| H | 6.27444000 | -3.68734500 | -2.25952800 |
| H | 6.47818800 | 2.29806400 | 1.01244500 |
| H | 7.10392100 | 1.12539900 | 2.19700000 |
| C | -3.80922900 | 0.09073500 | -1.06472100 |
| C | -4.28253900 | -0.86786500 | -0.14820000 |
| C | -5.47232800 | -1.58408800 | -0.44752800 |
| C | -6.20736700 | -1.28876000 | -1.62470800 |
| C | -5.74423600 | -0.28178700 | -2.47479700 |
| C | -4.56520100 | 0.38480200 | -2.20108200 |
| C | -3.59861600 | -1.17160200 | 1.07424700 |
| C | -5.90764300 | -2.63344400 | 0.40749500 |
| H | -6.30746500 | -0.04816600 | -3.37420300 |
| H | -4.19020200 | 1.13029600 | -2.89687300 |
| C | -4.01719600 | -2.16573900 | 1.89103200 |
| H | -2.72627700 | -0.58465800 | 1.34303400 |
| H | -3.48079600 | -2.38495900 | 2.81057900 |
| C | -5.17039200 | -2.95140700 | 1.57653700 |
| C | -7.38756600 | -2.04390000 | -1.91462000 |
| C | -7.79718400 | -3.04605200 | -1.10184600 |
| H | -7.94515600 | -1.79723200 | -2.81438200 |
| H | -8.68993800 | -3.61975200 | -1.33661200 |
| C | -7.06754200 | -3.38372100 | 0.08205700 |
| C | -7.45346000 | -4.43600000 | 0.91610900 |
| C | -6.72049600 | -4.74608900 | 2.05166000 |
| H | -8.33900400 | -5.01188200 | 0.66046100 |
| H | -7.03390700 | -5.56821700 | 2.68732700 |
| C | -5.59243700 | -4.01046800 | 2.38314800 |
| H | -5.02326600 | -4.25059800 | 3.27723100 |
| C | 3.54187800 | -6.99481400 | 0.85350300 |
| H | 4.12548100 | -7.20424700 | -0.04975400 |
| H | 4.22057000 | -7.04859200 | 1.70971700 |
| H | 2.79455200 | -7.78640700 | 0.96576400 |
| C | 6.60882300 | -5.46834800 | -1.06554600 |
| H | 6.28256400 | -5.89740100 | -0.11164300 |
| H | 6.34398500 | -6.17321200 | -1.85920700 |
| H | 7.69999800 | -5.38379200 | -1.04257900 |
| C | 8.62778600 | 2.37673100 | 1.29091300 |
| H | 8.83195900 | 2.86877900 | 0.33345200 |
| H | 8.59982900 | 3.14847600 | 2.06581600 |
| H | 9.46560400 | 1.70826300 | 1.51318900 |
| C | -0.54949800 | 7.39358300 | 0.46587700 |
| H | -1.40592900 | 7.43757800 | 1.14504400 |
| H | 0.13850500 | 8.19924000 | 0.74058800 |
| H | -0.91486700 | 7.58918500 | -0.54829400 |
| C | -3.22191700 | 5.86830500 | -1.93456200 |
| H | -2.84199800 | 6.57400500 | -2.67909700 |
| H | -4.29684500 | 5.74878000 | -2.10292300 |
| H | -3.08201100 | 6.31644800 | -0.94465400 |
| C | -4.05152200 | -4.77604000 | -1.68276100 |
| H | -4.91826900 | -4.26658700 | -2.11627000 |
| H | -3.82416900 | -5.64735300 | -2.30487900 |
| H | -4.34302600 | -5.12841800 | -0.68677900 |
| C | 7.37299600 | 5.20732200 | -1.08175800 |
| H | 7.73866600 | 5.11476000 | -0.05313800 |
| H | 8.15875400 | 4.84300300 | -1.74993500 |
| [281] H | 7.21775000 | 6.26990700 | -1.29287900 |

[282]

(c) Cartesian coordinates of conformer 2 of **PzPy** at the DFT/M06/6-311G(d,p)/CH$_2$Cl$_2$ level of theory.

| | | | |
|---|---|---|---|
| [283] H | 1.86437700 | 1.38034100 | 0.69190700 |
| [284] N | 2.18517700 | 2.32522900 | 0.50948300 |
| [285] C | 1.36997900 | 3.41948200 | 0.44161100 |
| [286] C | 3.48684300 | 2.67648400 | 0.31338900 |
| [287] C | 2.21773900 | 4.56415800 | 0.18992000 |
| [288] C | 3.51912400 | 4.11215500 | 0.13677900 |
| [289] N | 0.06819700 | 3.44865200 | 0.58055000 |
| [290] N | 4.52549000 | 1.87759800 | 0.24109100 |
| [291] C | -0.67253700 | 2.36621200 | 0.78155000 |
| [292] C | 4.43999900 | 0.55837300 | 0.33449600 |
| [293] C | -2.13503100 | 2.45222900 | 0.90244700 |
| [294] C | 5.64056900 | -0.28652200 | 0.25575000 |
| [295] C | -2.57045200 | 1.17252300 | 1.05605000 |

| | | | | |
|---|---|---:|---:|---:|
| [296] | C | 5.20411100 | -1.56267400 | 0.44650900 |
| [297] | C | -1.36552000 | 0.34454000 | 1.02042000 |
| [298] | C | 3.74305400 | -1.46329100 | 0.59047000 |
| [299] | N | 3.32462400 | -0.18079900 | 0.52508400 |
| [300] | N | -0.24629900 | 1.08656300 | 0.86993400 |
| [301] | N | -1.45595200 | -0.98111800 | 1.12257500 |
| [302] | N | 2.99767700 | -2.54781500 | 0.74417600 |
| [303] | C | 1.69072600 | -2.52094000 | 0.86645700 |
| [304] | C | -0.42693100 | -1.78199800 | 1.06318000 |
| [305] | C | 0.85413200 | -3.67025300 | 1.08480700 |
| [306] | C | -0.44808500 | -3.22944600 | 1.21522800 |
| [307] | N | 0.87552600 | -1.42793700 | 0.83916100 |
| [308] | H | 1.18337400 | -0.47815200 | 0.66251200 |
| [309] | S | 1.69738300 | 6.23529500 | 0.13839500 |
| [310] | S | 4.93025200 | 5.06309100 | -0.27465300 |
| [311] | S | 7.30070900 | 0.26246900 | 0.16674400 |
| [312] | S | -3.13676800 | 3.87651600 | 0.70454200 |
| [313] | S | 6.14758400 | -3.03105400 | 0.31348900 |
| [314] | S | -1.77617000 | -4.30397000 | 1.54205700 |
| [315] | S | 1.41144200 | -5.31300300 | 1.33659600 |
| [316] | C | 6.18366800 | 4.48413200 | 0.94420900 |
| [317] | C | 0.24793700 | 6.18417300 | -0.99831000 |
| [318] | H | 6.28131300 | 3.40071400 | 0.84976000 |
| [319] | H | 5.81169300 | 4.72307900 | 1.94418800 |
| [320] | H | -0.44949000 | 5.42797700 | -0.63277600 |
| [321] | H | 0.60979600 | 5.88900900 | -1.98689800 |
| [322] | C | -2.33568500 | 5.10596700 | 1.81750200 |
| [323] | H | -1.28299900 | 5.18536400 | 1.53927300 |
| [324] | H | -2.40591700 | 4.72637500 | 2.84056500 |
| [325] | C | -3.29393500 | -3.27482500 | 1.52322700 |
| [326] | C | 2.48149300 | -5.59197900 | -0.13656900 |
| [327] | H | -3.38435100 | -2.78698900 | 0.54914800 |
| [328] | H | -3.19859500 | -2.49645600 | 2.28190000 |
| [329] | H | 3.23392200 | -4.80028800 | -0.15970100 |
| [330] | H | 1.85235500 | -5.51417800 | -1.02720300 |
| [331] | C | 5.60477300 | -4.01061600 | 1.77477900 |
| [332] | C | 7.27996800 | 1.40246900 | -1.27866800 |
| [333] | H | 4.51546500 | -4.07922100 | 1.75249800 |
| [334] | H | 5.91107000 | -3.46963800 | 2.67430900 |
| [335] | H | 6.47653900 | 2.12618400 | -1.12807700 |
| [336] | H | 7.06509500 | 0.80981600 | -2.17201000 |
| [337] | C | -3.93018700 | 0.63544100 | 1.15167300 |
| [338] | C | -4.36085600 | -0.05327500 | 2.30096700 |
| [339] | C | -5.63009500 | -0.68907200 | 2.29190800 |
| [340] | C | -6.46911300 | -0.59489300 | 1.15065100 |
| [341] | C | -6.03204000 | 0.14514700 | 0.05120700 |
| [342] C | -4.78443600 | 0.74313400 | 0.05496600 |
| [343] C | -3.56314900 | -0.14417300 | 3.48745800 |
| [344] C | -6.05305600 | -1.45254900 | 3.41408800 |
| [345] H | -6.67505000 | 0.22881200 | -0.82067300 |
| [346] H | -4.43996600 | 1.27894400 | -0.82545200 |
| [347] C | -3.97084100 | -0.86312800 | 4.55944500 |
| [348] H | -2.61450700 | 0.38313600 | 3.51771300 |
| [349] H | -3.35069700 | -0.92310600 | 5.45006100 |
| [350] C | -5.21799300 | -1.56353400 | 4.55512700 |
| [351] C | -7.73071200 | -1.27098500 | 1.15896800 |
| [352] C | -8.12677200 | -2.00567600 | 2.22417200 |
| [353] H | -8.36369800 | -1.18442000 | 0.27975900 |
| [354] H | -9.08326600 | -2.52171800 | 2.21616700 |
| [355] C | -7.30068000 | -2.12920300 | 3.38694300 |
| [356] C | -7.67588800 | -2.90738800 | 4.48429400 |
| [357] C | -6.84928300 | -3.01632500 | 5.59259600 |
| [358] H | -8.62825100 | -3.43024100 | 4.45735500 |
| [359] H | -7.15675400 | -3.62672900 | 6.43581700 |
| [360] C | -5.63489100 | -2.34883500 | 5.63249100 |
| [361] H | -4.99107000 | -2.43135500 | 6.50408700 |
| [362] C | 3.10872600 | -6.96151800 | -0.00639300 |
| [363] H | 3.71377900 | -7.03422700 | 0.90406400 |
| [364] H | 3.76408100 | -7.15603400 | -0.86038100 |

| | | | | |
|---|---|---|---|---|
| [365] | H |  2.35294100 | -7.75249900 |  0.02462700 |
| [366] | C |  6.25658200 | -5.37270800 |  1.69098100 |
| [367] | H |  5.95842400 | -5.89774200 |  0.77663800 |
| [368] | H |  5.95345800 | -5.98662400 |  2.54427400 |
| [369] | H |  7.34885900 | -5.30193400 |  1.69954500 |
| [370] | C |  8.63189600 |  2.07482100 | -1.36377900 |
| [371] | H |  8.84204100 |  2.65426800 | -0.45795800 |
| [372] | H |  8.65690900 |  2.76158100 | -2.21500400 |
| [373] | H |  9.43951100 |  1.34775500 | -1.49534900 |
| [374] | C | -0.37307100 |  7.56284300 | -1.01541800 |
| [375] | H | -1.23890100 |  7.57395900 | -1.68384900 |
| [376] | H |  0.33218200 |  8.32247600 | -1.36692900 |
| [377] | H | -0.71592300 |  7.85440100 | -0.01658800 |
| [378] | C | -3.06169100 |  6.42269600 |  1.65638200 |
| [379] | H | -2.61038400 |  7.18187600 |  2.30181100 |
| [380] | H | -4.11924300 |  6.33853300 |  1.92570700 |
| [381] | H | -3.00310800 |  6.78234600 |  0.62315800 |
| [382] | C | -4.45972600 | -4.19808300 |  1.80806800 |
| [383] | H | -5.39392700 | -3.62940900 |  1.76058800 |
| [384] | H | -4.38973900 | -4.63649200 |  2.80937900 |
| [385] | H | -4.52433500 | -5.01237900 |  1.07949000 |
| [386] | C |  7.48325400 |  5.19622300 |  0.64284300 |
| [387] | H |  7.83570800 |  4.96456300 | -0.36833600 |

| [388] H | 8.25617000 | 4.87667900 | 1.34795400 |
|---|---|---|---|
| [389] H | 7.38054800 | 6.28265700 | 0.72622100 |

**Table S3.** Excited state energies, wavelengths, wavenumbers, oscillator strengths and composition coefficient of the Slater determinant and its percentage in terms of MO's involved in the excitation. Integer numbers indicate orbital numbers evaluated from the HOMO (0) and LUMO (0), e.g., 0→0 means HOMO to LUMO excitation.

| | 1 | 1 | | State: | Singlet | | | |
|---|---|---|---|---|---|---|---|---|
| | 1.8257 | eV | | 679.11 | nm | 14725.cm$^{-1}$ | f= | 0.2024 |
| 242 | → | 247 | -0.10496 | 2.2 | % | -3 | → | 1 |
| 243 | → | 247 | 0.10632 | 2.26 | % | -2 | → | 1 |
| 244 | → | 247 | 0.11593 | 2.69 | % | -1 | → | 1 |
| 245 | → | 246 | 0.58803 | 69.16 | % | 0 | → | 0 |
| 245 | → | 247 | -0.31507 | 19.85 | % | 0 | → | 1 |
| | 2 | 2 | | Excited | State: | Singlet | | |
| | 1.8653 | eV | | 664.69 | nm | 15045.cm$^{-1}$ | f= | 0.0938 |
| 242 | → | 246 | 0.14253 | 4.06 | % | -3 | → | 0 |
| 243 | → | 246 | -0.1536 | 4.72 | % | -2 | → | 0 |
| 245 | → | 246 | 0.30218 | 18.26 | % | 0 | → | 0 |
| 245 | → | 247 | 0.58589 | 68.65 | % | 0 | → | 1 |
| | 3 | 3 | | Excited | State: | Singlet | | |
| | 1.8929 | eV | | 654.98 | nm | 15268.cm$^{-1}$ | f= | 0.1034 |
| 242 | → | 247 | -0.15567 | 4.85 | % | -3 | → | 1 |
| 243 | → | 246 | 0.19632 | 7.71 | % | -2 | → | 0 |
| 244 | → | 246 | 0.6364 | 81 | % | -1 | → | 0 |
| | 4 | 4 | | Excited | State: | Singlet | | |
| | 1.9059 | eV | | 650.52 | nm | 15372.cm$^{-1}$ | f= | 0.0032 |
| 243 | → | 246 | 0.53081 | 56.35 | % | -2 | → | 0 |
| 243 | → | 247 | -0.39231 | 30.78 | % | -2 | → | 1 |
| 244 | → | 246 | -0.14284 | 4.08 | % | -1 | → | 0 |
| 245 | → | 246 | 0.12252 | 3 | % | 0 | → | 0 |
| | 5 | 5 | | Excited | State: | Singlet | | |
| | 1.9613 | eV | | 632.16 | nm | 15819.cm$^{-1}$ | f= | 0.0649 |
| 241 | → | 246 | 0.11908 | 2.84 | % | -4 | → | 0 |
| 242 | → | 246 | 0.22928 | 10.51 | % | -3 | → | 0 |
| 243 | → | 247 | -0.24261 | 11.77 | % | -2 | → | 1 |
| 244 | → | 247 | 0.574 | 65.9 | % | -1 | → | 1 |
| 245 | → | 246 | -0.10299 | 2.12 | % | 0 | → | 0 |
| | 6 | 6 | | Excited | State: | Singlet | | |
| | 2.0243 | eV | | 612.47 | nm | 16327.cm$^{-1}$ | f= | 0.0646 |
| 242 | → | 246 | 0.10504 | 2.21 | % | -3 | → | 0 |
| 243 | → | 246 | 0.37035 | 27.43 | % | -2 | → | 0 |
| 243 | → | 247 | 0.50673 | 51.36 | % | -2 | → | 1 |
| 244 | → | 247 | 0.21987 | 9.67 | % | -1 | → | 1 |

|  | 7 | 7 | Excited | State: | Singlet | | |
|---|---|---|---|---|---|---|---|
|  | 2.1026 | eV | 589.68 | nm | 16958.cm$^{-1}$ | f= | 0.0132 |
| 241 | → | 246 | 0.49011 | 48.04 | % | -4 → | 0 |
| 241 | → | 247 | 0.12268 | 3.01 | % | -4 → | 1 |
| 242 | → | 246 | -0.413 | 34.11 | % | -3 → | 0 |
| 242 | → | 247 | -0.2458 | 12.08 | % | -3 → | 1 |
|  | 8 | 8 | Excited | State: | Singlet | | |
|  | 2.1629 | eV | 573.23 | nm | 17445.cm$^{-1}$ | f= | 0.3145 |
| 241 | → | 246 | 0.38282 | 29.31 | % | -4 → | 0 |
| 241 | → | 247 | 0.22168 | 9.83 | % | -4 → | 1 |
| 242 | → | 246 | 0.43228 | 37.37 | % | -3 → | 0 |
| 244 | → | 247 | -0.24521 | 12.03 | % | -1 → | 1 |
| 245 | → | 247 | -0.15116 | 4.57 | % | 0 → | 1 |
|  | 9 | 9 | Excited | State: | Singlet | | |
|  | 2.208 | eV | 561.53 | nm | 17808.cm$^{-1}$ | f= | 0.1197 |
| 241 | → | 246 | 0.24262 | 11.77 | % | -4 → | 0 |
| 241 | → | 247 | -0.37694 | 28.42 | % | -4 → | 1 |
| 242 | → | 246 | -0.13380 | 3.58 | % | -3 → | 0 |
| 242 | → | 247 | 0.50309 | 50.62 | % | -3 → | 1 |
| 244 | → | 246 | 0.10116 | 2.05 | % | -1 → | 0 |
|  | 10 | 10 | Excited | State: | Singlet | | |
|  | 2.2263 | eV | 556.91 | nm | 17956.cm$^{-1}$ | f= | 0.262 |
| 240 | → | 246 | 0.14968 | 4.48 | % | -5 → | 0 |
| 241 | → | 246 | -0.11188 | 2.5 | % | -4 → | 0 |
| 241 | → | 247 | 0.51583 | 53.22 | % | -4 → | 1 |
| 242 | → | 246 | -0.17662 | 6.24 | % | -3 → | 0 |
| 242 | → | 247 | 0.33854 | 22.92 | % | -3 → | 1 |
| 244 | → | 247 | 0.13407 | 3.59 | % | -1 → | 1 |
| 245 | → | 246 | 0.12184 | 2.97 | % | 0 → | 0 |
|  | 11 | 11 | Excited | State: | Singlet | | |
|  | 2.4194 | eV | 512.45 | nm | 19514.cm$^{-1}$ | f= | 0.1174 |
| 239 | → | 246 | 0.11627 | 2.7 | % | -6 → | 0 |
| 240 | → | 246 | 0.62181 | 77.33 | % | -5 → | 0 |
| 240 | → | 247 | -0.21706 | 9.42 | % | -5 → | 1 |
| 241 | → | 247 | -0.11888 | 2.83 | % | -4 → | 1 |
| 244 | → | 246 | -0.11726 | 2.75 | % | -1 → | 0 |
|  | 12 | 12 | Excited | State: | Singlet | | |
|  | 2.499 | eV | 496.13 | nm | 20156.cm$^{-1}$ | f= | 0.0423 |
| 239 | → | 247 | 0.10121 | 2.05 | % | -6 → | 1 |
| 240 | → | 246 | 0.22451 | 10.08 | % | -5 → | 0 |
| 240 | → | 247 | 0.64844 | 84.09 | % | -5 → | 1 |
|  | 13 | 13 | Excited | State: | Singlet | | |
|  | 2.8898 | eV | 429.04 | nm | 23308.cm$^{-1}$ | f= | 0.0002 |

| | | | | | | | | |
|---|---|---|---|---|---|---|---|---|
| 238 | → | 246 | 0.3652 | 26.67 | % | -7 | → | 0 |
| 238 | → | 247 | -0.2853 | 16.28 | % | -7 | → | 1 |
| 239 | → | 246 | 0.43568 | 37.96 | % | -6 | → | 0 |
| 239 | → | 247 | -0.23448 | 11 | % | -6 | → | 1 |
| | 14 | 14 | Excited | State: | Singlet | | | |
| | 2.9274 | eV | 423.52 | nm | 23612.cm$^{-1}$ | f= | 0.0254 | |
| 238 | → | 246 | -0.35809 | 25.65 | % | -7 | → | 0 |
| 238 | → | 247 | 0.31466 | 19.8 | % | -7 | → | 1 |
| 239 | → | 246 | 0.49217 | 48.45 | % | -6 | → | 0 |
| 240 | → | 246 | -0.10105 | 2.04 | % | -5 | → | 0 |
| | 15 | 15 | Excited | State: | Singlet | | | |
| | 2.9653 | eV | 418.11 | nm | 23917.cm$^{-1}$ | f= | 0.0007 | |
| 238 | → | 246 | 0.19334 | 7.48 | % | -7 | → | 0 |
| 239 | → | 246 | 0.18151 | 6.59 | % | -6 | → | 0 |
| 239 | → | 247 | 0.63560 | 80.8 | % | -6 | → | 1 |
| | 16 | 16 | Excited | State: | Singlet | | | |
| | 3.0231 | eV | 410.12 | nm | 24383.cm$^{-1}$ | f= | 0.0545 | |
| 238 | → | 246 | 0.41509 | 34.46 | % | -7 | → | 0 |
| 238 | → | 247 | 0.53906 | 58.12 | % | -7 | → | 1 |
| 239 | → | 247 | -0.11498 | 2.64 | % | -6 | → | 1 |
| | 17 | 17 | Excited | State: | Singlet | | | |
| | 3.1105 | eV | 398.6 | nm | 25088.cm$^{-1}$ | f= | 0.013 | |
| 237 | → | 246 | 0.64165 | 82.34 | % | -8 | → | 0 |
| 237 | → | 247 | 0.17029 | 5.8 | % | -8 | → | 1 |
| 245 | → | 248 | 0.11222 | 2.52 | % | 0 | → | 2 |
| | 18 | 18 | Excited | State: | Singlet | | | |
| | 3.1998 | eV | 387.48 | nm | 25808.cm$^{-1}$ | f= | 0.0051 | |
| 237 | → | 246 | -0.1989 | 7.91 | % | -8 | → | 0 |
| 237 | → | 247 | 0.6288 | 79.08 | % | -8 | → | 1 |
| 245 | → | 248 | 0.18498 | 6.84 | % | 0 | → | 2 |
| | 19 | 19 | Excited | State: | Singlet | | | |
| | 3.2599 | eV | 380.33 | nm | 26293.cm$^{-1}$ | f= | 0.1659 | |
| 232 | → | 246 | 0.15653 | 4.9 | % | -13 | → | 0 |
| 234 | → | 246 | 0.24646 | 12.15 | % | -11 | → | 0 |
| 236 | → | 246 | -0.10191 | 2.08 | % | -9 | → | 0 |
| 237 | → | 247 | -0.19572 | 7.66 | % | -8 | → | 1 |
| 245 | → | 248 | 0.55359 | 61.29 | % | 0 | → | 2 |
| | 20 | 20 | Excited | State: | Singlet | | | |
| | 3.2823 | eV | 377.74 | nm | 26473.cm$^{-1}$ | f= | 0.1095 | |
| 231 | → | 246 | -0.10232 | 2.09 | % | -14 | → | 0 |
| 232 | → | 246 | 0.43675 | 38.15 | % | -13 | → | 0 |
| 233 | → | 246 | -0.15653 | 4.9 | % | -12 | → | 0 |
| 234 | → | 246 | 0.25514 | 13.02 | % | -11 | → | 0 |
| 235 | → | 246 | -0.13889 | 3.86 | % | -10 | → | 0 |
| 236 | → | 246 | -0.24258 | 11.77 | % | -9 | → | 0 |
| 245 | → | 248 | -0.29774 | 17.73 | % | 0 | → | 2 |
| | 21 | 21 | | Excited | State: | Singlet | | |
| | 3.3058 | eV | 375.05 | nm | 26663.cm$^{-1}$ | f= | 0.0411 | |
| 232 | → | 246 | -0.17343 | 6.02 | % | -13 | → | 0 |
| 233 | → | 246 | 0.15822 | 5.01 | % | -12 | → | 0 |
| 234 | → | 246 | 0.23236 | 10.8 | % | -11 | → | 0 |
| 234 | → | 247 | -0.21199 | 8.99 | % | -11 | → | 1 |
| 235 | → | 246 | 0.38587 | 29.78 | % | -10 | → | 0 |
| 235 | → | 247 | 0.23629 | 11.17 | % | -10 | → | 1 |
| 236 | → | 246 | -0.23372 | 10.93 | % | -9 | → | 0 |
| 236 | → | 247 | -0.21267 | 9.05 | % | -9 | → | 1 |
| | 22 | 22 | | Excited | State: | Singlet | | |
| | 3.3293 | eV | 372.4 | nm | 26853.cm$^{-1}$ | f= | 0.003 | |
| 231 | → | 247 | -0.11839 | 2.8 | % | -14 | → | 1 |
| 232 | → | 247 | 0.48347 | 46.75 | % | -13 | → | 1 |
| 233 | → | 247 | -0.17397 | 6.05 | % | -12 | → | 1 |
| 234 | → | 247 | 0.26455 | 14 | % | -11 | → | 1 |
| 235 | → | 247 | -0.14045 | 3.95 | % | -10 | → | 1 |
| 236 | → | 247 | -0.32265 | 20.82 | % | -9 | → | 1 |
| | 23 | 23 | | Excited | State: | Singlet | | |
| | 3.4053 | eV | 364.09 | nm | 27466.cm$^{-1}$ | f= | 0.0142 | |
| 233 | → | 246 | -0.12019 | 2.89 | % | -12 | → | 0 |
| 233 | → | 247 | -0.13473 | 3.63 | % | -12 | → | 1 |
| 234 | → | 246 | -0.13944 | 3.89 | % | -11 | → | 0 |
| 235 | → | 246 | 0.20063 | 8.05 | % | -10 | → | 0 |
| 235 | → | 247 | -0.28424 | 16.16 | % | -10 | → | 1 |
| 236 | → | 246 | -0.31993 | 20.47 | % | -9 | → | 0 |
| 236 | → | 247 | 0.18394 | 6.77 | % | -9 | → | 1 |
| 244 | → | 248 | 0.37855 | 28.66 | % | -1 | → | 2 |
| | 24 | 24 | | Excited | State: | Singlet | | |
| | 3.414 | eV | 363.16 | nm | 27536.cm$^{-1}$ | f= | 0.3069 | |
| 235 | → | 246 | -0.11131 | 2.48 | % | -10 | → | 0 |
| 235 | → | 247 | 0.24502 | 12.01 | % | -10 | → | 1 |
| 236 | → | 246 | 0.18245 | 6.66 | % | -9 | → | 0 |
| 236 | → | 247 | -0.15625 | 4.88 | % | -9 | → | 1 |
| 244 | → | 248 | 0.55735 | 62.13 | % | -1 | → | 2 |
| | 25 | 25 | | Excited | State: | Singlet | | |
| | 3.4516 | eV | 359.21 | nm | 27839.cm$^{-1}$ | f= | 0.0328 | |
| 230 | → | 246 | 0.11369 | 2.59 | % | -15 | → | 0 |
| 232 | → | 246 | 0.31958 | 20.43 | % | -13 | → | 0 |
| 232 | → | 247 | -0.2621 | 13.74 | % | -13 | → | 1 |
| 233 | → | 246 | 0.15749 | 4.96 | % | -12 | → | 0 |
| 233 | → | 247 | -0.11473 | 2.63 | % | -12 | → | 1 |
| 234 | → | 246 | -0.29431 | 17.32 | % | -11 | → | 0 |
| 234 | → | 247 | 0.22383 | 10.02 | % | -11 | → | 1 |
| 235 | → | 246 | 0.14651 | 4.29 | % | -10 | → | 0 |
| 235 | → | 247 | 0.12334 | 3.04 | % | -10 | → | 1 |
| 236 | → | 247 | -0.18087 | 6.54 | % | -9 | → | 1 |
| | 26 | 26 | Excited | State: | Singlet | | | |
| | 3.5641 | eV | 347.87 | nm | 28746.cm$^{-1}$ | f= | 0.0005 | |
| 232 | → | 246 | 0.17145 | 5.88 | % | -13 | → | 0 |
| 234 | → | 247 | -0.25110 | 12.61 | % | -11 | → | 1 |
| 236 | → | 246 | 0.17840 | 6.37 | % | -9 | → | 0 |
| 243 | → | 248 | 0.50554 | 51.11 | % | -2 | → | 2 |
| 243 | → | 249 | -0.22373 | 10.01 | % | -2 | → | 3 |
| | 27 | 27 | Excited | State: | Singlet | | | |
| | 3.577 | eV | 346.61 | nm | 28851.cm$^{-1}$ | f= | 0.0789 | |
| 232 | → | 246 | -0.26232 | 13.76 | % | -13 | → | 0 |
| 234 | → | 247 | 0.27285 | 14.89 | % | -11 | → | 1 |
| 235 | → | 246 | -0.24475 | 11.98 | % | -10 | → | 0 |
| 236 | → | 246 | -0.30261 | 18.31 | % | -9 | → | 0 |
| 243 | → | 248 | 0.33614 | 22.6 | % | -2 | → | 2 |
| 243 | → | 249 | -0.15033 | 4.52 | % | -2 | → | 3 |
| 245 | → | 249 | 0.10818 | 2.34 | % | 0 | → | 3 |
| | 28 | 28 | Excited | State: | Singlet | | | |
| | 3.5981 | eV | 344.58 | nm | 29021.cm$^{-1}$ | f= | 0.1812 | |
| 232 | → | 247 | 0.1124 | 2.53 | % | -13 | → | 1 |
| 234 | → | 246 | 0.12915 | 3.34 | % | -11 | → | 0 |
| 234 | → | 247 | 0.11931 | 2.85 | % | -11 | → | 1 |
| 235 | → | 246 | 0.22328 | 9.97 | % | -10 | → | 0 |
| 236 | → | 246 | 0.12551 | 3.15 | % | -9 | → | 0 |
| 236 | → | 247 | 0.27569 | 15.2 | % | -9 | → | 1 |
| 245 | → | 249 | 0.50068 | 50.14 | % | 0 | → | 3 |
| | 29 | 29 | Excited | State: | Singlet | | | |
| | 3.6098 | eV | 343.47 | nm | 29115.cm$^{-1}$ | f= | 0.0017 | |
| 230 | → | 246 | 0.1229 | 3.02 | % | -15 | → | 0 |
| 232 | → | 247 | 0.30356 | 18.43 | % | -13 | → | 1 |
| 233 | → | 246 | 0.23056 | 10.63 | % | -12 | → | 0 |
| 234 | → | 246 | -0.23388 | 10.94 | % | -11 | → | 0 |
| 234 | → | 247 | -0.14844 | 4.41 | % | -11 | → | 1 |
| 235 | → | 246 | -0.19532 | 7.63 | % | -10 | → | 0 |
| 235 | → | 247 | 0.31544 | 19.9 | % | -10 | → | 1 |
| 236 | → | 246 | -0.18197 | 6.62 | % | -9 | → | 0 |
| 236 | → | 247 | 0.20092 | 8.07 | % | -9 | → | 1 |
| 243 | → | 248 | -0.11338 | 2.57 | % | -2 | → | 2 |
| | 30 | 30 | Excited | State: | Singlet | | | |
| | 3.633 | eV | 341.28 | nm | 29301.cm$^{-1}$ | f= | 0.0084 | |
| 229 | → | 246 | -0.10313 | 2.13 | % | -16 | → | 0 |
| 231 | → | 246 | 0.10215 | 2.09 | % | -14 | → | 0 |
| 231 | → | 247 | -0.10437 | 2.18 | % | -14 | → | 1 |
| 232 | → | 247 | -0.1199 | 2.88 | % | -13 | → | 1 |
| 233 | → | 246 | 0.14569 | 4.25 | % | -12 | → | 0 |
| 234 | → | 247 | -0.21327 | 9.1 | % | -11 | → | 1 |
| 235 | → | 246 | -0.20293 | 8.24 | % | -10 | → | 0 |
| 235 | → | 247 | -0.24241 | 11.75 | % | -10 | → | 1 |
| 236 | → | 246 | -0.12973 | 3.37 | % | -9 | → | 0 |
| 236 | → | 247 | -0.24093 | 11.61 | % | -9 | → | 1 |
| 245 | → | 249 | 0.35165 | 24.73 | % | 0 | → | 3 |
| 245 | → | 250 | -0.10616 | 2.25 | % | 0 | → | 4 |
| | 31 | 31 | Excited | State: | Singlet | | | |
| | 3.6514 | eV | 339.56 | nm | 29450.cm$^{-1}$ | f= | 0.0036 | |
| 239 | → | 248 | -0.11083 | 2.46 | % | -6 | → | 2 |
| 240 | → | 248 | 0.1604 | 5.15 | % | -5 | → | 2 |
| 240 | → | 249 | 0.1054 | 2.22 | % | -5 | → | 3 |
| 241 | → | 248 | -0.11016 | 2.43 | % | -4 | → | 2 |
| 242 | → | 248 | 0.32104 | 20.61 | % | -3 | → | 2 |
| 244 | → | 249 | 0.44257 | 39.17 | % | -1 | → | 3 |
| 244 | → | 250 | -0.10732 | 2.3 | % | -1 | → | 4 |
| 245 | → | 249 | 0.11514 | 2.65 | % | 0 | → | 3 |
| 245 | → | 250 | 0.19568 | 7.66 | % | 0 | → | 4 |
| | 32 | 32 | Excited | State: | Singlet | | | |
| | 3.6669 | eV | 338.11 | nm | 29576.cm$^{-1}$ | f= | 0.0584 | |
| 231 | → | 246 | 0.36863 | 27.18 | % | -14 | → | 0 |
| 231 | → | 247 | -0.31377 | 19.69 | % | -14 | → | 1 |
| 232 | → | 247 | -0.12708 | 3.23 | % | -13 | → | 1 |
| 233 | → | 246 | 0.15565 | 4.85 | % | -12 | → | 0 |
| 233 | → | 247 | -0.30389 | 18.47 | % | -12 | → | 1 |
| 234 | → | 246 | 0.13716 | 3.76 | % | -11 | → | 0 |
| 236 | → | 247 | 0.17065 | 5.82 | % | -9 | → | 1 |
| | 33 | 33 | Excited | State: | Singlet | | | |
| | 3.7054 | eV | 334.6 | nm | 29886.cm$^{-1}$ | f= | 0.0316 | |
| 233 | → | 247 | -0.14809 | 4.39 | % | -12 | → | 1 |
| 239 | → | 248 | 0.15215 | 4.63 | % | -6 | → | 2 |
| 239 | → | 249 | 0.12642 | 3.2 | % | -6 | → | 3 |
| 240 | → | 248 | -0.24531 | 12.04 | % | -5 | → | 2 |
| 240 | → | 249 | -0.10993 | 2.42 | % | -5 | → | 3 |
| 241 | → | 248 | 0.12558 | 3.15 | % | -4 | → | 2 |
| 242 | → | 248 | 0.37866 | 28.68 | % | -3 | → | 2 |
| 244 | → | 249 | 0.15761 | 4.97 | % | -1 | → | 3 |
| 244 | → | 250 | 0.17119 | 5.86 | % | -1 | → | 4 |
| 245 | → | 249 | -0.15855 | 5.03 | % | 0 | → | 3 |
| 245 | → | 250 | -0.29378 | 17.26 | % | 0 | → | 4 |
| | 34 | 34 | Excited | State: | Singlet | | | |
| | 3.7198 | eV | 333.31 | nm | 30002.cm$^{-1}$ | f= | 0.003 | |
| 241 | → | 248 | 0.12368 | 3.06 | % | -4 | → | 2 |
| 242 | → | 248 | -0.41099 | 33.78 | % | -3 | → | 2 |
| 242 | → | 249 | 0.10762 | 2.32 | % | -3 | → | 3 |
| 244 | → | 249 | 0.46345 | 42.96 | % | -1 | → | 3 |
| 245 | → | 250 | -0.13625 | 3.71 | % | 0 | → | 4 |
| | 35 | 35 | Excited | State: | Singlet | | | |
| | 3.7683 | eV | 329.02 | nm | 30393.cm$^{-1}$ | f= | 0.0721 | |
| 233 | → | 246 | -0.19835 | 7.87 | % | -12 | → | 0 |
| 233 | → | 247 | -0.20136 | 8.11 | % | -12 | → | 1 |
| 235 | → | 247 | 0.10895 | 2.37 | % | -10 | → | 1 |
| 241 | → | 248 | 0.43887 | 38.52 | % | -4 | → | 2 |
| 241 | → | 249 | -0.23173 | 10.74 | % | -4 | → | 3 |
| 245 | → | 249 | 0.14417 | 4.16 | % | 0 | → | 3 |
| 245 | → | 250 | 0.12108 | 2.93 | % | 0 | → | 4 |
| | 36 | 36 | Excited | State: | Singlet | | | |
| | 3.7972 | eV | 326.52 | nm | 30626.cm$^{-1}$ | f= | 0.1291 | |
| 227 | → | 246 | 0.15369 | 4.72 | % | -18 | → | 0 |
| 228 | → | 246 | -0.24588 | 12.09 | % | -17 | → | 0 |
| 228 | → | 247 | 0.17131 | 5.87 | % | -17 | → | 1 |
| 229 | → | 246 | 0.43227 | 37.37 | % | -16 | → | 0 |
| 234 | → | 246 | -0.15999 | 5.12 | % | -11 | → | 0 |
| 234 | → | 247 | -0.15275 | 4.67 | % | -11 | → | 1 |
| 241 | → | 248 | -0.21458 | 9.21 | % | -4 | → | 2 |
| | 37 | 37 | Excited | State: | Singlet | | | |
| | 3.815 | eV | 325 | nm | 30769.cm$^{-1}$ | f= | 0.2503 | |
| 227 | → | 246 | 0.1462 | 4.27 | % | -18 | → | 0 |
| 228 | → | 246 | -0.10025 | 2.01 | % | -17 | → | 0 |
| 229 | → | 246 | 0.32136 | 20.65 | % | -16 | → | 0 |
| 229 | → | 247 | -0.28841 | 16.64 | % | -16 | → | 1 |
| 230 | → | 247 | 0.11414 | 2.61 | % | -15 | → | 1 |
| 233 | → | 246 | 0.24206 | 11.72 | % | -12 | → | 0 |
| 233 | → | 247 | 0.1628 | 5.3 | % | -12 | → | 1 |
| 234 | → | 247 | 0.11116 | 2.47 | % | -11 | → | 1 |
| 241 | → | 248 | 0.29752 | 17.7 | % | -4 | → | 2 |
| 241 | → | 249 | -0.10245 | 2.1 | % | -4 | → | 3 |
| | 38 | 38 | Excited | State: | Singlet | | | |
| | 3.8416 | eV | 322.74 | nm | 30985.cm$^{-1}$ | f= | 0.0043 | |
| 228 | → | 246 | 0.34568 | 23.9 | % | -17 | → | 0 |
| 228 | → | 247 | 0.21542 | 9.28 | % | -17 | → | 1 |
| 229 | → | 247 | 0.23181 | 10.75 | % | -16 | → | 1 |
| 230 | → | 246 | -0.16693 | 5.57 | % | -15 | → | 0 |
| 230 | → | 247 | -0.15205 | 4.62 | % | -15 | → | 1 |
| 231 | → | 246 | 0.26656 | 14.21 | % | -14 | → | 0 |
| 231 | → | 247 | 0.21718 | 9.43 | % | -14 | → | 1 |
| 233 | → | 247 | 0.16453 | 5.41 | % | -12 | → | 1 |
| | 39 | 39 | Excited | State: | Singlet | | | |
| | 3.8472 | eV | 322.27 | nm | 31030.cm$^{-1}$ | f= | 0.4159 | |
| 228 | → | 246 | 0.328 | 21.52 | % | -17 | → | 0 |
| 228 | → | 247 | -0.16764 | 5.62 | % | -17 | → | 1 |
| 229 | → | 246 | 0.10847 | 2.35 | % | -16 | → | 0 |
| 229 | → | 247 | -0.17369 | 6.03 | % | -16 | → | 1 |
| 230 | → | 246 | -0.12256 | 3 | % | -15 | → | 0 |
| 230 | → | 247 | 0.15309 | 4.69 | % | -15 | → | 1 |
| 231 | → | 246 | 0.1557 | 4.85 | % | -14 | → | 0 |
| 231 | → | 247 | -0.23152 | 10.72 | % | -14 | → | 1 |
| 233 | → | 246 | -0.26512 | 14.06 | % | -12 | → | 0 |
| 234 | → | 246 | -0.16537 | 5.47 | % | -11 | → | 0 |
| 235 | → | 247 | 0.18136 | 6.58 | % | -10 | → | 1 |
| | 40 | 40 | Excited | State: | Singlet | | | |
| | 3.9011 | eV | 317.82 | nm | 31464.cm$^{-1}$ | f= | 0.3093 | |
| 228 | → | 246 | -0.22016 | 9.69 | % | -17 | → | 0 |
| 228 | → | 247 | 0.192 | 7.37 | % | -17 | → | 1 |
| 229 | → | 246 | -0.1708 | 5.83 | % | -16 | → | 0 |
| 230 | → | 246 | 0.22006 | 9.69 | % | -15 | → | 0 |
| 231 | → | 246 | 0.2639 | 13.93 | % | -14 | → | 0 |
| 231 | → | 247 | -0.13017 | 3.39 | % | -14 | → | 1 |
| 233 | → | 246 | -0.18052 | 6.52 | % | -12 | → | 0 |
| 233 | → | 247 | 0.32368 | 20.95 | % | -12 | → | 1 |
| 234 | → | 247 | 0.10233 | 2.09 | % | -11 | → | 1 |
| 235 | → | 246 | 0.13761 | 3.79 | % | -10 | → | 0 |
| 240 | → | 248 | 0.11594 | 2.69 | % | -5 | → | 2 |
| | 41 | 41 | Excited | State: | Singlet | | | |
| | 3.9454 | eV | 314.25 | nm | 31822.cm$^{-1}$ | f= | 0.1371 | |
| 227 | → | 247 | 0.1671 | 5.58 | % | -18 | → | 1 |
| 228 | → | 247 | -0.30919 | 19.12 | % | -17 | → | 1 |
| 229 | → | 247 | 0.3912 | 30.61 | % | -16 | → | 1 |
| 231 | → | 246 | -0.10641 | 2.26 | % | -14 | → | 0 |
| 231 | → | 247 | -0.24546 | 12.05 | % | -14 | → | 1 |
| 233 | → | 247 | 0.18437 | 6.8 | % | -12 | → | 1 |
| 242 | → | 249 | -0.17535 | 6.15 | % | -3 | → | 3 |
| 243 | → | 249 | 0.14202 | 4.03 | % | -2 | → | 3 |
| | 42 | 42 | Excited | State: | Singlet | | | |
| | 3.9613 | eV | 312.99 | nm | 31950.cm$^{-1}$ | f= | 0.0895 | |
| 228 | → | 247 | -0.20419 | 8.34 | % | -17 | → | 1 |
| 229 | → | 247 | 0.14629 | 4.28 | % | -16 | → | 1 |
| 233 | → | 247 | 0.1181 | 2.79 | % | -12 | → | 1 |
| 240 | → | 248 | -0.23424 | 10.97 | % | -5 | → | 2 |
| 242 | → | 249 | 0.41336 | 34.17 | % | -3 | → | 3 |
| 243 | → | 248 | -0.11758 | 2.77 | % | -2 | → | 2 |
| 243 | → | 249 | -0.33901 | 22.99 | % | -2 | → | 3 |
| | 43 | 43 | Excited | State: | Singlet | | | |

|     |     |     | 3.9799   | eV | 311.53   | nm      | 32100.cm$^{-1}$ | f= | 0.2322 |   |
|-----|-----|-----|----------|----|----------|---------|-----------------|----|--------|---|
| 228 | →   | 246 | -0.20374 |    | 8.3      | %       | -17             | →  | 0      |   |
| 228 | →   | 247 | -0.30348 |    | 18.42    | %       | -17             | →  | 1      |   |
| 230 | →   | 247 | 0.11296  |    | 2.55     | %       | -15             | →  | 1      |   |
| 231 | →   | 246 | 0.32315  |    | 20.89    | %       | -14             | →  | 0      |   |
| 231 | →   | 247 | 0.36593  |    | 26.78    | %       | -14             | →  | 1      |   |
| 233 | →   | 247 | -0.13298 |    | 3.54     | %       | -12             | →  | 1      |   |
| 240 | →   | 248 | 0.14232  |    | 4.05     | %       | -5              | →  | 2      |   |

|     |     |     | 44       | 44 | Excited  | State:  | Singlet         |    |        |   |
|-----|-----|-----|----------|----|----------|---------|-----------------|----|--------|---|
|     |     |     | 3.9991   | eV | 310.03   | nm      | 32255.cm$^{-1}$ | f= | 0.0258 |   |
| 226 | →   | 246 | 0.10625  |    | 2.26     | %       | -19             | →  | 0      |   |
| 239 | →   | 248 | 0.20207  |    | 8.17     | %       | -6              | →  | 2      |   |
| 240 | →   | 248 | 0.47322  |    | 44.79    | %       | -5              | →  | 2      |   |
| 241 | →   | 249 | -0.11496 |    | 2.64     | %       | -4              | →  | 3      |   |
| 242 | →   | 249 | 0.20303  |    | 8.24     | %       | -3              | →  | 3      |   |
| 243 | →   | 248 | -0.10368 |    | 2.15     | %       | -2              | →  | 2      |   |
| 243 | →   | 249 | -0.23299 |    | 10.86    | %       | -2              | →  | 3      |   |
| 244 | →   | 250 | 0.10757  |    | 2.31     | %       | -1              | →  | 4      |   |
| 245 | →   | 250 | -0.17711 |    | 6.27     | %       | 0               | →  | 4      |   |

|     |     |     | 45       | 45 | Excited  | State:  | Singlet         |    |        |   |
|-----|-----|-----|----------|----|----------|---------|-----------------|----|--------|---|
|     |     |     | 4.0757   | eV | 304.2    | nm      | 32873.cm$^{-1}$ | f= | 0.088  |   |
| 224 | →   | 246 | -0.10431 |    | 2.18     | %       | -21             | →  | 0      |   |
| 226 | →   | 246 | 0.30333  |    | 18.4     | %       | -19             | →  | 0      |   |
| 226 | →   | 247 | -0.20447 |    | 8.36     | %       | -19             | →  | 1      |   |
| 228 | →   | 246 | -0.14214 |    | 4.04     | %       | -17             | →  | 0      |   |
| 228 | →   | 247 | 0.10643  |    | 2.27     | %       | -17             | →  | 1      |   |
| 229 | →   | 246 | -0.12311 |    | 3.03     | %       | -16             | →  | 0      |   |
| 230 | →   | 246 | -0.29116 |    | 16.95    | %       | -15             | →  | 0      |   |
| 230 | →   | 247 | 0.19753  |    | 7.8      | %       | -15             | →  | 1      |   |
| 242 | →   | 249 | 0.21597  |    | 9.33     | %       | -3              | →  | 3      |   |
| 243 | →   | 248 | 0.13822  |    | 3.82     | %       | -2              | →  | 2      |   |
| 243 | →   | 249 | 0.21311  |    | 9.08     | %       | -2              | →  | 3      |   |

|     |     |     | 46       | 46 | Excited  | State:  | Singlet         |    |        |   |
|-----|-----|-----|----------|----|----------|---------|-----------------|----|--------|---|
|     |     |     | 4.0875   | eV | 303.32   | nm      | 32968.cm$^{-1}$ | f= | 0.0232 |   |
| 227 | →   | 246 | 0.51178  |    | 52.38    | %       | -18             | →  | 0      |   |
| 227 | →   | 247 | 0.32114  |    | 20.63    | %       | -18             | →  | 1      |   |
| 228 | →   | 246 | 0.11832  |    | 2.8      | %       | -17             | →  | 0      |   |
| 230 | →   | 246 | 0.18216  |    | 6.64     | %       | -15             | →  | 0      |   |
| 242 | →   | 249 | 0.12464  |    | 3.11     | %       | -3              | →  | 3      |   |

|     |     |     | 47       | 47 | Excited  | State:  | Singlet         |    |        |   |
|-----|-----|-----|----------|----|----------|---------|-----------------|----|--------|---|
|     |     |     | 4.0975   | eV | 302.58   | nm      | 33049.cm$^{-1}$ | f= | 0.1033 |   |
| 226 | →   | 246 | -0.14828 |    | 4.4      | %       | -19             | →  | 0      |   |
| 226 | →   | 247 | 0.1614   |    | 5.21     | %       | -19             | →  | 1      |   |
| 227 | →   | 246 | -0.12628 |    | 3.19     | %       | -18             | →  | 0      |   |
| 230 | →   | 247 | -0.16799 |    | 5.64     | %       | -15             | →  | 1      |   |
| 242 | →   | 248 | 0.12245  |    | 3        | %       | -3              | →  | 2      |   |

| | | | | | | | | |
|---|---|---|---|---|---|---|---|---|
| 242 | → | 249 | 0.36464 | 26.59 | % | -3 | → | 3 |
| 243 | → | 248 | 0.18635 | 6.95 | % | -2 | → | 2 |
| 243 | → | 249 | 0.38255 | 29.27 | % | -2 | → | 3 |
| | | 48 | 48 | Excited | State: | Singlet | | |
| | 4.1497 | eV | 298.78 | nm | 33469.cm$^{-1}$ | f= | 0.048 | |
| 224 | → | 246 | -0.11109 | 2.47 | % | -21 | → | 0 |
| 226 | → | 246 | 0.14092 | 3.97 | % | -19 | → | 0 |
| 226 | → | 247 | -0.31717 | 20.12 | % | -19 | → | 1 |
| 227 | → | 246 | -0.14117 | 3.99 | % | -18 | → | 0 |
| 227 | → | 247 | -0.11573 | 2.68 | % | -18 | → | 1 |
| 228 | → | 246 | 0.13366 | 3.57 | % | -17 | → | 0 |
| 229 | → | 246 | 0.21897 | 9.59 | % | -16 | → | 0 |
| 229 | → | 247 | 0.12772 | 3.26 | % | -16 | → | 1 |
| 230 | → | 246 | 0.42393 | 35.94 | % | -15 | → | 0 |
| 230 | → | 247 | 0.11159 | 2.49 | % | -15 | → | 1 |
| 233 | → | 246 | -0.10886 | 2.37 | % | -12 | → | 0 |
| | | 49 | 49 | Excited | State: | Singlet | | |
| | 4.1748 | eV | 296.98 | nm | 33672.cm$^{-1}$ | f= | 0.0705 | |
| 224 | → | 246 | 0.10389 | 2.16 | % | -21 | → | 0 |
| 225 | → | 246 | -0.1243 | 3.09 | % | -20 | → | 0 |
| 226 | → | 246 | -0.38051 | 28.96 | % | -19 | → | 0 |
| 228 | → | 247 | 0.2007 | 8.06 | % | -17 | → | 1 |
| 229 | → | 247 | 0.13908 | 3.87 | % | -16 | → | 1 |
| 230 | → | 247 | 0.43408 | 37.69 | % | -15 | → | 1 |
| | | 50 | 50 | Excited | State: | Singlet | | |
| | 4.1829 | eV | 296.41 | nm | 33737.cm$^{-1}$ | f= | 0.0077 | |
| 230 | → | 247 | 0.12446 | 3.1 | % | -15 | → | 1 |
| 240 | → | 248 | 0.14163 | 4.01 | % | -5 | → | 2 |
| 240 | → | 249 | -0.28544 | 16.3 | % | -5 | → | 3 |
| 241 | → | 248 | 0.2212 | 9.79 | % | -4 | → | 2 |
| 241 | → | 249 | 0.54081 | 58.5 | % | -4 | → | 3 |
| | | 51 | 51 | Excited | State: | Singlet | | |
| | 4.2547 | eV | 291.4 | nm | 34317.cm$^{-1}$ | f= | 0.011 | |
| 224 | → | 246 | -0.14434 | 4.17 | % | -21 | → | 0 |
| 224 | → | 247 | -0.1841 | 6.78 | % | -21 | → | 1 |
| 226 | → | 246 | 0.32393 | 20.99 | % | -19 | → | 0 |
| 226 | → | 247 | 0.45871 | 42.08 | % | -19 | → | 1 |
| 229 | → | 246 | 0.10194 | 2.08 | % | -16 | → | 0 |
| 230 | → | 246 | 0.12548 | 3.15 | % | -15 | → | 0 |
| 230 | → | 247 | 0.18519 | 6.86 | % | -15 | → | 1 |
| | | 52 | 52 | Excited | State: | Singlet | | |
| | 4.273 | eV | 290.15 | nm | 34465.cm$^{-1}$ | f= | 0.016 | |
| 227 | → | 247 | 0.18268 | 6.67 | % | -18 | → | 1 |
| 229 | → | 247 | -0.1041 | 2.17 | % | -16 | → | 1 |
| 239 | → | 249 | 0.11201 | 2.51 | % | -6 | → | 3 |
| 240 | → | 249 | 0.53526 | 57.3 | % | -5 | → | 3 |
| 241 | → | 248 | 0.15578 | 4.85 | % | -4 | → | 2 |
| 241 | → | 249 | 0.27919 | 15.59 | % | -4 | → | 3 |
| | 53 | 53 | Excited | State: | Singlet | | | |
| | 4.2992 | eV | 288.39 | nm | 34675.cm$^{-1}$ | f= | 0.0005 | |
| 223 | → | 246 | -0.11558 | 2.67 | % | -22 | → | 0 |
| 226 | → | 247 | -0.12693 | 3.22 | % | -19 | → | 1 |
| 227 | → | 246 | -0.29499 | 17.4 | % | -18 | → | 0 |
| 227 | → | 247 | 0.47573 | 45.26 | % | -18 | → | 1 |
| 229 | → | 246 | 0.11689 | 2.73 | % | -16 | → | 0 |
| 229 | → | 247 | -0.20117 | 8.09 | % | -16 | → | 1 |
| 240 | → | 249 | -0.20434 | 8.35 | % | -5 | → | 3 |
| | 54 | 54 | Excited | State: | Singlet | | | |
| | 4.3505 | eV | 284.99 | nm | 35089.cm$^{-1}$ | f= | 0.0024 | |
| 244 | → | 251 | -0.33849 | 22.92 | % | -1 | → | 5 |
| 245 | → | 251 | 0.59913 | 71.79 | % | 0 | → | 5 |
| | 55 | 55 | Excited | State: | Singlet | | | |
| | 4.3721 | eV | 283.58 | nm | 35263.cm$^{-1}$ | f= | 0.0293 | |
| 223 | → | 246 | 0.21068 | 8.88 | % | -22 | → | 0 |
| 224 | → | 246 | 0.17011 | 5.79 | % | -21 | → | 0 |
| 224 | → | 247 | 0.126 | 3.18 | % | -21 | → | 1 |
| 225 | → | 246 | 0.52697 | 55.54 | % | -20 | → | 0 |
| 225 | → | 247 | 0.24168 | 11.68 | % | -20 | → | 1 |
| | 56 | 56 | Excited | State: | Singlet | | | |
| | 4.3951 | eV | 282.09 | nm | 35450.cm$^{-1}$ | f= | 0.0189 | |
| 239 | → | 248 | 0.16811 | 5.65 | % | -6 | → | 2 |
| 242 | → | 250 | 0.10448 | 2.18 | % | -3 | → | 4 |
| 244 | → | 250 | 0.49806 | 49.61 | % | -1 | → | 4 |
| 245 | → | 250 | 0.43704 | 38.2 | % | 0 | → | 4 |
| | 57 | 57 | Excited | State: | Singlet | | | |
| | 4.4223 | eV | 280.36 | nm | 35668.cm$^{-1}$ | f= | 0.0077 | |
| 223 | → | 246 | -0.32745 | 21.44 | % | -22 | → | 0 |
| 223 | → | 247 | 0.29249 | 17.11 | % | -22 | → | 1 |
| 224 | → | 246 | 0.10817 | 2.34 | % | -21 | → | 0 |
| 225 | → | 247 | 0.39476 | 31.17 | % | -20 | → | 1 |
| 227 | → | 247 | -0.18178 | 6.61 | % | -18 | → | 1 |
| 239 | → | 248 | -0.1246 | 3.11 | % | -6 | → | 2 |
| 244 | → | 250 | 0.10724 | 2.3 | % | -1 | → | 4 |
| | 58 | 58 | Excited | State: | Singlet | | | |
| | 4.4637 | eV | 277.76 | nm | 36002.cm$^{-1}$ | f= | 0.1527 | |
| 223 | → | 246 | -0.12106 | 2.93 | % | -22 | → | 0 |
| 239 | → | 248 | 0.49181 | 48.38 | % | -6 | → | 2 |
| 244 | → | 250 | -0.3603 | 25.96 | % | -1 | → | 4 |
| 245 | → | 250 | 0.19879 | 7.9 | % | 0 | → | 4 |
|  | 59 | 59 | Excited | State: | Singlet |  |  |  |
|  | 4.5262 | eV | 273.92 | nm | 36507.cm$^{-1}$ | f= | 0.0019 |  |
| 222 | → | 246 | -0.24955 | 12.46 | % | -23 | → | 0 |
| 223 | → | 246 | 0.38981 | 30.39 | % | -22 | → | 0 |
| 223 | → | 247 | 0.28781 | 16.57 | % | -22 | → | 1 |
| 224 | → | 246 | -0.25051 | 12.55 | % | -21 | → | 0 |
| 225 | → | 246 | -0.13494 | 3.64 | % | -20 | → | 0 |
| 225 | → | 247 | 0.21733 | 9.45 | % | -20 | → | 1 |
|  | 60 | 60 | Excited | State: | Singlet |  |  |  |
|  | 4.5584 | eV | 271.99 | nm | 36766.cm$^{-1}$ | f= | 0.0097 |  |
| 222 | → | 246 | 0.1752 | 6.14 | % | -23 | → | 0 |
| 222 | → | 247 | -0.29194 | 17.05 | % | -23 | → | 1 |
| 223 | → | 247 | 0.3372 | 22.74 | % | -22 | → | 1 |
| 224 | → | 246 | 0.10309 | 2.13 | % | -21 | → | 0 |
| 224 | → | 247 | -0.29773 | 17.73 | % | -21 | → | 1 |
| 225 | → | 246 | 0.10769 | 2.32 | % | -20 | → | 0 |
| 225 | → | 247 | -0.17158 | 5.89 | % | -20 | → | 1 |
| 238 | → | 248 | -0.12114 | 2.93 | % | -7 | → | 2 |
| 239 | → | 249 | 0.12262 | 3.01 | % | -6 | → | 3 |
|  | 61 | 61 | Excited | State: | Singlet |  |  |  |
|  | 4.5949 | eV | 269.83 | nm | 37060.cm$^{-1}$ | f= | 0.0049 |  |
| 222 | → | 246 | 0.30758 | 18.92 | % | -23 | → | 0 |
| 222 | → | 247 | 0.16372 | 5.36 | % | -23 | → | 1 |
| 223 | → | 246 | 0.22923 | 10.51 | % | -22 | → | 0 |
| 223 | → | 247 | -0.10307 | 2.12 | % | -22 | → | 1 |
| 225 | → | 246 | -0.2672 | 14.28 | % | -20 | → | 0 |
| 225 | → | 247 | 0.16869 | 5.69 | % | -20 | → | 1 |
| 235 | → | 248 | -0.11688 | 2.73 | % | -10 | → | 2 |
| 236 | → | 248 | -0.20876 | 8.72 | % | -9 | → | 2 |
| 236 | → | 249 | -0.10854 | 2.36 | % | -9 | → | 3 |
| 237 | → | 248 | 0.13647 | 3.72 | % | -8 | → | 2 |
| 239 | → | 249 | 0.20507 | 8.41 | % | -6 | → | 3 |
|  | 62 | 62 | Excited | State: | Singlet |  |  |  |
|  | 4.6099 | eV | 268.95 | nm | 37182.cm$^{-1}$ | f= | 0.0051 |  |
| 222 | → | 246 | 0.17781 | 6.32 | % | -23 | → | 0 |
| 222 | → | 247 | 0.11607 | 2.69 | % | -23 | → | 1 |
| 225 | → | 247 | -0.17262 | 5.96 | % | -20 | → | 1 |
| 238 | → | 248 | 0.51794 | 53.65 | % | -7 | → | 2 |
| 238 | → | 249 | -0.25195 | 12.7 | % | -7 | → | 3 |
| 239 | → | 249 | -0.10279 | 2.11 | % | -6 | → | 3 |
|  | 63 | 63 | Excited | State: | Singlet |  |  |  |
|  | 4.6177 | eV | 268.49 | nm | 37245.cm$^{-1}$ | f= | 0.0057 |  |
| 222 | → | 246 | 0.32055 | 20.55 | % | -23 | → | 0 |
| 222 | → | 247 | 0.13371 | 3.58 | % | -23 | → | 1 |
| 223 | → | 246 | 0.11142 | 2.48 | % | -22 | → | 0 |
| 223 | → | 247 | 0.14084 | 3.97 | % | -22 | → | 1 |
| 235 | → | 248 | 0.1812 | 6.57 | % | -10 | → | 2 |
| 235 | → | 249 | 0.12948 | 3.35 | % | -10 | → | 3 |
| 236 | → | 248 | 0.30075 | 18.09 | % | -9 | → | 2 |
| 236 | → | 249 | 0.17706 | 6.27 | % | -9 | → | 3 |
| 238 | → | 248 | -0.13303 | 3.54 | % | -7 | → | 2 |
| 239 | → | 248 | 0.1438 | 4.14 | % | -6 | → | 2 |
| 239 | → | 249 | -0.25951 | 13.47 | % | -6 | → | 3 |
| | 64 | 64 | Excited | State: | Singlet | | | |
| | 4.6628 | eV | 265.9 | nm | 37608.cm$^{-1}$ | f= | 0.0157 | |
| 222 | → | 247 | -0.16126 | 5.2 | % | -23 | → | 1 |
| 223 | → | 247 | -0.16526 | 5.46 | % | -22 | → | 1 |
| 225 | → | 247 | 0.10572 | 2.24 | % | -20 | → | 1 |
| 235 | → | 248 | 0.20384 | 8.31 | % | -10 | → | 2 |
| 235 | → | 249 | 0.11897 | 2.83 | % | -10 | → | 3 |
| 236 | → | 248 | 0.27141 | 14.73 | % | -9 | → | 2 |
| 236 | → | 249 | 0.15953 | 5.09 | % | -9 | → | 3 |
| 238 | → | 248 | 0.19543 | 7.64 | % | -7 | → | 2 |
| 238 | → | 249 | -0.13904 | 3.87 | % | -7 | → | 3 |
| 239 | → | 249 | 0.31688 | 20.08 | % | -6 | → | 3 |
| 240 | → | 250 | -0.11839 | 2.8 | % | -5 | → | 4 |
| 242 | → | 250 | 0.12547 | 3.15 | % | -3 | → | 4 |
| | 65 | 65 | Excited | State: | Singlet | | | |
| | 4.687 | eV | 264.53 | nm | 37803.cm$^{-1}$ | f= | 0.0143 | |
| 215 | → | 247 | 0.10914 | 2.38 | % | -30 | → | 1 |
| 222 | → | 246 | -0.12594 | 3.17 | % | -23 | → | 0 |
| 222 | → | 247 | 0.38196 | 29.18 | % | -23 | → | 1 |
| 223 | → | 247 | 0.23155 | 10.72 | % | -22 | → | 1 |
| 225 | → | 246 | 0.10076 | 2.03 | % | -20 | → | 0 |
| 225 | → | 247 | -0.19447 | 7.56 | % | -20 | → | 1 |
| 239 | → | 249 | 0.23539 | 11.08 | % | -6 | → | 3 |
| 240 | → | 250 | -0.10853 | 2.36 | % | -5 | → | 4 |
| 242 | → | 250 | 0.23523 | 11.07 | % | -3 | → | 4 |
| 243 | → | 250 | -0.17965 | 6.45 | % | -2 | → | 4 |
| | 66 | 66 | Excited | State: | Singlet | | | |
| | 4.7078 | eV | 263.36 | nm | 37971.cm$^{-1}$ | f= | 0.0285 | |
| 222 | → | 247 | -0.1497 | 4.48 | % | -23 | → | 1 |
| 239 | → | 249 | -0.25753 | 13.26 | % | -6 | → | 3 |
| 242 | → | 250 | 0.46898 | 43.99 | % | -3 | → | 4 |
| 243 | → | 250 | -0.36848 | 27.16 | % | -2 | → | 4 |
| | 67 | 67 | Excited | State: | Singlet | | | |
| | 4.7755 | eV | 259.63 | nm | 38516.cm$^{-1}$ | f= | 0.0006 | |
| 222 | → | 247 | -0.1144 | 2.62 | % | -23 | → | 1 |
| 224 | → | 246 | -0.3067 | 18.81 | % | -21 | → | 0 |
| 225 | → | 246 | 0.15187 | 4.61 | % | -20 | → | 0 |
| 226 | → | 246 | -0.14207 | 4.04 | % | -19 | → | 0 |
| 237 | → | 248 | 0.49303 | 48.62 | % | -8 | → | 2 |
| 237 | → | 249 | -0.19798 | 7.84 | % | -8 | → | 3 |
| | 68 | 68 | Excited | State: | Singlet | | | |
| | 4.7917 | eV | 258.75 | nm | 38647.cm$^{-1}$ | f= | 0.0033 | |
| 222 | → | 246 | -0.22876 | 10.47 | % | -23 | → | 0 |
| 224 | → | 246 | 0.38696 | 29.95 | % | -21 | → | 0 |
| 224 | → | 247 | -0.19227 | 7.39 | % | -21 | → | 1 |
| 225 | → | 246 | -0.12497 | 3.12 | % | -20 | → | 0 |
| 226 | → | 246 | 0.16447 | 5.41 | % | -19 | → | 0 |
| 237 | → | 248 | 0.33921 | 23.01 | % | -8 | → | 2 |
| 237 | → | 249 | -0.13993 | 3.92 | % | -8 | → | 3 |
| 239 | → | 249 | -0.14365 | 4.13 | % | -6 | → | 3 |
| | 69 | 69 | Excited | State: | Singlet | | | |
| | 4.8282 | eV | 256.79 | nm | 38942.cm$^{-1}$ | f= | 0.0245 | |
| 222 | → | 247 | -0.22719 | 10.32 | % | -23 | → | 1 |
| 223 | → | 247 | 0.11423 | 2.61 | % | -22 | → | 1 |
| 224 | → | 246 | 0.17363 | 6.03 | % | -21 | → | 0 |
| 224 | → | 247 | 0.46127 | 42.55 | % | -21 | → | 1 |
| 225 | → | 247 | -0.21501 | 9.25 | % | -20 | → | 1 |
| 226 | → | 247 | 0.1986 | 7.89 | % | -19 | → | 1 |
| 240 | → | 250 | 0.11197 | 2.51 | % | -5 | → | 4 |
| | 70 | 70 | Excited | State: | Singlet | | | |
| | 4.8397 | eV | 256.18 | nm | 39035.cm$^{-1}$ | f= | 0.2106 | |
| 224 | → | 247 | -0.11464 | 2.63 | % | -21 | → | 1 |
| 239 | → | 249 | 0.14249 | 4.06 | % | -6 | → | 3 |
| 240 | → | 250 | 0.54278 | 58.92 | % | -5 | → | 4 |
| 241 | → | 250 | -0.33759 | 22.79 | % | -4 | → | 4 |
| | 71 | 71 | Excited | State: | Singlet | | | |
| | 4.8593 | eV | 255.15 | nm | 39193.cm$^{-1}$ | f= | 0.0165 | |
| 242 | → | 250 | 0.37301 | 27.83 | % | -3 | → | 4 |
| 243 | → | 250 | 0.49408 | 48.82 | % | -2 | → | 4 |
| 243 | → | 252 | -0.18031 | 6.5 | % | -2 | → | 6 |
| 243 | → | 253 | -0.10172 | 2.07 | % | -2 | → | 7 |
| 243 | → | 254 | -0.15957 | 5.09 | % | -2 | → | 8 |
| | 72 | 72 | Excited | State: | Singlet | | | |
| | 4.8838 | eV | 253.87 | nm | 39390.cm$^{-1}$ | f= | 0.0274 | |
| 242 | → | 250 | 0.17568 | 6.17 | % | -3 | → | 4 |
| 243 | → | 250 | 0.26733 | 14.29 | % | -2 | → | 4 |
| 243 | → | 252 | 0.31296 | 19.59 | % | -2 | → | 6 |
| 243 | → | 253 | 0.17012 | 5.79 | % | -2 | → | 7 |
| 243 | → | 254 | 0.28728 | 16.51 | % | -2 | → | 8 |
| 243 | → | 257 | 0.10769 | 2.32 | % | -2 | → | 11 |
| 244 | → | 251 | -0.12539 | 3.14 | % | -1 | → | 5 |
| 244 | → | 253 | 0.12409 | 3.08 | % | -1 | → | 7 |
| 245 | → | 253 | 0.10596 | 2.25 | % | 0 | → | 7 |

| | 73 | | 73 | | Excited | State: | Singlet | | |
|---|---|---|---|---|---|---|---|---|---|
| | 4.9283 | | eV | | 251.58 | nm | 39749.cm$^{-1}$ | f= | 0.0213 |
| 242 | → | 251 | 0.15555 | 4.84 | % | | -3 | → | 5 |
| 242 | → | 252 | 0.10575 | 2.24 | % | | -3 | → | 6 |
| 242 | → | 253 | 0.21964 | 9.65 | % | | -3 | → | 7 |
| 242 | → | 255 | 0.10524 | 2.22 | % | | -3 | → | 9 |
| 242 | → | 256 | 0.11779 | 2.77 | % | | -3 | → | 10 |
| 243 | → | 252 | 0.1561 | 4.87 | % | | -2 | → | 6 |
| 243 | → | 254 | 0.11782 | 2.78 | % | | -2 | → | 8 |
| 244 | → | 251 | 0.41576 | 34.57 | % | | -1 | → | 5 |
| 245 | → | 251 | 0.23748 | 11.28 | % | | 0 | → | 5 |
| 245 | → | 255 | 0.11363 | 2.58 | % | | 0 | → | 9 |
| | 74 | | 74 | | Excited | State: | Singlet | | |
| | 4.9463 | | eV | | 250.66 | nm | 39895.cm$^{-1}$ | f= | 0.0313 |
| 220 | → | 246 | 0.22935 | 10.52 | % | | -25 | → | 0 |
| 220 | → | 247 | -0.15755 | 4.96 | % | | -25 | → | 1 |
| 221 | → | 247 | -0.10737 | 2.31 | % | | -24 | → | 1 |
| 234 | → | 248 | -0.16245 | 5.28 | % | | -11 | → | 2 |
| 241 | → | 254 | -0.21415 | 9.17 | % | | -4 | → | 8 |
| 242 | → | 253 | 0.22336 | 9.98 | % | | -3 | → | 7 |
| 242 | → | 256 | 0.12193 | 2.97 | % | | -3 | → | 10 |
| 243 | → | 252 | -0.11298 | 2.55 | % | | -2 | → | 6 |
| 243 | → | 253 | -0.18124 | 6.57 | % | | -2 | → | 7 |
| 244 | → | 252 | -0.19634 | 7.71 | % | | -1 | → | 6 |
| 245 | → | 252 | -0.10454 | 2.19 | % | | 0 | → | 6 |
| 245 | → | 253 | 0.10452 | 2.18 | % | | 0 | → | 7 |
| | 75 | | 75 | | Excited | State: | Singlet | | |
| | 4.962 | | eV | | 249.86 | nm | 40022.cm$^{-1}$ | f= | 0.0192 |
| 217 | → | 246 | 0.12353 | 3.05 | % | | -28 | → | 0 |
| 217 | → | 247 | 0.12557 | 3.15 | % | | -28 | → | 1 |
| 219 | → | 246 | 0.10575 | 2.24 | % | | -26 | → | 0 |
| 220 | → | 246 | 0.33437 | 22.36 | % | | -25 | → | 0 |
| 220 | → | 247 | -0.19165 | 7.35 | % | | -25 | → | 1 |
| 221 | → | 246 | 0.12912 | 3.33 | % | | -24 | → | 0 |
| 221 | → | 247 | -0.14621 | 4.28 | % | | -24 | → | 1 |
| 241 | → | 254 | 0.3063 | 18.76 | % | | -4 | → | 8 |
| 243 | → | 253 | 0.1149 | 2.64 | % | | -2 | → | 7 |
| 244 | → | 252 | 0.16192 | 5.24 | % | | -1 | → | 6 |

**References**

[1] [1] A. C. Ferrari, D. M. Basko, *Nat. Nanotech.* **2013**, *8*, 235.

[2] [2] A.C. Ferrari, J. C. Meyer, V. Scardaci, C. Casiraghi, M. Lazzeri, F. Mauri, S. Piscanec, D. Jiang, K. S. Novoselov, S. Roth, A. K. Geim, *Phys. Rev. Lett.* **2006**, *97*, 187401.

[3] [3] A. C. Ferrari, J. Robertson, *Phys. Rev. B* **2001**, *64*, 075414.

[4] [4] A. C. Ferrari, J. Robertson, *Phys. Rev. B* **2000**, *61*, 14095.

[5] [5] H. Kataura, Y. Kumazawa, Y. Maniwa, Y. Ohtsuka, R. Sen, S. Suzuki, Y. Achiba, *Carbon* **2000**, *38*, 1691.

[6] [6] M. Takase, H. Ajiki, Y. Mizumoto, K. Komeda, M. Nara, H. Nabika, S. Yasuda, H. Ishihara, K. Murakoshi, *Nat. Photon.* **2013**, *7*, 550.

[7] [7] M. S. Strano, *J. Am. Chem. Soc*. **2003**, *125*, 16148.

[8] [8] R. B Weisman, S. M. Bachilo, *Nano Lett*. **2003**, *3*, 1235.

[9] [9] J.S. Soares, L.G. Cancado, E.B. Barros, A. Jorio, *Physica Status Solidi B*, **2010**, *247*, 2835.